\documentclass[twocolumn,preprintnumbers]{aastex63} 
\usepackage{afterpage}
\usepackage{capt-of}
\usepackage{diagbox}

\usepackage[figuresright]{rotating}
\usepackage{amsmath,natbib}
\usepackage{amssymb,amsmath,latexsym,graphics, graphicx,epsfig,multirow,comment,appendix,feyn,slashed,xcolor,afterpage, makecell} 
\usepackage{booktabs}
\usepackage{tabularx}

\newcommand{\dm}{\text{DM} }
\newcommand{\g}{\gamma}
\newcommand{\msun}{M_{\odot}}
\newcommand{\jfac}{$J$-factor}
\newcommand{\nstars}{n_\mathrm{stars}}
\newcommand{\nstarssel}{n_\mathrm{stars}^\mathrm{sel}}

\newcommand {\be} {\begin {equation}}
\newcommand {\ee} {\end {equation}} 

\newcommand {\bes} {\begin {equation*}}
\newcommand {\ees} {\end {equation*}}

\newcolumntype{L}[1]{>{\raggedright\let\newline\\\arraybackslash\hspace{0pt}}m{#1}}
\newcolumntype{C}[1]{>{\centering\let\newline\\\arraybackslash\hspace{0pt}}m{#1}}
\newcolumntype{R}[1]{>{\raggedleft\let\newline\\\arraybackslash\hspace{0pt}}m{#1}}

\newcommand{\es}[2] {\begin{equation} \label{#1} \begin{split} #2 \end{split} \end{equation}}

\makeatletter
\newcommand\footnoteref[1]{\protected@xdef\@thefnmark{\ref{#1}}\@footnotemark}
\makeatother

\usepackage{xspace}

\DeclareRobustCommand{\Sec}[1]{Sec.~\ref{#1}}
\DeclareRobustCommand{\Secs}[2]{Secs.~\ref{#1} and \ref{#2}}

\DeclareRobustCommand{\Tab}[1]{Table~\ref{#1}}

\DeclareRobustCommand{\Fig}[1]{Fig.~\ref{#1}}

\DeclareRobustCommand{\Eq}[1]{Eq.~(\ref{#1})}
\DeclareRobustCommand{\Eqs}[2]{Eqs.~(\ref{#1}) and (\ref{#2})}

\DeclareRobustCommand{\errorbars}[3]{{#1}^{+#2}_{-#3}}

\newcommand{\beq}{\begin{equation}}
\newcommand{\eeq}{\end{equation}}

\newcommand {\p}{\partial}
\newcommand {\ov}{\overline}

\begin{document}

\title{
Dark Matter Density Profiles in Dwarf Galaxies: \\
Linking Jeans Modeling Systematics and Observation.
}

\author{Laura J. Chang}
\affil{Department of Physics, Princeton University, Princeton, NJ 08544}

\author{Lina Necib}
\affil{Walter Burke Institute for Theoretical Physics,
California Institute of Technology, Pasadena, CA 91125}

\begin{abstract}
The distribution of dark matter in dwarf galaxies can have important implications on our understanding of galaxy formation as well as the particle physics properties of dark matter. However, accurately characterizing the dark matter content of dwarf galaxies is challenging due to limited data and complex dynamics that are difficult to accurately model. In this paper, we apply spherical Jeans modeling to simulated stellar kinematic data of spherical, isotropic dwarf galaxies with the goal of identifying the future observational directions that can improve the accuracy of the inferred dark matter distributions in the Milky Way dwarf galaxies. We explore how the dark matter inference is affected by the location and number of observed stars as well as the line-of-sight velocity measurement errors. We use mock observation to demonstrate the difficulty in constraining the inner core/cusp of the dark matter distribution with datasets of fewer than 10,000 stars. We also demonstrate the need for additional measurements to make robust estimates of the expected dark matter annihilation signal strength. For the purpose of deriving robust indirect detection constraints, we identify Ursa Major II, Ursa Minor, and Draco as the systems that would most benefit from additional stars being observed. 
\\
\end{abstract}

\section{Introduction} 
\label{sec:intro}

The standard $\Lambda$CDM model, consisting of the cosmological constant $\Lambda$ and cold dark matter (CDM), has had remarkable success at predicting physics on large scales, e.g., the cosmic microwave background~\citep{Aghanim:2018eyx} and the large-scale distribution of matter in the Universe~\citep{Tegmark:2002cy,Tegmark:2008au,Hlozek:2011pc}, but faces several small-scale challenges~\citep{Bullock:2017xww}. Among these challenges is the ``core-cusp problem''~\citep{1994ApJ...427L...1F,1994Natur.370..629M}---$\Lambda$CDM predicts that, in the absence of baryonic physics, dark matter (DM) halos universally follow a Navarro-Frenk-White (NFW) density profile~\citep{Navarro:1996gj}, which steeply rises as $\rho\propto r^{-1}$ towards central regions. However, a number of measurements of rotation curves and stellar dynamics have suggested that the DM distribution in the centers of dwarf galaxies may be more consistent with having a constant density core~\citep[e.g.,][]{1994ApJ...427L...1F,1994Natur.370..629M,2000ApJ...537L...9S,2003ApJ...583..732S,2005AJ....129.2119S,2008AJ....136.2563W,2011AJ....142...24O,2015AJ....149..180O}. In this paper, we apply spherical Jeans modeling to individual stars in simulated dwarf galaxies to characterize the observational regimes in which the method can robustly distinguish a cored halo from a cuspy one.

If the DM halos of dwarf galaxies truly are cored, one potential way to explain the apparent discrepancy is through baryonic physics. During baryonic contraction, the central density of a galaxy increases due to the infall of dissipative baryons, deepening the potential well and dragging DM into the central region, which leads to the formation of a DM core---this happens primarily in Milky Way-sized galaxies \citep{1986ApJ...301...27B}. On smaller scales, stellar feedback can lead to core formation due to the ejection of baryons \citep{1996MNRAS.283L..72N,2005MNRAS.356..107R,2006Natur.442..539M,2012MNRAS.421.3464P}. 

While there is qualitative agreement in the simulation literature surrounding the formation of cores in dwarf galaxy-sized DM halos, there is considerable scatter in the quantitative results from various works. Recent studies of hydrodynamic simulations have shown that lower mass dwarfs ($M_{*} \lesssim 10^6 \msun$) have cuspy DM halos, while efficient core formation from stellar feedback turns on around $M_* \sim 10^9 \msun$; for galaxies slightly more massive than the Milky Way, the DM halo reverts back to a cuspy distribution \citep[see, e.g.][]{diCintio2014, Tollet2016, Lazar2020}. \cite{2016MNRAS.459.2573R} correlated the presence of cores to an active stellar formation history in isolated simulated dwarf galaxies. Similarly, simulations with a lower density threshold for star formation, for example Auriga \citep{2017MNRAS.467..179G} and APOSTLE \citep{2016MNRAS.457.1931S}, find that cores do not form at dwarf galaxy sizes \citep{2019MNRAS.486.4790B}. Core formation thus depends on the baryonic feedback model, and while present observations are inconsistent with low star formation thresholds \citep{2019MNRAS.486..655D,2019MNRAS.488.2387B}, reliable observational evidence for cusps or cusps in dwarf galaxies has important implications for understanding stellar feedback and galaxy formation.

A different approach to resolving the core-cusp problem is to modify the particle model of DM itself---for example, models of self-interacting dark matter (SIDM) notably predict the formation of central cores in the DM density profiles of low-mass galaxies~\citep{spergel2000}. There has been extensive work in the literature studying halo formation in SIDM \citep[e.g.][]{Tulin:2017ara,Fitts:2018ycl,Despali:2018zpw,Robles:2019mfq}. In addition to SIDM, other theories of DM can also predict different halo properties from the $\Lambda$CDM prediction, e.g., theories of dissipative DM have been shown to lead to the formation of halos with inner density profiles that are more steeply cusped than NFW halos \citep{jacob2020}. The inner profiles of dwarf galaxy DM halos can therefore encode information about the particle physics that governs the DM.

Whether the Milky Way dwarf galaxies truly all reside in cored or cuspy halos, or there is a large scatter in the inner density profile shapes, there would be important consequences for our understanding of the underlying baryonic and DM physics. At present, there is a lack of consensus in the dwarf galaxy literature on whether the stellar data favors cuspy or cored DM distributions. One specific example is the case of Sculptor, one of the more extensively analyzed dwarf galaxies in the mass modeling literature. Sculptor has been observed to have two chemo-dynamically distinct subpopulations of stars with different half-light radii, which can be leveraged to constrain the DM density at two different radii. \cite{2008ApJ...681L..13B} applied separate Jeans analyses to the two stellar components and found that either a cored halo or an NFW halo were statistically consistent with their data. \cite{2011ApJ...742...20W} applied a mass estimator to the data for the two components and concluded that their analysis ruled out an NFW profile at $\gtrsim99\%$ significance. \cite{2012MNRAS.419..184A} used a separable distribution function method and found strong statistical preference for a cored DM profile, while \cite{Strigari:2014yea} found that with a more flexible distribution function model, the statistical preference went away and the data was consistent with an NFW halo.

Aside from addressing the core-cusp problem, robustly inferring the DM density distribution in dwarf galaxies is also important in the context of DM indirect detection. Indirect detection is the process in which DM annihilates or decays into Standard Model (SM) particles, and the resulting SM particles are subsequently detected. The probability of detecting such a signal is maximized in regions of the sky with high DM density, such as the centers of dwarf galaxies or the Milky Way Galactic Center (GC). Indeed, an excess of $\sim\mathrm{GeV}$ photons was detected near the GC by the \emph{Fermi} Large Area Telescope~\citep{Atwood:2009ez}, which could be interpreted as a signal of DM annihilation \citep[e.g.][]{Goodenough:2009gk, Daylan:2014rsa, Calore:2014xka, TheFermi-LAT:2015kwa}. However, DM analyses near the GC are complicated by bright and complex astrophysical backgrounds, and it is important to have complementary search targets, some of which have excluded or placed the DM interpretation of the excess under tension~\citep[e.g.,][]{Ackermann:2015zua,Fermi-LAT:2016uux,Lisanti:2017qlb,Chang:2018bpt,Calore:2018sdx,Hoof:2018hyn,DiMauro:2019frs}. 

Some of the complementary targets studied in the indirect detection literature have been the Milky Way halo at high latitudes~\citep{Chang:2018bpt,Zechlin:2017uzo,Ackermann:2012rg}, galaxy groups~\citep{Lisanti:2017qlb,Lisanti:2017qoz}, Andromeda~\citep{DiMauro:2019frs}, and stacked dwarf galaxies~\citep[e.g.,][]{Ackermann:2011wa,2011PhRvL.107x1303G,Ackermann:2015zua,Fermi-LAT:2016uux,Calore:2018sdx,Hoof:2018hyn}. In particular, dwarf galaxies are generally considered to be the most robust search targets within the indirect detection literature, because they are expected to have little astrophysical background emission~\citep{Gallagher:2003nx,Grcevich:2009}. 

In general, the expected signal flux from DM annihilation is proportional to the so-called astrophysical {\jfac}, which is defined as the integrals over the solid angle $\Omega$ and along the line of sight $s$ of the DM density squared,
\begin{align}
J &= \int ds \int d\Omega ~ \rho^2(s,\Omega)\,, 
\label{eq:jfac}
\end{align}
where $\rho$ is the DM density. The robustness of any dwarf galaxy-based indirect detection constraint on DM annihilation is dependent on accurately estimating the {\jfac}s of the analyzed dwarf galaxies, and therefore dependent on accurately inferring their DM density distributions. 

Finally, reliably reconstructing the total DM mass in dwarf galaxies also has important scientific ramifications. As we demonstrate in this paper, this is related to---but can be separate from---accurately inferring the full DM density distribution, because while the density and enclosed mass distributions are directly related, there can be cases where the total mass is accurately estimated even if the shape of the density distribution is not fully reconstructed. Obtaining accurate estimates of the total DM mass in dwarf galaxies plays a key role in determining the low-mass end of the stellar-to-halo mass relation (SHMR)~(see \cite{Wechsler:2018pic} for a review of the galaxy-halo connection). Studies on simulations have found that galaxy formation is significantly suppressed in DM halos with virial mass below $\sim 10^8\,M_\odot$~\citep[e.g.,][]{2017MNRAS.467.2019R,alej2020detailed}, leading \cite{2013ApJ...770...57B,2017MNRAS.464.3108G} to propose scatter at the low-mass end of the SHMR. A more accurate determination of the DM halo mass in the smallest dwarf galaxies would help empirically anchor the SHMR for the smallest systems, for which the uncertainty on the relation between galaxies and their DM halos is the largest.

In this paper, we apply spherical Jeans modeling \citep{1915MNRAS..76...70J,1980MNRAS.190..873B,1985AJ.....90.1027M,1992ApJ...391..531D} to simulated dwarf galaxy kinematic datasets, varying over properties of the mock observations such as the total number of observed stars, the measurement error on line-of-sight velocities, as well as the locations of the observed stars (e.g., whether they are primarily in the central region of the dwarf or farther out). We choose to focus on spherical isotropic dwarf galaxies in equilibrium. By studying the limitations of the Jeans analysis method even in this simplified scenario, we are able to identify which observational advancements are more likely to make an impact on our ability to accurately reconstruct the properties of dwarf galaxy DM halos in the near future. 

This paper is organized as follows. In Section~\ref{sec:pipeline}, we present details on the Jeans modeling method and simulated datasets used in this study. In Section~\ref{sec:results}, we explore the effects of the number of observed stars (\Sec{sec:n_stars}), the measurement errors of the line-of-sight velocities (\Sec{sec:errors}), and the locations of the observed stars (\Sec{sec:location}) on the DM inference. \Sec{sec:degeneracies} explores the impact of degeneracies between model parameters. \Sec{sec:indirect} recasts our results into the context of {\jfac}s for indirect detection, where we emphasize the dependence of the indirect detection results on the priors chosen in the Jeans analysis and discuss recommendations for future observations. We summarize our main conclusions in \Sec{sec:conclusions}. 

\section{Methods}
\label{sec:pipeline}
In this section, we describe the Jeans modeling procedure we employ (Sec.~\ref{sec:jeans}), the models we use to parameterize the distributions of the DM (Sec.~\ref{sec:dm_profiles}) and the stars (Sec.~\ref{sec:stars_profiles}), the specifics of the mock data we generate (Sec.~\ref{sec:mocks}), and the priors we assume for the model parameters throughout our analysis (Sec.~\ref{sec:priors}). We use the public code \textsc{StarSampler}\footnote{\url{https://github.com/maoshenl/StarSampler}} to generate our mock data.

\subsection{Jeans Modeling}
\label{sec:jeans}

We summarize the standard procedure for inferring the velocity dispersion profile of the stars in a dwarf galaxy from measurements of their line-of-sight velocities, following the derivations of \cite{1982MNRAS.200..361B,BT2}. We start with the collisionless Boltzmann equation, 
\be \label{eq:boltzmann}
\frac{\p f}{\p t} + \vec{v} \frac{\p f}{\p \vec{x}}  - \frac{\p \Phi}{\p \vec{x}}  . \frac{\p f}{\p \vec{v}}  = 0\,, 
\ee
where $f$ is the phase-space density of a stellar tracer population, a function of the position $\vec{x}$ and velocity $\vec{v}$ of each star, and $\Phi$ is the gravitational potential of the dwarf galaxy.
Multiplying \Eq{eq:boltzmann} by velocity component $v_j$ and integrating over all velocities, we have
\be
\frac{\p}{\p t} (\nu \ov{v_j}) + \frac{\p}{\p x_i} (\nu \ov{v_i v_j} ) + \nu \frac{\p \Phi}{\p x_j} = 0\,, 
\ee
where we have defined $\nu = \int d \vec{v}^3 f(\vec{x}, \vec{v})$, the spatial number density of the tracer stars. 
Assuming the system is spherically symmetric and is in steady state (and therefore the $\partial/\partial t$ term is negligible), we have
\be\label{eq:spherical}
\frac{\p}{\p r} (\nu \sigma_r^2) + \nu \left( \frac{\p \Phi}{\p r} + \frac{2 \sigma_r^2 - \sigma_\theta^2 - \sigma_\phi^2 }{r} \right) = 0\,,
\ee
where $\sigma_i^2$ is the square of the $i^\mathrm{th}$ component of the velocity dispersion, i.e., $\sigma_i^2 = \langle v_i^2 \rangle - \langle v_i \rangle^2$, for $i \in \{ r, \theta, \phi \}$.

We can then define the velocity anisotropy, 
\be \label{eq:beta}
\beta (r) = 1 - \frac{\sigma_\theta^2 + \sigma_\phi^2}{2 \sigma_r^2}\,,
\ee
and explicitly write the potential as
\be
\Phi = -\frac{G M(<r)}{r}\,,
\ee
where $G$ is the gravitational constant and $M(<r)$ is the enclosed mass within radius $r$. Plugging these quantities back into \Eq{eq:spherical}, we end up with the following first-order differential equation for $\nu \sigma_r^2$:
\be \label{eq:nusigma}
 \frac{1}{\nu} \left[ \frac{\partial}{ \partial r} (\nu \sigma_r^2) + \frac{2 \beta(r)}{r} ( \nu \sigma_r^2)  \right] = - \frac{G M(<r)}{r^2}\,.
\ee
The generic solution to \Eq{eq:nusigma} takes the form 
\be \label{eq:sigma_spherical}
\nu(r) \sigma_r^2(r) = \frac{1}{g(r)} \int_r^\infty \frac{G M(<\tilde{r}) \nu(\tilde{r})}{\tilde{r}^2} g(\tilde{r}) d \tilde{r}\,,
\ee
where the new function $g(r)$ is defined as
\be \label{eq:g_function}
g(r) = \exp \left(2 \int \frac{\beta(r)}{r} dr \right)\,.
\ee

The enclosed mass $M(<r)$ in \Eq{eq:sigma_spherical} can be related to the overall density distribution by
\be \label{eq:mass}
M(<r) = 4 \pi \int_0^r \rho(s) s^2 ds\,,
\ee
where, again, we have assumed spherical symmetry of the system. While both the stars and DM contribute to the mass density distribution, i.e., $\rho = \rho_{\rm{DM}} + \rho_{\rm{stars}}$, we expect the density of DM to dominate, and therefore make the approximation $\rho \approx \rho_{\rm{DM}}$. This is a valid approximation due to the large mass-to-light ratios of dwarf galaxies, $M_{\rm{halo}}/M_* \approx 10^2$--$10^5$~\citep{2017MNRAS.464.3108G}. 

In practice, typically only projected radii and line-of-sight velocities are measured, and therefore \Eq{eq:sigma_spherical} needs to be projected along the line of sight. To do so, we use the Abel transform, defined for a spherically-symmetric function as
\be
S(R) = 2 \int_R^{+\infty} \frac{s(r) r dr}{\sqrt{r^2 - R^2}}\,,
\ee where $s(r)$ is the function in three-dimensional spherical coordinates, $R$ is the projected radius, and $S(R)$ is the resulting projected function. 
Projecting \Eq{eq:sigma_spherical} along the line of sight leads to the equation \citep{1982MNRAS.200..361B,2010MNRAS.401.2433M} 
\be \label{eq:sigma_p}
\sigma_p^2(R) I(R) = 2 \int_R^{\infty} \left( 1 - \beta (r ) \frac{R^2}{r^2} \right) \frac{\nu(r) \sigma_r^2 (r) r}{\sqrt{r^2 - R^2}} dr\,,
\ee
where $\sigma_p$ is the projected velocity dispersion profile and $I(R)$ is the projected number density distribution of the tracer stars, given by
\be \label{eq:surface_brightness}
I(R) = 2 \int_R^\infty \frac{\nu(r) r dr}{ \sqrt{r^2 - R^2}}\,.
\ee
Throughout the remainder of this paper, $I(R)$ is referred to as the surface brightness profile or light profile.

Using \Eq{eq:sigma_p}, we build a likelihood function to fit the observed data and extract information on the dark matter distribution. In the literature, the analysis has been performed in either a binned \citep[e.g.,][]{Strigari:2006rd,2011MNRAS.418.1526C} or unbinned \citep[e.g.,][]{2008ApJ...678..614S} fashion. In this work, we will focus on the unbinned analysis. The unbinned Gaussian likelihood function is given by \citep{2008ApJ...678..614S}
\be \label{eq:likelihood}
\mathcal{L} = \prod_{i=1}^{N_{\rm{stars}}} \frac{(2\pi)^{-1/2}}{ \sqrt{ \sigma_p^2 (R_i) + \Delta_{v_i}^2}} \exp \left[ - \frac{1}{2} \left( \frac{(v_i - \ov{v})^2}{\sigma_p^2(R_i) + \Delta_{v_i}^2} \right) \right],
\ee
 where $\ov{v}$ is the mean velocity for the population of tracer stars, and for star $i$, $v_i$ is the measured line-of-sight velocity, $\sigma_p(R_i)$ is the intrinsic velocity dispersion at the projected radius $R_i$, and $\Delta_{v_i}$ is the velocity measurement error. In our analysis, we choose closed-form parameterizations for the stellar and dark matter distributions, thereby reducing the number of integrals that need to be performed when calculating the likelihood. 

It is important to emphasize the interplay between the intrinsic velocity dispersion and the measurement error in \Eq{eq:likelihood}---if the measurement errors are subdominant to the intrinsic velocity dispersion of the system, it is not expected that improvements to the line-of-sight velocity measurements would drastically improve the quality of the fit. This will be further discussed in \Sec{sec:errors}. 

The degeneracy between the velocity anisotropy, $\beta(r)$, and the enclosed mass profile, $M(<r)$, is a known complication in Jeans modeling \citep[e.g.,][]{1990AJ.....99.1548M,Wilkinson:2001ut,Lokas:2003ks,DeLorenzi:2008zq,2017MNRAS.471.4541R,Genina:2019job}. It can be seen from \Eq{eq:sigma_p} that $\beta(r)$ and $\sigma_r^2(r)$ are degenerate with each other, which, combined with \Eq{eq:sigma_spherical}, implies that $\beta(r)$ is degenerate with $M(<r)$. Unfortunately, $\beta(r)$ can only be measured with full 3D velocity information, which is not yet available for the majority of the stars in dwarf galaxies. It is therefore common in Jeans analyses to assume a parametric model for $\beta(r)$ and fit for it in conjunction with fitting for $M(<r)$~\citep[e.g.,][]{2005MNRAS.363..705M,2013MNRAS.429.3079M,Geringer-Sameth:2014yza,Bonnivard:2015xpq,2015arXiv150408273M}. The effect on dynamical mass modeling estimates when the assumed $\beta(r)$ model does not match the true velocity anisotropy distribution has been studied in~\cite{2017ApJ...835..193E}. In this work, we choose to focus entirely on isotropic datasets and models in order to understand the limitations of the Jeans modeling procedure even in the absence of additional complications due to velocity anisotropy, i.e., we assume 
\begin{equation} \label{eq:beta_0}
    \beta(r) = 0\,.
\end{equation}

\subsection{Dark Matter Profile}
\label{sec:dm_profiles}

Using \textsc{StarSampler}, we generate the tracer stars in a DM potential which follows the Hernquist/Zhao profile \citep{1990ApJ...356..359H,1996MNRAS.278..488Z}
\begin{equation} \label{eq:zhao}
    \rho_{\dm}^{\rm{Zhao}} (r) = \rho_0  \left( \frac{r}{r_s} \right)^{-\gamma} \left[1 + \left( \frac{r}{r_s} \right)^\alpha \right] ^{(\gamma - \beta)/\alpha }\,,
\end{equation}
where $\alpha, \beta, \gamma$ are the slopes of the distribution, $\rho_0$ is the overall normalization of the density profile, and $r_s$ is the scale radius---in particular, $\gamma$ sets the asymptotic inner slope of the distribution. This model has five free parameters, which introduces too many degenerate degrees of freedom into the model to effectively constrain the parameters (we discuss the role of degeneracies in \Sec{sec:degeneracies}). We therefore simplify the DM profile by setting $\alpha = 1$ and $\beta = 3$, which reduces \Eq{eq:zhao} to a generalized Navarro-Frenk-White (gNFW) distribution with inner slope parameter $\gamma$, defined as \citep{Navarro:1996gj}
 \begin{equation} \label{eq:gnfw}
    \rho_{\dm}^{\rm{gNFW}} (r) = \rho_0 \left( \frac{r}{r_s} \right)^{-\gamma} \left(1 +\frac{r}{r_s}\right) ^{-(3-\gamma)}\,.
\end{equation}

While we use the gNFW distribution to model the DM profile in our fiducial analysis setup, we additionally consider the special cases where the inner slope $\gamma=0$ or 1. The case of $\gamma=1$ corresponds to the standard, cuspy Navarro-Frenk-White (NFW) profile
\begin{equation} \label{eq:nfw}
    \rho_{\dm}^{\rm{NFW}} (r) = \rho_0 \left( \frac{r}{r_s} \right)^{-1} \left(1 +\frac{r}{r_s} \right) ^{-2}\,,
\end{equation}
whereas the case of $\gamma=0$ leads to a constant density central core. We refer to this distribution as the cored NFW (NFWc) distribution, given by
\begin{equation} \label{eq:nfw_cored}
     \rho_{\dm}^{\rm{NFWc}} (r) = \rho_0 \left( 1 + \frac{r}{r_s} \right)^{-3}\,.
\end{equation}
The profiles defined by Eqs.~(\ref{eq:gnfw})--(\ref{eq:nfw_cored}) give rise to closed-form enclosed mass distributions, which we list in Appendix~\ref{sec:enclosed_mass} for reference.

\subsection{Light Profile}
\label{sec:stars_profiles}

\begin{table*}[t]
    \centering
    \hspace*{-3.5cm}
    \begin{tabular}{|l|c|c|c|c|c|c|c|c|l|}
    \hline
      &  & $\rho_0 $  & $r_s$ & $\gamma$ & $M_{200}$ & $c_{200}$ & $\sigma_p$\\
      &   & $ [M_{\odot}/\rm{kpc}^3]$  &  [kpc] &  &  $[M_\odot]$ &  & [km/s]\\
    \hline
    \hline
    Cusp & I &  $6.4 \times 10^7$ & 1 & 1 & $1.9\times 10^9$ & $25.8$ & $14.6$ \\ 
              & II &  $6.4 \times 10^7$ & 0.2 & 1 & $1.5\times10^7$ & $25.8$ & $2.9$ \\ 
    \hline
    Core &  III & $6.4 \times 10^7$  & 1 & 0 & $1.4\times 10^9$ & $23.6$ & $9.5$ \\
             &  IV & $6.4 \times 10^7$  & 0.2 & 0 & $1.1\times 10^7$ & $23.6$ & $1.9$ \\
    \hline

    \end{tabular}
    \caption{DM halo parameters and properties of the datasets generated in this work. $\rho_0$, $r_s$, and $\gamma$ are the true values of the normalization, scale length, and inner slope input into \Eq{eq:gnfw}. $M_{200}$ is defined as the enclosed mass at $r_{200}$, the radius within which the average density is equal to 200 times the critical density of the Universe at redshift $z=0$, derived from the true density distribution. We adopt a generalized definition of the concentration $c_{200}\equiv r_{200}/r_s$ for all of our parameter sets. $\sigma_p$ is the median line-of-sight velocity dispersion across all the datasets generated for each set of parameters (10 realizations each for sample sizes of 20, 100, 1000, and 10,000 stars, resulting in a total of 40 datasets).}
    \label{tab:mock_data}
\end{table*}  

Using \textsc{StarSampler}, we can model the stellar density distribution also as a Hernquist/Zhao profile
\begin{equation} \label{eq:zhao_stars}
\nu(r) = \rho_* \left( \frac{r}{r_*} \right)^{-{\gamma_*}} \left[1 + \left( \frac{r}{r_*} \right)^{\alpha_*} \right] ^{({\gamma_*} - {\beta_*})/{\alpha_*} }\,.
\end{equation}
In this paper, we generate stars following a Plummer profile, which is a specific case of \Eq{eq:zhao_stars}. For ease of comparison across different samples, the stellar mocks are all generated with the same level of ``embeddedness'' in their respective DM halos by setting the scale radius of the tracers, $r_*$, to be equal to the scale radius of the DM distribution, $r_s$. 

Correspondingly, in our Jeans analysis, we model the stellar density $\nu(r)$ as a 3d Plummer profile \citep{1911MNRAS..71..460P}, defined as
\be \label{eq:plummer}
\nu(r) =  \frac{3 L}{4 \pi a^3} \left( 1 + \frac{r^2}{a^2} \right)^{-5/2}\,,
\ee
where $L$ is the total luminosity and $a$ is scale length of the distribution. \Eq{eq:plummer} has the same form as \Eq{eq:zhao_stars}, with $\alpha_* = 2$, $\beta_* = 5$, $\gamma_* = 0$, $r_*=a$, and $\rho_*=3M/(4\pi a^3)$.\footnote{In practice, when we generate our samples, we follow the examples of \textsc{StarSampler} and \cite{2011ApJ...742...20W} and set $\gamma_* = 0.1$ rather than $\gamma_*=0$ for ease of comparison. We do not expect it to affect the results.}  The surface brightness profile (or light profile), which is the projection of $\nu(r)$ along the line of sight, is then given by the closed-form expression
\be \label{eq:surface_density}
I(R) =  \frac{L}{\pi a^2} \left( 1 + \frac{R^2}{a^2} \right)^{-2}\,.
\ee
Because we have assumed the contribution of the stellar tracers to the gravitational potential is negligible, changing the value of $L$ in \Eqs{eq:plummer}{eq:surface_density} does not meaningfully affect the result of the Jeans modeling.

\subsection{Mock Data}
\label{sec:mocks}

We generate datasets with four different sets of DM halo parameters (summarized in \Tab{tab:mock_data}). Our parameter choices span different halo masses and either an inner cusp or inner core in the DM density profile while maintaining approximately the same halo concentration. Due to the large amount of scatter in the theoretical predictions for the subhalo mass-concentration relation, we choose not to focus on a specific mass-concentration model; however, the concentrations of our simulated halos are consistent with theoretical predictions in the literature for the relevant mass range~\citep{Pieri:2009je,Sanchez-Conde:2013yxa,Moline:2016pbm}. 

Parameter sets I and III correspond to $M_\mathrm{200}\sim10^9\,M_\odot$ halos, while sets II and IV correspond to smaller halos with mass $M_{200}\sim10^7\,M_\odot$. We emphasize that we have chosen to study $M_{200}\sim10^7\,M_\odot$ halos for demonstrative purposes, to study how the effect of the measurement error on the line-of-sight velocities impacts less massive halos differently from more massive ones. We have adopted a generalized definition of the halo concentration, $c_{200}\equiv r_{200}/r_s$, for all of the parameter sets that we generate, where $r_{200}$ is the radius within which the average density is 200 times the critical density of the Universe at redshift $z=0$. The virial mass $M_{200}$ is subsequently defined as the enclosed mass at $r_{200}$. 

For each set of DM parameters, we generate 10 realizations each of datasets with 20, 100, 1000, and 10,000 stars, respectively. The chosen sample sizes are meant to provide comparison with current measurements of ultrafaint dwarfs and classical dwarfs (see \Tab{tab:J_n_stars} for comparison), as well as projections for how future measurements might improve the quality of the DM inference. For our fiducial analyses, we assume a measurement error of $\Delta v=2\,\mathrm{km/s}$ on the line-of-sight velocity. This is comparable to the typical errors in current measurements (see, e.g., references within Table~\ref{tab:J_n_stars}). We explore the effect of increasing or decreasing the measurement error relative to our fiducial value of $2\,\mathrm{km/s}$ in Sec.~\ref{sec:errors}. We show representative distributions of the projected and 3d galactocentric radii in our generated stellar datasets in Figure~\ref{fig:hist_stars} of the Appendix. Throughout this paper, we will use $R$ to denote the projected radius and $r$ to denote the 3d galactocentric radius.

\begin{table}[t]
    \begin{center}
    \hspace*{-1.6cm}
    \begin{tabular}{|c|c|c|}
    \hline
    Parameter  & Prior  \\
    \hline
    \hline
    $\log_{10}(a/\mathrm{kpc})$  & $[-3,3]$ \\
    $\log_{10}(L/L_\odot)$  & $[-2,5]$ \\
    \hline
    \hline
    $\ln(\rho_0/(M_\odot\,\mathrm{kpc}^{-3}))$ & $[5,30]$\\
    $\ln(r_s/\mathrm{kpc})$ & $[-10,10]$\\
    $\overline{v}/(\mathrm{km\,s}^{-1})$ & $[-100,100]$\\
    $\gamma$ & $[-1,5]$\\
    \hline
    \end{tabular}
    \end{center}
    \caption{Prior ranges for the stellar and DM parameters used in our analysis. We implement uniform priors within each of the listed prior ranges. The ranges listed here for $\log_{10}(a)$ and $\log_{10}(L)$ are used in the initial light profile fit; in the full Jeans scan, we set the prior ranges for $\log_{10}(a)$ and $\log_{10}(L)$ to be the middle 95\% containment range of the posterior for each parameter from the initial fit (see Appendix~\ref{sec:light_profile_fit} for more discussion on the light profile fit). }
    \label{tab:priors}
    \end{table}  

\subsection{Parameters and Priors}
\label{sec:priors}

We perform our Jeans modeling procedure in two stages. First, we perform a fit to only the positions of the stars. We describe this light profile fitting procedure in Appendix~\ref{sec:light_profile_fit}. We do so because the light profile is generally much better constrained than the stellar kinematics. We can then use the results from the initial fit to set the prior range on the light profile parameters in our full Jeans fit. We conservatively set the prior ranges on the light profile parameters in the full scan to be the middle 95\% containment range of the posterior probability distributions output from the initial fit. In both stages, we use the \textsc{PyMultinest} module (introduced in \cite{Buchner:2014nha}), which interfaces with the nested sampling Monte Carlo library \textsc{Multinest}~\citep{Feroz:2008xx}, to sample the relevant likelihood.\footnote{We use $n_\mathrm{live}=100$ live points in the nested sampling procedure throughout this paper, but have verified that increasing to $n_\mathrm{live}=1000$ does not affect the results.} 

We summarize the priors for all of the parameters in our model in \Tab{tab:priors}. We choose a wide prior of $[-1,5]$ for the parameter $\gamma$, which sets the inner slope of the gNFW distribution. The lower edge is chosen to be at $-1$ such that there is sufficient range for convergence at $\gamma=0$ while not allowing for larger negative values of $\gamma$, which are unphysical. We note that values of $\gamma\geq3$ are also unphysical, as they lead to divergent enclosed mass at finite radius; we have verified that excluding these values from our prior range has negligible effect on our results (see Appendix~\ref{sec:prior_tests} for more detailed discussion on varying priors).

\begin{figure*}[t]
    \centering
    \includegraphics[width=0.8\textwidth]{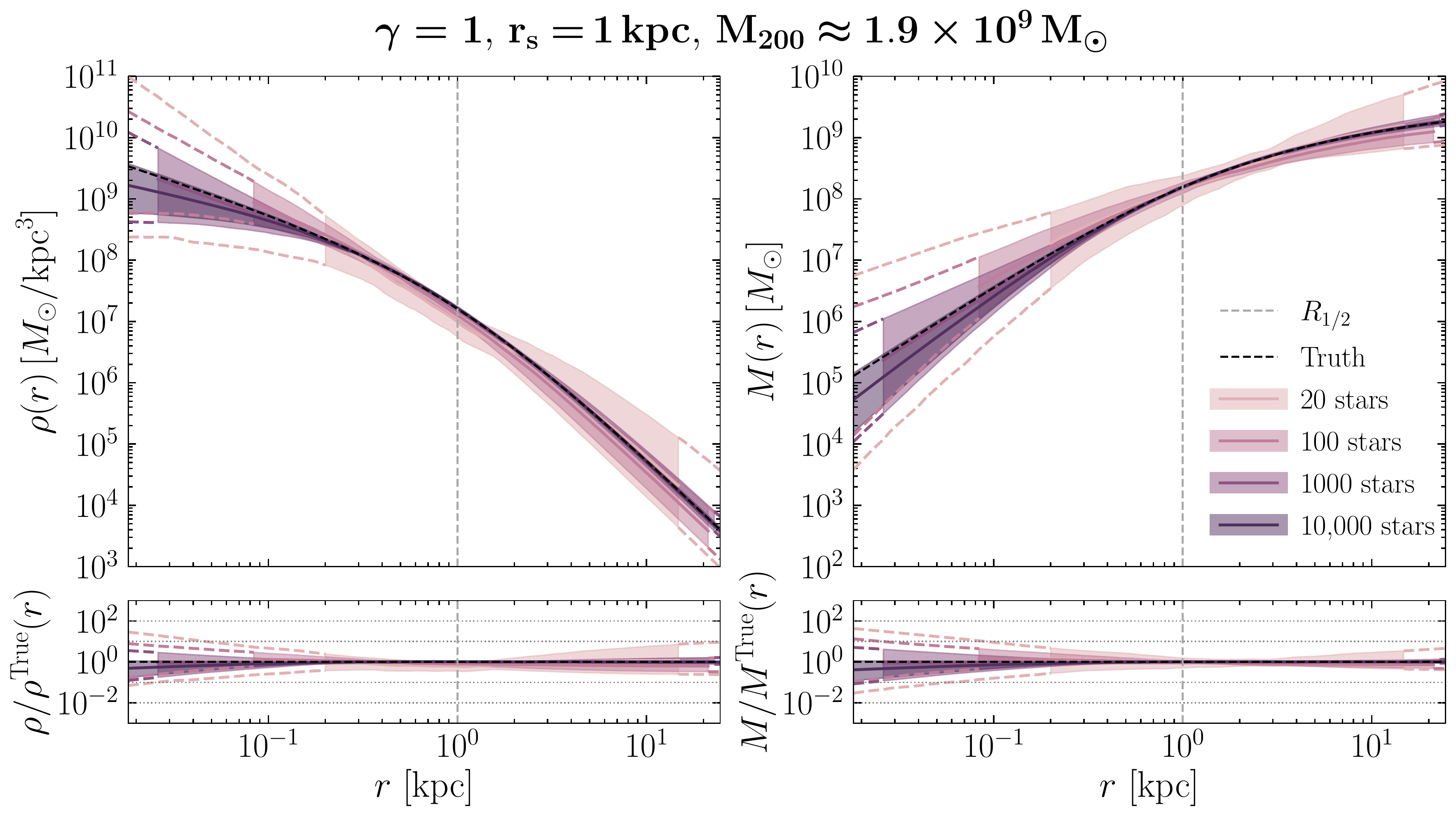}
    \caption{Inferred DM density profiles $\rho(r)$ (left panels) and corresponding enclosed mass profiles $M(r)$ (right panels) for parameter set I---from lightest to darkest color, we plot the results for samples with 20, 100, 1000, and 10,000 stars. We show the full reconstructed distributions in the top panels as well as the fractional (relative to truth) distributions in the bottom panels. For each sample size, the solid line denotes the median (across our 10 independent realizations) of the median recovered profiles, while the shaded band shows the median of the 68\% containment regions, plotted from the innermost to outermost star across all 10 datasets for that sample size. For the samples with fewer than 10,000 stars, we additionally extrapolate the median 68\% containment regions over the full radial range, shown bracketed by each pair of dashed lines in the color corresponding to the sample size. Across all sample sizes, the typical inferred density profile and enclosed mass profile are consistent within uncertainty with the true distributions. Increasing the observed sample size reduces the uncertainty on the recovered profiles, as expected.}
    \label{fig:rho_Menc_gamma_1}
\end{figure*}

In our fiducial model, there are a total of six free parameters: two for the light profile, three for the DM density distribution parameterized as a gNFW profile, and one for the mean stellar velocity. In our discussion on characterizing the inner slope of the DM distribution, we additionally perform fits assuming either an NFW or cored NFW distribution, and compare the Bayesian evidence between the two models---in these fits, there are a total of five free parameters. 

\section{Results}
\label{sec:results}

We now apply the analysis pipeline described in Sec.~\ref{sec:pipeline} to the simulated stellar samples described in \Sec{sec:mocks} and summarized in Table~\ref{tab:mock_data}. Our main figures of merit for evaluating the success or limitations of our analyses are: {\it(i)} the overall recovered DM density profile, {\it(ii)} the recovered enclosed DM mass, which we quantify as the recovered virial mass $M_{200}$, and {\it(iii)} the recovered inner slope of the DM density profile, i.e., the parameter $\gamma$ in \Eq{eq:gnfw}. Of the figures of merit, {\it(i)} has important implications on the inferred astrophysical {\jfac}s (\Eq{eq:jfac}) which are used in indirect DM searches, {\it(ii)} is crucial for empirically probing the SHMR down to low halo masses, while {\it(iii)} can shed light on the particle physics properties of the DM as well as baryonic feedback and galaxy formation mechanisms.

We explore how several factors in the analysis influence the accuracy of the inferred DM profiles, focusing primarily on the effects of variations on the specifics of the analyzed datasets. In \Sec{sec:n_stars}, we study how the total number of observed stars influences the inferred DM profile. In \Sec{sec:errors}, we study the role of the line-of-sight velocity measurement errors; we explore how the magnitude of the error differently impacts the DM inference in dwarf galaxies with different halo masses. In \Sec{sec:location}, we study the effect of the locations of observed stars on the inferred DM profile. In \Sec{sec:degeneracies}, we explore how the presence of degeneracies between the DM profile parameters affects the inference of the inner slope $\gamma$. 

\subsection{Increase in Sample Size}
\label{sec:n_stars}

\begin{figure*}[t]
    \centering
     \includegraphics[width=0.8\textwidth]{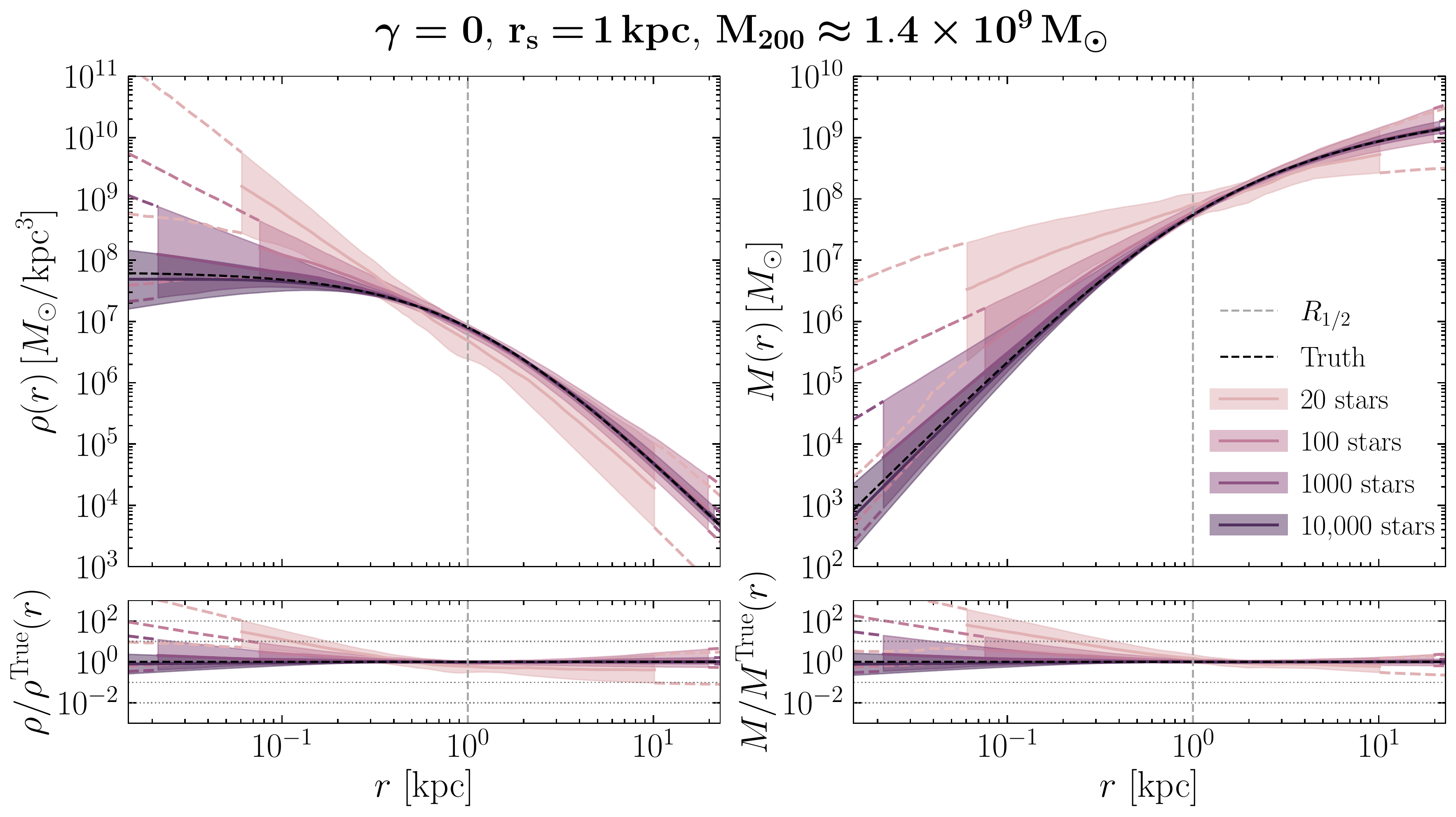}
    \caption{Same as Figure~\ref{fig:rho_Menc_gamma_1}, except for parameter set III. For sample sizes of 20 and 100 stars, the inferred density distribution is typically biased towards a steeper inner profile than the true distribution; however, the inferred virial mass is still consistent with the true virial mass (values listed in Table~\ref{tab:mvir_deltav}). For sample sizes with 1000 and 10,000 stars, both the inferred density distribution and enclosed mass profile are consistent within uncertainty with the true distributions across the measured radial range.}
    \label{fig:rho_Menc_gamma_0}
\end{figure*}

\begin{figure*}[t]
    \centering
    \includegraphics[width=0.95\textwidth]{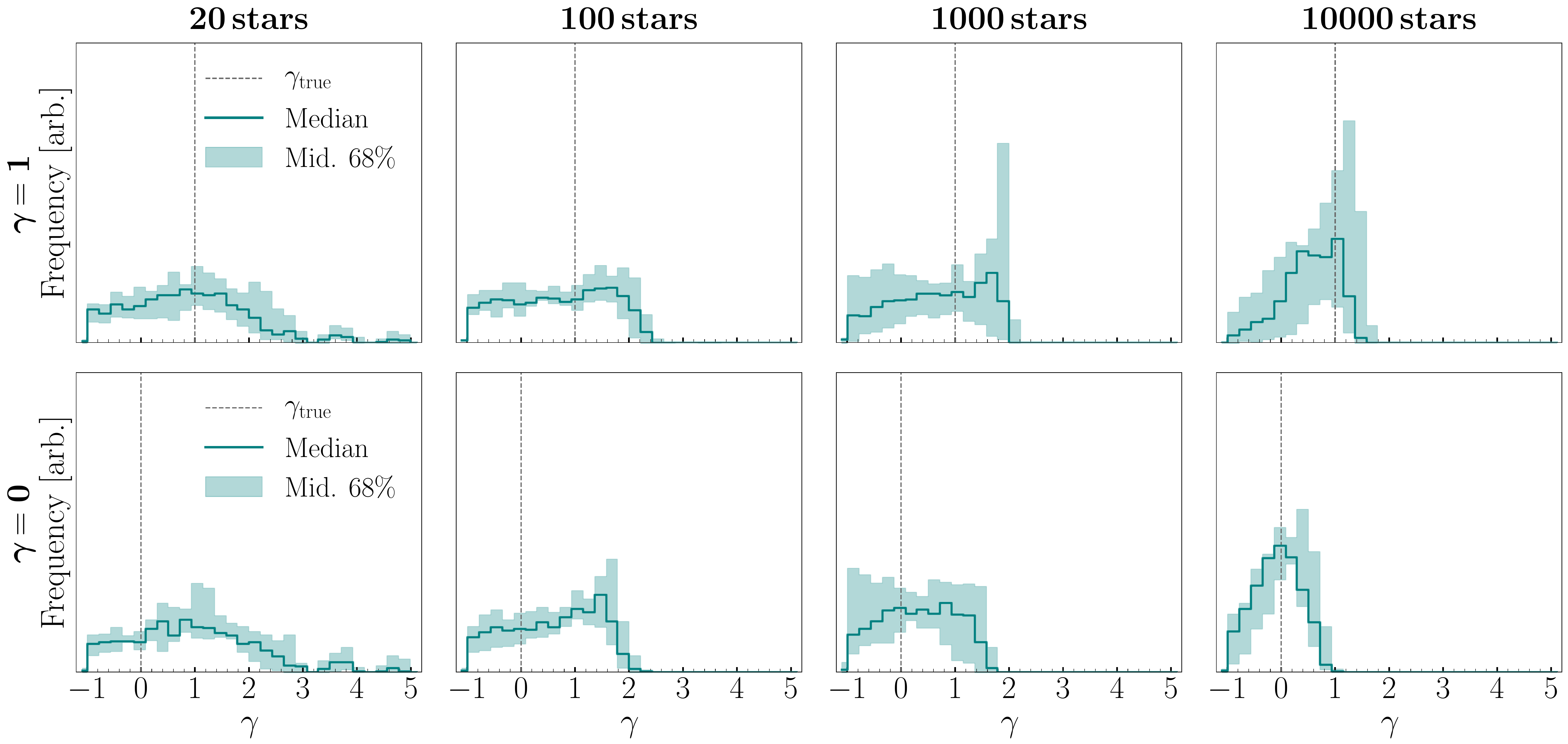}
    \caption{Posterior distributions for the inner slope $\gamma$. The top row corresponds to the scans shown in \Fig{fig:rho_Menc_gamma_1} (parameter set I), while the bottom row corresponds to the scans shown in \Fig{fig:rho_Menc_gamma_0} (parameter set III). The lines(bands) show the median(middle 68\%) in each $\gamma$ bin across the 10 realizations. In both cases, the inner slope is generally poorly constrained for the smaller samples, with the median posterior distribution only peaking near the true value of $\gamma$ (vertical dashed line in each panel) for the largest sample size of 10,000 stars. All panels in this figure share the same vertical scale.}
    \label{fig:gamma_post_1_0}
\end{figure*}

Our first question of interest is how the number of observed stars in a dwarf galaxy affects the DM inference. In Figure~\ref{fig:rho_Menc_gamma_1}, we show the inferred DM density profiles $\rho(r)$ and corresponding enclosed mass profiles $M(r)$ for parameter set I (which has $\gamma=1$), for the four different sample sizes---from lightest to darkest color, we plot the results for 20, 100, 1000, and 10,000 stars. For a given sample size, we run each of our 10 realizations through the analysis pipeline and obtain the resulting posterior density and enclosed mass profiles. Each solid line in Fig.~\ref{fig:rho_Menc_gamma_1} shows the median across the 10 realizations of the median recovered profiles, while the shaded band depicts the median of the 68\% containment regions across the realizations. The solid line and shaded band for each sample size are plotted from the innermost to outermost star across the 10 generated datasets for that sample size; outside of the data range for the smaller samples, we extrapolate the results and outline the 68\% containment region with dashed lines in the color corresponding to each sample size. The extrapolation down to smaller radii is particularly important in understanding the implications for indirect detection, which we discuss in Section~\ref{sec:indirect}. The vertical dashed gray line indicates the projected half-light radius, $R_{1/2}$, which for a Plummer profile is equal to the scale radius $a$.

We find that, for all sample sizes in parameter set I ($\gamma=1$), the typical inferred density profile and enclosed mass profile are consistent within uncertainty with the true distributions over the full range of measured radii. This can be seen from the fact that the dashed black lines in the top panels of Figure~\ref{fig:rho_Menc_gamma_1}, indicating the true distributions, are contained within the bands for all of the sample sizes, as well as the fact that all the bands in the bottom panels overlap with the horizontal dashed black line. Additionally, we find that increasing the observed sample size reduces the uncertainty on the inferred density and enclosed mass profiles, as is to be expected. For all sample sizes and parameter sets, we list the median across our 10 realizations of the median and $\pm1\sigma$ values of the inferred virial mass, $M_{200}$, in Table~\ref{tab:mvir_deltav}.

We show the analogous results for parameter set III ($\gamma=0$) in Figure~\ref{fig:rho_Menc_gamma_0}. In this case, for sample sizes of 20 stars and 100 stars, the inferred density distribution is typically biased towards a steeper inner profile than the true distribution, which has an inner slope of $\gamma=0$, while for the datasets with 1000 and 10,000 stars, the typical inferred density profiles are consistent with the true distribution within uncertainty. Importantly, across all of the sample sizes, we obtain an accurate estimate for the total mass of the system, with the uncertainties on the estimate reduced as the sample size is increased (values listed in the fourth column of the corresponding panel in Tab.~\ref{tab:mvir_deltav}). 

This suggests that while the inferred density distribution may not always accurately represent the true underlying distribution, the virial mass estimate remains fairly robust. Namely, if the inferred density profile is biased high in the inner region of the dwarf, (as seen in the $r\lesssim R_{1/2}$ region for the smaller samples from parameter set III), this is compensated for by the density profile being biased low in the outer region. We note that because the outer slope of the density profile is not a free parameter in the fit, the outer profile is uniquely determined by the scale radius and overall normalization. Our likelihood (\Eq{eq:likelihood}) depends directly on the enclosed mass distribution of the system rather than the density distribution, and therefore it is not surprising that the fit is successful at recovering the total mass of the system even when it fails to accurately reproduce the inner density profile.

Figs.~\ref{fig:rho_Menc_gamma_1}--\ref{fig:rho_Menc_gamma_0} demonstrate that the inner regions of the inferred DM density profiles can be biased and/or poorly constrained, especially for the smaller datasets. We can further assess how well the inner density profile is recovered by directly examining the posterior probability distribution of the parameter in our model which sets the asymptotic inner slope, $\gamma$. In the top row of Figure~\ref{fig:gamma_post_1_0}, we show histograms of the posterior $\gamma$ values corresponding to the scans shown in \Fig{fig:rho_Menc_gamma_1} (parameter set I), i.e., for a true inner slope of $\gamma=1$. The lines(bands) show the median(middle 68\%) in each bin across the 10 realizations. The inner slope is generally poorly constrained for the smaller samples, with the median posterior distribution only peaking around the true value of $\gamma=1$ for the largest sample size of 10,000 stars---notably, even in this case, there is typically non-negligible posterior probability at $\gamma=0$, so we would not be able to exclude an incorrect inner slope value of 0 at high significance. We also draw attention to the fact that, for the samples with 100 and 1000 stars, although the posterior distributions are fairly flat and poorly constrained, the posterior probability sharply drops off above $\gamma\sim2$. This is important because the enclosed mass for a gNFW profile (\Eq{eq:M_gNFW}) diverges at finite $r$ for $\gamma\geq3$. For the most statistics-limited samples containing only 20 stars, the fit cannot fully exclude unphysical values of $\gamma\geq3$.

In the bottom row of Figure~\ref{fig:gamma_post_1_0}, we show the results for parameter set III, which has a true inner slope of $\gamma=0$. The results are qualitatively similar: the posterior distributions of $\gamma$ tend to be poorly constrained for the smaller sample sizes, and we are only able to recover the true value of the inner slope for the 10,000-star samples. In this case, for the largest sample size, we would be able to exclude an incorrect inner slope value of 1 at high significance. However, for datasets with $\lesssim 1000$ stars from both parameter sets---on par with the existing dwarf galaxy measurements---we cannot determine whether the underlying halo has an inner slope of $\gamma=0$ or $\gamma=1$ in a statistically significant manner, consistent with previous Jeans modeling-based results in the literature~\cite[e.g.,][]{2009ApJ...704.1274W,Read:2018pft,Genina:2019job}. We further note that, for the smaller sample sizes, the fact that the posterior $\gamma$ distributions are unconstrained implies that the results are highly sensitive to the choice of priors on $\gamma$, and we therefore choose to present the full posterior distributions rather than to quote recovered median values or quantiles. 

A separate method for quantifying the ability of this procedure to distinguish whether the underlying DM distribution has an inner cusp ($\gamma=1$) or core ($\gamma=0$) is to compare the statistical preference for a cuspy DM model over a cored DM model, or vice versa. In particular, we analyze the same datasets as before, this time fixing the value of $\gamma$ in our model to either 1 or 0 in \Eq{eq:gnfw}. The resulting models respectively correspond to the standard NFW distribution (\Eq{eq:nfw}) or the cored NFW distribution (\Eq{eq:nfw_cored}). We then calculate the Bayes factor (BF) in preference for a model in which $\gamma$ is fixed to the true value for the given dataset, relative to a model in which $\gamma$ is fixed to the alternative value, i.e.,
\be \label{eq:BF}
\mathrm{BF}=\frac{\mathrm{Pr}(d|\gamma=\gamma_\mathrm{true})}{\mathrm{Pr}(d|\gamma=\gamma_\mathrm{alt.})}\,.
\ee
On the Jeffreys scale, as amended by \cite{10.2307/2291091}, $\mathrm{BF}<3.2$ is ``not worth more than a bare mention,'' $\mathrm{BF}\in[3.2,10)$ provides substantial evidence, $\mathrm{BF}\in[10,100)$ provides strong evidence, and $\mathrm{BF}\geq100$ provides decisive evidence.

In Table~\ref{tab:BF_gamma_1}, we list for parameter set I the median and $\pm1\sigma$ (second column) as well as the minimum (third column) and maximum (fourth column) BF values in preference for the true value of $\gamma=1$ across the 10 datasets. For the smaller samples, the BF values are generally indeterminate, which is consistent with the relatively unconstrained posterior distributions shown in the top row of \Fig{fig:gamma_post_1_0}. For a sample size of 10,000 stars, the median BF is also indeterminate, although we find that there is one realization for which there is decisive evidence, and two realizations for which there is strong evidence, in favor of a model with a cusp. This is consistent with the rightmost panel in the top row of \Fig{fig:gamma_post_1_0}, in which the average posterior probability is non-negligible at $\gamma=0$ and there is significant variation in the height of the peak at $\gamma\sim1$ across realizations. Although there is significant scatter in the BF values between realizations, we emphasize that the BF in preference for the cored model over the cuspy one is always less than 10---the minimum benchmark for claiming statistically significant preference for a cored DM profile---and therefore, even in cases where we are unable to robustly identify the presence of a cusp, we would not falsely claim the presence of a core.

\begin{table}[t]
    \renewcommand{\arraystretch}{1.2}
    \begin{center}
    \begin{tabular}{|c|c|c|c|}
    \multicolumn{4}{c}{$\mathbf{I.\,\,}\boldsymbol{\gamma}\,\mathbf{=1,\,r_s=1\,kpc}$}\vspace{1.5mm} \\
    \hline
    $n_\mathrm{stars}$ & BF$_{1,0}=\frac{\mathrm{Pr}(d|\gamma=1)}{\mathrm{Pr}(d|\gamma=0)}$ & min(BF$_{1,0}$) & max(BF$_{1,0}$) \\
    \hline
    20 & $1.27_{-0.52}^{+0.70}$ & 0.44 & 2.60 \\
    100 & $1.18_{-0.40}^{+0.74}$ & 0.56 & 3.05 \\
    1000 & $1.77_{-0.96}^{+1.57}$ & 0.67 & 6.19 \\
    10,000 & $2.02_{-1.52}^{+54.45}$ & 0.22 & 225.70\\
    \hline
    \end{tabular}
    \end{center}
    \caption{Values of the Bayes Factor (BF) from fitting parameter set I with a model assuming a cusp ($\gamma=1$) relative to a model assuming a core ($\gamma=0$). The second column lists the median and lower/upper 1$\sigma$, while the third(fourth) column lists the minimum(maximum) BF value across the 10 datasets for each sample size. Of the 10,000-star samples, two realizations have $10\leq\mathrm{BF}_{1,0}<100$, providing strong evidence, and one realization has $\mathrm{BF}_{1,0}\geq100$, providing decisive evidence in favor of a cusp over a core.}
    \label{tab:BF_gamma_1}
\end{table}
\begin{table}[t]
    \begin{center}
    \renewcommand{\arraystretch}{1.2}
    \begin{tabular}{|c|c|c|c|}
    \multicolumn{4}{c}{$\mathbf{III.\,\,}\boldsymbol{\gamma}\,\mathbf{=0,\,r_s=1\,kpc}$}\vspace{1.5mm} \\
    \hline
    $n_\mathrm{stars}$ & BF$_{0,1}=\frac{\mathrm{Pr}(d|\gamma=0)}{\mathrm{Pr}(d|\gamma=1)}$ & min(BF$_{0,1}$) & max(BF$_{0,1}$) \\
    \hline
    20 & $0.58_{-0.06}^{+0.25}$ & 0.50 & 1.42 \\
    100 & $0.58_{-0.16}^{+0.19}$ & 0.27 & 1.78 \\
    1000 & $0.96_{-0.60}^{+2.83}$ & 0.17 & 6.78 \\
    10,000 & $256.74_{-235.60}^{+837.48}$ & 2.98 & 46971.85\\
    \hline
    \end{tabular}
    \end{center}
    \caption{Same as Table~\ref{tab:BF_gamma_1}, but for parameter set III, in this case comparing a model assuming a core ($\gamma=0$) to a model assuming a cusp ($\gamma=1$). Of the 10,000-star samples, seven realizations have BF$_{0,1}\geq100$, providing decisive evidence in favor of a cored distribution over a cuspy one.}
    \label{tab:BF_gamma_0}
\end{table}

We list the analogous results for parameter set III in Table~\ref{tab:BF_gamma_0}. In this case, the median BF for a sample size of 10,000 stars is decisively in favor of a model with a core. This is also consistent with the posterior distribution shown in the bottom rightmost panel of \Fig{fig:gamma_post_1_0}, which is peaked at $\gamma\sim0$, sharply drops near $\gamma\sim1$, and has relatively little spread across realizations. Importantly, across all sample sizes and realizations for parameter set I(III), for which the true DM profile is cuspy(cored), the BF in preference for a cored(cuspy) profile over a cuspy(cored) one is always less than 10. This demonstrates that, even when we are unable to recover statistical evidence for the true inner DM profile, we would not erroneously claim evidence for the wrong inner profile.

\begin{figure*}[t]
    \centering
    \includegraphics[width=0.95\textwidth]{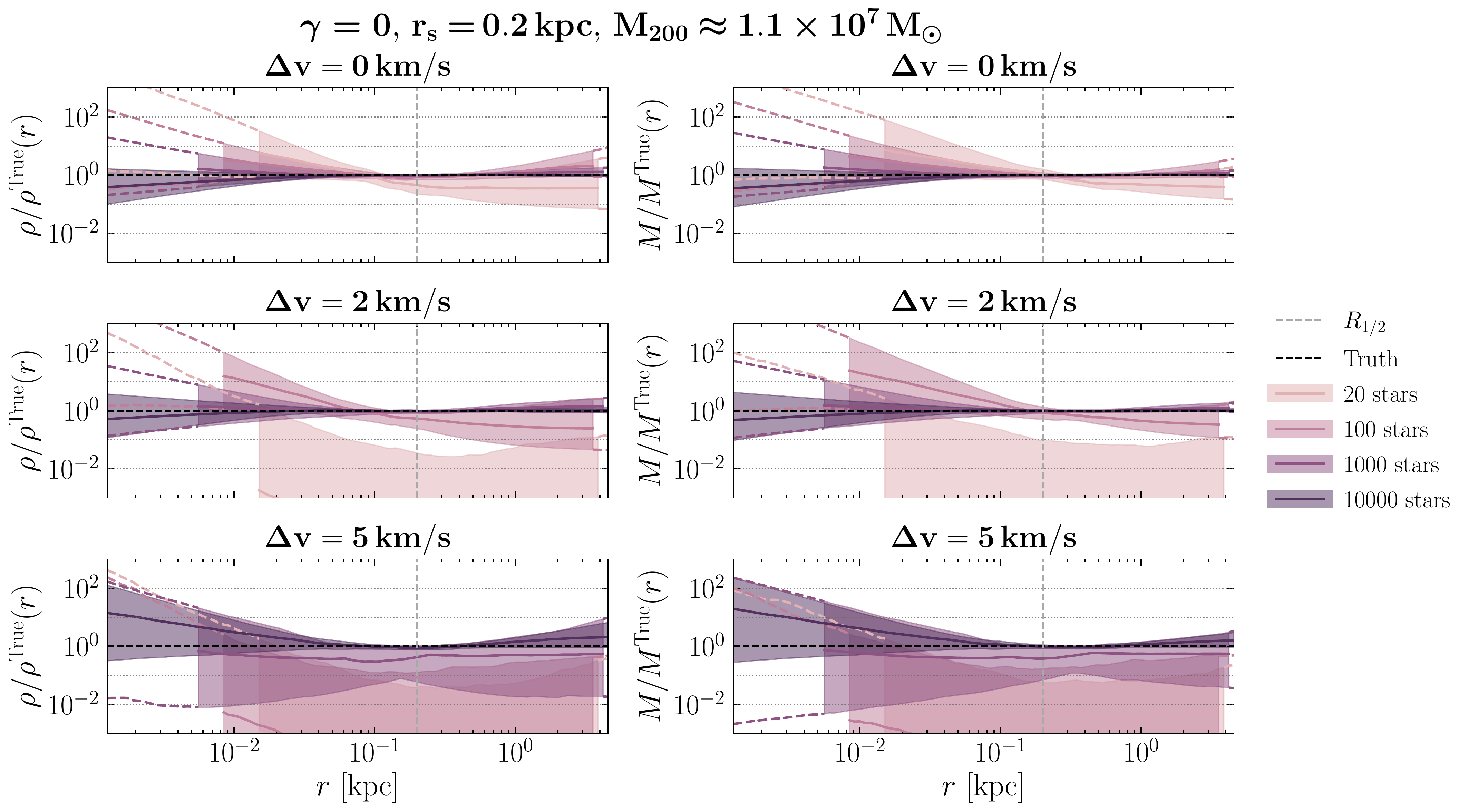}
    \caption{Fractional recovered density profiles (left panels) and enclosed mass profiles (right panels) for parameter set IV, varying over the line-of-sight velocity measurement error $\Delta v$ as well as the sample size. For each sample size, the solid line denotes the median (across our 10 independent realizations) of the median fractional recovered profiles, while the shaded band shows the median of the 68\% containment regions, plotted over the maximal radial range across all 10 datasets for that sample size. For the samples with fewer than 10,000 stars, we additionally extrapolate the median 68\% containment regions over the full radial range, shown bracketed by each pair of dashed lines in the color corresponding to the sample size. Varying $\Delta v$ has a particularly drastic effect on the smaller samples---for a sample size of 20 stars, a measurement error of $\Delta v=2\,\mathrm{km/s}$ is insufficient for recovering the DM density and enclosed mass profiles.}
    \label{fig:fractional_rho_Menc_gamma_0_rs_0p2}
\end{figure*}

Thus far, we have demonstrated that, for datasets with $\lesssim 1000$ measured stars---on par with the current measurements---we can robustly recover the total enclosed DM mass, but we cannot accurately reconstruct the inner profile or constrain the inner slope of the DM density distribution, even within our simplified framework. We have also tested samples with 5000 stars and found that the posterior $\gamma$ distributions were typically flat as well, demonstrating that in order to constrain $\gamma$ in our setup, a sample size of $\sim10,000$ stars is truly needed. In Section~\ref{sec:degeneracies}, we explore how degeneracies between DM model parameters contribute to the difficulty of recovering $\gamma$. In Section~\ref{sec:indirect}, we investigate how the limitations on being able to accurately reconstruct the full density profile---which we emphasize is related to, but separate from, the issue of constraining the posterior distribution of $\gamma$---may affect the results of indirect detection analyses.

\subsection{Velocity Uncertainties}
\label{sec:errors}

\begin{figure*}[t]
    \centering
    \includegraphics[width=0.95\textwidth]{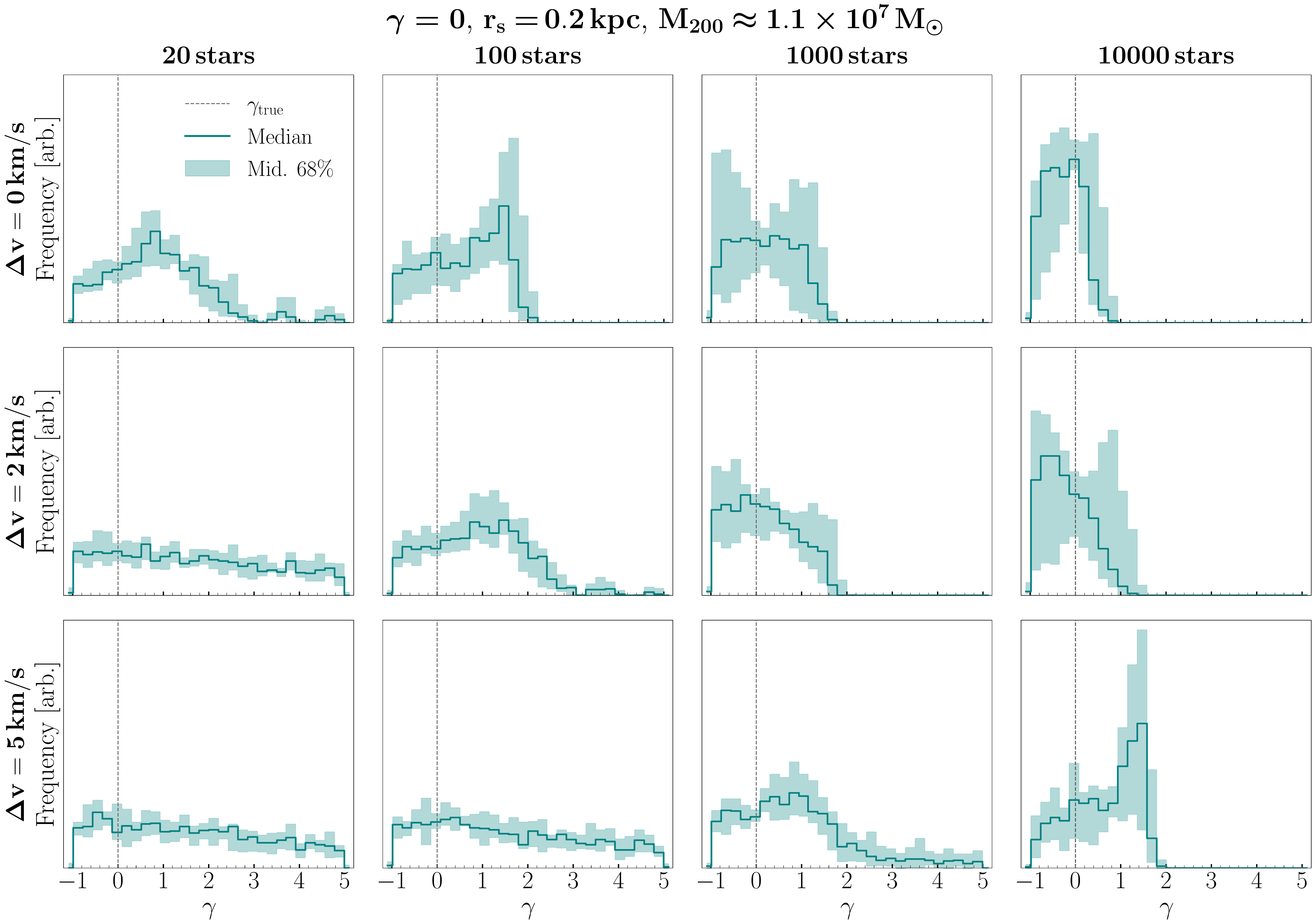}
    \caption{Posterior distributions for the inner slope $\gamma$, varying over the line-of-sight velocity measurement error $\Delta v$ as well as the sample size, shown for parameter set IV. These posteriors correspond to the results shown in Fig.~\ref{fig:fractional_rho_Menc_gamma_0_rs_0p2}. The lines(bands) show the median(middle 68\%) in each $\gamma$ bin across the 10 realizations. Varying $\Delta v$ has a drastic effect on the inference of $\gamma$ in this case; in contrast, for the more massive halo of parameter set III, varying the measurement error has negligible effect (Fig.~\ref{fig:gamma_post_all_verr_gamma_0_rs_1}). All panels in this figure share the same vertical scale.}
    \label{fig:gamma_post_all_verr_gamma_0_rs_0p2}
\end{figure*}

Looking towards future measurements, it is important to understand how increasingly precise measurements of line-of-sight velocities might affect our ability to reconstruct DM halo properties. To address this, we generate simulated datasets assuming different values of measurement error $\Delta v$ (uniform across all generated stars), and repeat our analysis setting $\Delta_{v_i}=\Delta v$ for all stars in \Eq{eq:likelihood}. We compare our fiducial results, which assume a measurement error of $\Delta v=2\,\mathrm{km/s}$, to results assuming a more conservative value of $\Delta v=5\,\mathrm{km/s}$, as well as results in the limit of perfect measurements, $\Delta v=0\,\mathrm{km/s}$.\footnote{These values are chosen for reasonable comparison to current spectrographs such as Keck/DEIMOS~\citep[e.g.][]{2007ApJ...670..313S,Martin:2007ic}, Magellan/IMACS~\citep[e.g.][]{2017ApJ...838...11S,2017ApJ...838....8L,2020ApJ...892..137S}, VLT/GIRAFFES+FLAMES~\citep[e.g.][]{2002Msngr.110....1P,2018ApJ...857..145L}, and APOGEE~\citep[e.g.][]{2017AJ....154...94M,2020AJ....160..120J}.} If the intrinsic velocity dispersion of a system is much larger than $\sim 5\,\mathrm{km/s}$, we do not expect varying $\Delta v$ in the range of 0--5 km/s to have a significant effect on the analysis results. On the other hand, if the intrinsic velocity dispersion is $\lesssim 5\,\mathrm{km/s}$, we expect the results to be dependent on the value of $\Delta v$, especially if the sample size is small. For parameter sets I and III discussed in Section~\ref{sec:n_stars}, the intrinsic velocity dispersion is $\sim10$--$15$ km/s. Parameter sets II and IV have the same DM inner slope and concentration as parameter sets I and III, respectively, but are approximately 100 times less massive and have an intrinsic velocity dispersion of $\sim2$--3 km/s. 

In Figure~\ref{fig:fractional_rho_Menc_gamma_0_rs_0p2}, we show the inferred fractional DM density and enclosed mass profiles for parameter set IV.\footnote{We choose to present parameter set IV here because it has the smallest intrinsic velocity dispersion out of all of our parameter sets, and therefore is most drastically affected by increasing $\Delta v$.} From top to bottom, the rows correspond to $\Delta v=0,\,2,\,5\,\mathrm{km/s}$. The results are consistent with our intuition: because the typical intrinsic velocity dispersion for this set of systems is $\sim2\,\mathrm{km/s}$, a measurement error $\Delta v\gtrsim2\,\mathrm{km/s}$ has a drastic effect on the inferred results, especially when combined with limited sample size. A value of $\Delta v=5\,\mathrm{km/s}$ results in an inferred virial mass of $M_{200}\sim0$ for both the 20- and 100-star samples (see Tab.~\ref{tab:mvir_deltav}). For the 20-star samples, even our fiducial choice of $\Delta v=2\,\mathrm{km/s}$ results in essentially no DM being recovered. This can be understood as the measurement error being large enough that the observed velocity dispersion can be statistically consistent with the complete absence of DM.

For the larger sample sizes, with 1000 and 10,000 stars, the Jeans analysis is able to recover the correct density profile even when the measurement errors are of the same order as the dispersion of the system. This can be attributed to the fact that with large enough statistics, the analysis can distinguish the radially-dependent velocity dispersion $\sigma_p(R)$ from the radially-independent measurement error. These results indicate that in order to obtain accurate virial mass estimates for the dwarf galaxies with fewer than $\sim1000$ observed stars (see Tab.~\ref{tab:J_n_stars} for some examples of observed dwarf galaxies), it is crucial that the measurement error on the line-of-sight velocities be subdominant to the intrinsic velocity dispersion.

For parameter set III, which has the same DM inner slope and concentration as parameter set IV but is 100 times more massive, varying the measurement error has negligible effect on the inferred DM halo properties as expected (shown in Fig.~\ref{fig:fractional_rho_Menc_gamma_0_rs_1} of the Appendix). The results for parameter sets I and II (which have $\gamma=1$) are qualitatively similar to the results for parameters sets III and IV (which have $\gamma=0$), although quantitatively different due to slightly larger values of the intrinsic velocity dispersion; we present those results in Figs.~\ref{fig:fractional_rho_Menc_gamma_1_rs_1}--\ref{fig:fractional_rho_Menc_gamma_1_rs_0p2}. 

Similarly, $\Delta v$ affects the recovery of the inner slope more for the less massive halos than for the more massive ones. Figure~\ref{fig:gamma_post_all_verr_gamma_0_rs_0p2} shows the posterior $\gamma$ distributions corresponding to the scans shown in Fig.~\ref{fig:fractional_rho_Menc_gamma_0_rs_0p2}. As $\Delta v$ is increased, $\gamma$ becomes increasingly unconstrained for the smaller sample sizes, whereas for the 10,000-star samples, increasing $\Delta v$ appears to lead to a bias in the best-fit value of $\gamma$. For the more massive halo with the same inner slope (parameter set III), the posterior $\gamma$ distributions are mostly insensitive to these variations in the measurement error (shown in Fig.~\ref{fig:gamma_post_all_verr_gamma_0_rs_1}). The corresponding posterior $\gamma$ distributions for parameter sets I and II are shown in Figs.~\ref{fig:gamma_post_all_verr_gamma_1_rs_1}--\ref{fig:gamma_post_all_verr_gamma_1_rs_0p2}, and are qualitatively similar to the cases of parameter sets III and IV, respectively.

\subsection{Location of stars}
\label{sec:location}

\begin{figure*}[t]
    \centering
    \includegraphics[width=0.8\textwidth]{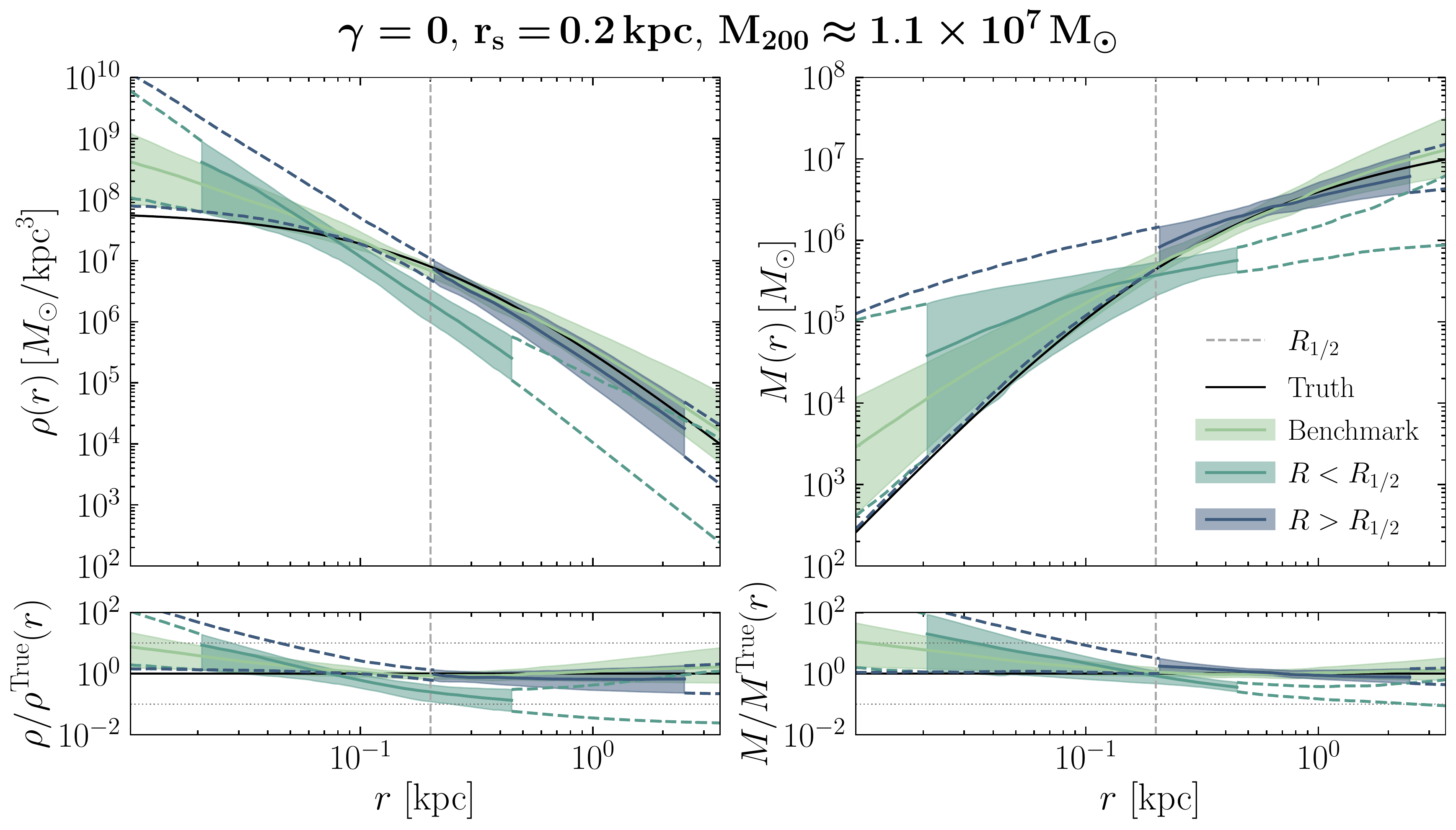}
    \caption{Inferred DM density profiles $\rho(r)$ (left panels) and corresponding enclosed mass profiles $M(r)$ (right panels) for parameter set IV, starting with a sample size of $\nstars=100$ stars (resulting in selected samples of $\nstarssel\sim50$ stars), with spatial selection functions applied. From lightest to darkest color, the results are for the benchmark datasets, the datasets keeping only stars with $R<R_{1/2}$, and the datasets keeping only stars with $R>R_{1/2}$. For each selection function, the solid line denotes the median (across our 10 independent realizations) of the median recovered profiles, while the shaded band shows the median of the 68\% containment regions; these are plotted from the median $r_\mathrm{min}$ to the median $r_\mathrm{max}$ across the 10 realizations, where $r_\mathrm{min}$($r_\mathrm{max}$) is the galactocentric radius of the innermost(outermost) star. We additionally extrapolate the median 68\% containment regions over the full radial range of the benchmark samples, shown bracketed by each pair of dashed lines in the color corresponding to the selection function. As expected, compared to the benchmark scenario, when the measured stars are all in the outer region of the dwarf, the DM profile is poorly constrained at small radii. Conversely, when the measured stars are all in the inner region of the dwarf, the DM profile is poorly constrained at larger radii. Moreover, the inner density profile is also less well-constrained for the $R<R_{1/2}$ case than for the benchmark scenario, suggesting that in order to constrain the inner DM profile, it is important to have measured stars across the full radial distribution, and not only in the inner region.}
    \label{fig:rho_Menc_gamma_0_rs_0p2_drop_stars}
\end{figure*}

In this section, we explore how the DM inference depends not only on how many stars are measured, but also on where the observed stars are within the dwarf galaxy. To study this effect, we start with our datasets of initial size $\nstars$ and apply the following selection functions, then repeat our analysis on the resulting datasets (where $R_{1/2}$ is the projected half-light radius):
\begin{itemize}
\item Inner stars analysis: keep only the stars in the inner region, with projected radius $R<R_{1/2}$.
\item Outer stars analysis: keep only the stars in the outer region, with projected radius $R>R_{1/2}$.
\end{itemize}
To account for the $\sim50\%$ change in the number of stars from implementing these selection functions, we compare the results to ``benchmark'' results on datasets with $\nstars/2$ stars which are also generated from the original $\nstars$-star datasets, subsampled uniformly to preserve the radial probability distribution of the original dataset. In doing so, we can compare the results for datasets that have approximately equal numbers ($\sim\nstars/2$) but distinct spatial distributions of stars. 

As before, we generate 10 independent datasets for each selection function. In Figure~\ref{fig:hist_gamma_0_drop_stars} of the Appendix, we show the distributions of the projected radius $R$ as well as the 3d radius $r$ for parameter set III with $\nstars=100$ (which is qualitatively representative of the distributions for all the parameter sets and sample sizes), for the three different selection functions. We note that, because we implement the selection function on the projected radius, and $r\geq R\,$ for all values of $R$, the $R<R_{1/2}$ datasets extend slightly beyond a 3d radius of $r=R_{1/2}$. We test the effect of selection functions on datasets with initial sizes of $\nstars=100,1000,\,\mathrm{and}\,10,000$ stars for each of the four parameter sets. For the purpose of studying the effects of spatial distributions in the cleanest setup, the studies presented in this section have been performed assuming $\Delta v=0$ km/s.

In Figure~\ref{fig:rho_Menc_gamma_0_rs_0p2_drop_stars}, we show the recovered DM density and enclosed mass profiles for the three different selection functions, for a particularly demonstrative example. This example is for parameter set IV, with an initial sample size of $\nstars=100$ stars; after applying each of the selection functions, we end up with a selected sample size of $\nstarssel\sim 50$ stars.  From lightest to darkest color, we show the results for the benchmark, $R<R_{1/2}$, and $R>R_{1/2}$ datasets. Like before, the solid lines denote the median across the 10 realizations of the median recovered profiles, while the shaded bands depict the median of the 68\% containment ranges across the realizations. For ease of presentation, we choose in this case to show the solid line and band for each selection function from the median $r_\mathrm{min}$ to the median $r_\mathrm{max}$ across the realizations, where $r_\mathrm{min}(r_\mathrm{max})$ is the galactocentric distance of the innermost(outermost) star in each individual realization. Beyond this range, we extrapolate the median 68\% containment ranges, shown by each pair of dashed lines in the color corresponding to the selection function.

As expected, when the measured stars are all in the outer region of the dwarf, the DM profile is poorly constrained at small radii compared to the benchmark scenario. Conversely, when the measured stars are all in the inner region of the dwarf, the DM profile is poorly constrained at larger radii. Interestingly, for the $R<R_{1/2}$ samples in this example, the DM profile is also typically less well-constrained at small radii; additionally, the density profile is biased high at small radii and low at large radii, to the extent that the total enclosed mass is also biased low (the recovered virial mass is $M_{200}\sim\errorbars{0.2}{0.7}{0.1}\times 10^7\msun$, while the true value is $M_{200}\sim1.1\times10^7\msun$). These biases, as well as the larger uncertainties on the DM profile in both the inner and outer regions, are present in spite of there being approximately twice as many stars within the half-light radius in the $R<R_{1/2}$ datasets as in the benchmark datasets. In this particular example, the posterior $\gamma$ distribution is unconstrained for all three selection functions due to the small size of the dataset, so we do not recover a corresponding bias in $\gamma$. 

The specific behavior of the results for the $R<R_{1/2}$ samples noted in this example is not generic to all the variations we have tested---in particular, for the datasets with larger selected sample size $\nstarssel$, the bias in the DM density profile is less severe, and in some cases the median 68\% containment band on the inner density profile is slightly narrower than in the benchmark case. This can be seen in Figure~\ref{fig:rho_Menc_gamma_0_rs_0p2_drop_stars_100} of the Appendix, which is the same as \Fig{fig:rho_Menc_gamma_0_rs_0p2_drop_stars}, except for an initial sample size of $\nstars=1000$, i.e., for spatially selected datasets of size $\nstarssel\sim500$. 

We can quantitatively compare the performance of the different selection functions, for different sample sizes $\nstarssel$, by comparing the recovered virial mass estimates as well as the recovered {\jfac}s (discussed more in \Sec{sec:indirect}), both detailed in Table~\ref{tab:mvir_Jfacs_drop_stars}. Across our four parameter sets, the results on spatial selection functions are the following:

\begin{itemize}
    \item Inner stars analysis ($R<R_{1/2}$)
    \begin{itemize}
        \item For the smallest sample size $\nstarssel\sim50$, for all parameter sets, the inferred virial mass is systematically underestimated (inconsistent with the true value within $1\sigma$ uncertainty for three of the four parameter sets). This becomes less severe as the sample size is increased, but across all four parameter sets for the larger sample sizes $\nstarssel\sim500$ and $\nstarssel\sim5000$, the uncertainty on the estimated virial mass is consistently larger than for either the $R>R_{1/2}$ datasets or the benchmark case, demonstrating that to achieve an accurate virial mass estimate, it is important to have measurements of outer stars.
        \item The behavior of the posterior $\gamma$ distribution varies across different sample sizes and different parameter sets---in some cases, the posterior $\gamma$ distribution is biased high when the selection function is applied; in other cases, it is unchanged from the posterior distribution in the benchmark case. In all cases, the $R<R_{1/2}$ selection function does not improve the ability of the method to accurately constrain $\gamma$, relative to the benchmark case. Therefore, for the purpose of constraining $\gamma$, additional stars need to be measured across all radii. 
        \item As we will discuss in \Sec{sec:indirect}, for the smallest sample size $\nstarssel\sim50$, for all parameter sets, the uncertainty on the {\jfac} estimate is larger than in the benchmark case. For the larger sample sizes, the uncertainty on the {\jfac} estimate is comparable to or slightly ($\mathcal{O}$(0.1 dex)) smaller than in the benchmark case. 
    \end{itemize}
    \vspace{0.9em}
    \item Outer stars analysis ($R>R_{1/2}$)
    \begin{itemize}
        \item For all parameter sets and all sample sizes $\nstarssel$, the estimated virial mass is consistent with the true value, and the uncertainty on the virial mass estimate is comparable to or slightly smaller than in the benchmark case, demonstrating that having measurements of inner stars is not crucial to the recovery of the virial mass.
        \item For all parameter sets and all sample sizes $\nstarssel$, the posterior $\gamma$ distribution is comparable to (when the benchmark posterior distribution is already unconstrained) or less constrained than in the benchmark case. 
        \item As we will discuss in \Sec{sec:indirect}, for all parameter sets and all sample sizes $\nstarssel$, the uncertainty on the {\jfac} estimate is comparable to or larger than in the benchmark case, indicating that having measurements of inner stars is important for the purpose of constraining {\jfac}s.
    \end{itemize}
\end{itemize}

While the $R>R_{1/2}$ datasets perform slightly better in terms of the uncertainty on the recovered virial mass relative to the two other selection functions, the improvement is marginal (see Table~\ref{tab:mvir_Jfacs_drop_stars} for values). Therefore, based on the overall performance at inferring the full DM density profile and the inner slope $\gamma$, especially for the smallest samples, we find that it is crucial to have measurements of stars across the full radial distribution of the dwarf galaxy. Doing so allows the fit to anchor the DM distribution across the full radial range, and consistently results in comparable or better performance at accurately reconstructing both the inner and outer profile of the DM distribution, relative to the cases when the data consists purely of stars in either the inner or outer region of the system. 

\subsection{Role of Degeneracies}
\label{sec:degeneracies}

\begin{figure}[t]
    \centering
     \includegraphics[width=0.47\textwidth]{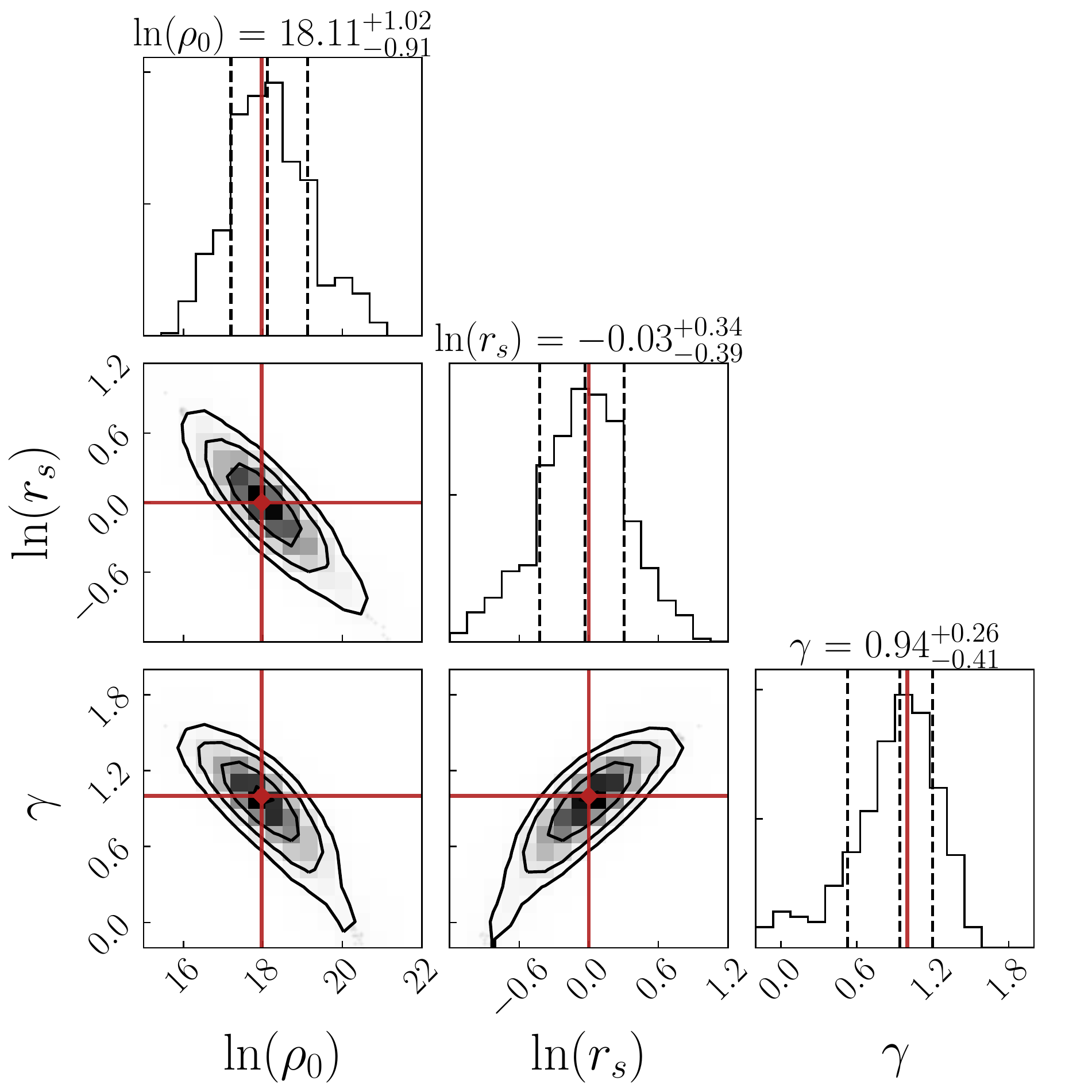}
    \caption{An example triangle plot of the posterior DM parameters from a scan of a 10,000-star sample from parameter set I, with $\Delta v=0$ km/s. While the parameters are converged about their true values (red lines), there are significant degeneracies between pairs of parameters.}
    \label{fig:corner_10k}
\end{figure}

\begin{figure}[thb]
    \centering
     \includegraphics[width=0.47\textwidth]{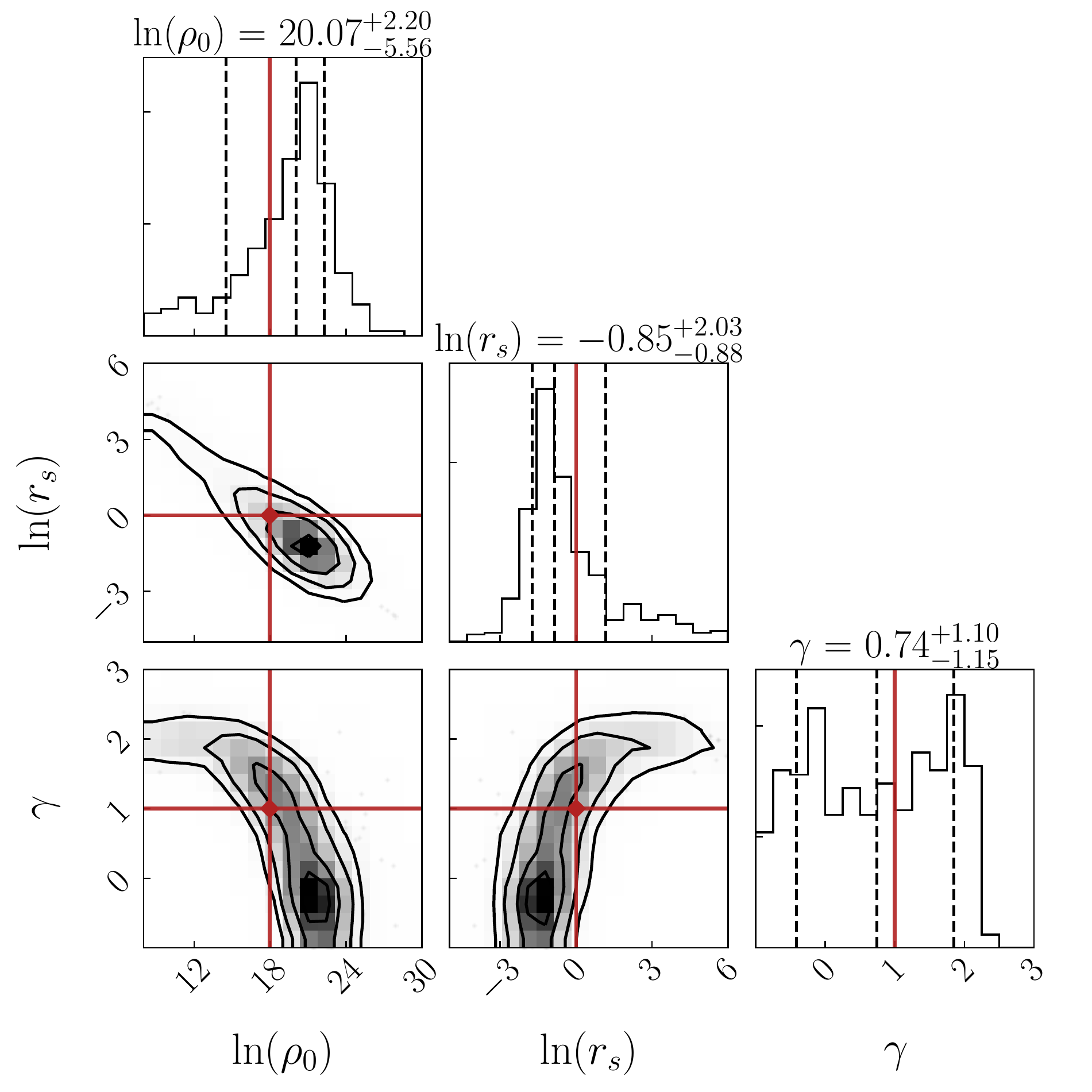}
    \caption{An example triangle plot of the posterior DM parameters from a scan of a 100-star sample from parameter set I, with $\Delta v=0$ km/s. Compared to the posteriors from the larger sample size shown in Figure~\ref{fig:corner_10k}, the parameters are much more poorly constrained in this case (note the wider axis ranges in this case compared to Fig.~\ref{fig:corner_10k}), and $\gamma$ is unconstrained at low values.}
    \label{fig:corner_100}
\end{figure}

\begin{figure*}[t]
    \centering
     \includegraphics[width=0.95\textwidth]{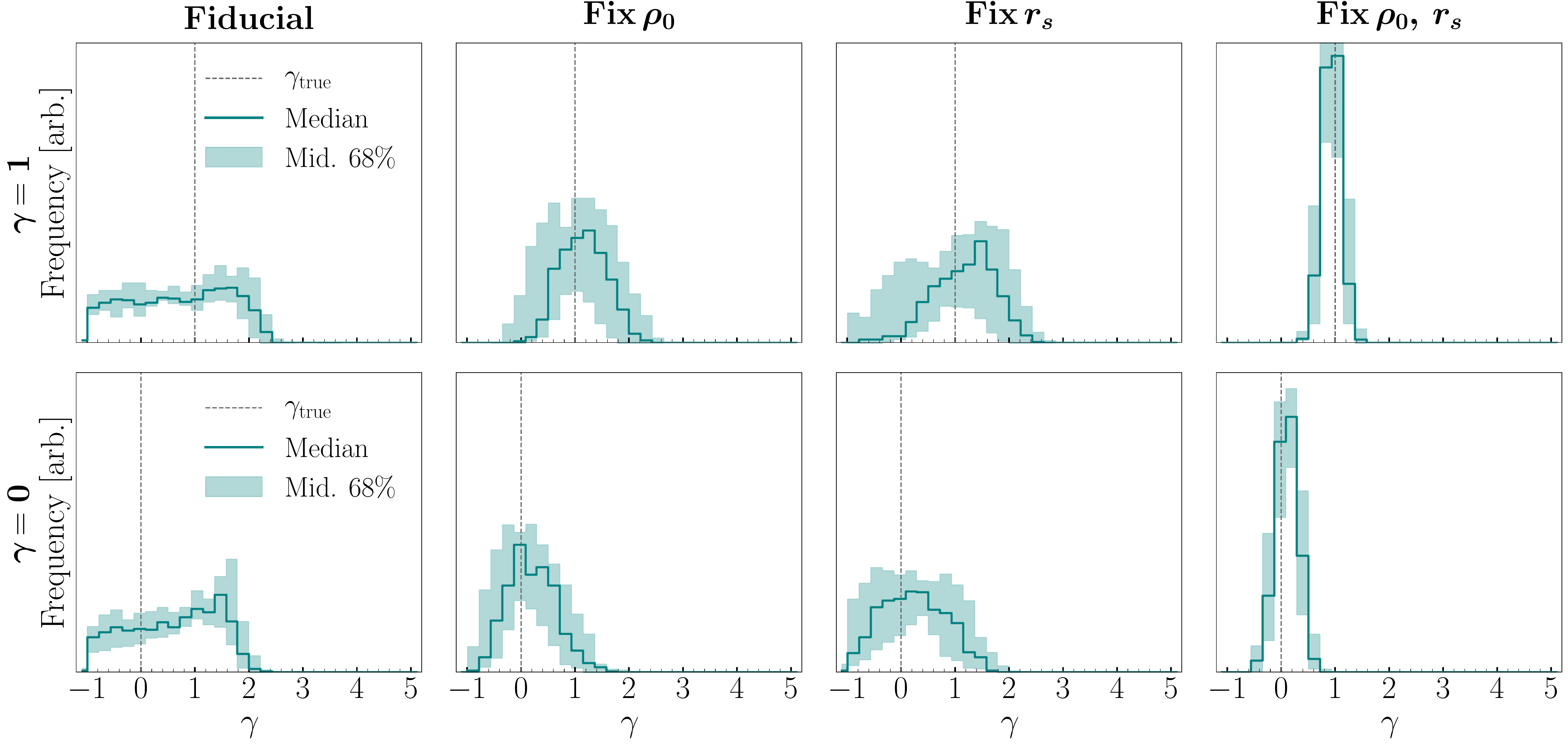}
    \caption{Posterior $\gamma$ distributions for 100-star samples from parameter sets I (top row) and III (bottom row), with $\Delta v=2$ km/s. The lines(bands) show the median(middle 68\%) in each $\gamma$ bin across the 10 realizations. We show the results for our fiducial setup (first column), fixing $\rho_0$ to its true value (second column), fixing $r_s$ to its true value (third column), or fixing both $\rho_0$ and $r_s$ to their respective true values (fourth column). Breaking the degeneracies between $\rho_0$, $r_s$, and $\gamma$ by holding $\rho_0$ and/or $r_s$ fixed gives rise to improved constraining power on $\gamma$.}
    \label{fig:fix_params_rs_1}
\end{figure*}

One of the factors that limits the accurate recovery of the inner slope of the DM density profile is degeneracy between the different halo parameters---different combinations of the normalization $\rho_0$, scale radius $r_s$, and inner slope $\gamma$ can result in similar enclosed mass profiles, and therefore are equally valid descriptors of the kinematic data. This is manifest in Figure~\ref{fig:corner_10k}, an example triangle plot of the posterior halo parameters from analyzing a single 10,000-star dataset. In this clean example, the fit converges near the true values of $\rho_0$, $r_s$, and $\gamma$, but there are clear degeneracies between each of the pairs of parameters. Such degeneracies make it increasingly difficult to constrain the value of $\gamma$ as the sample size is decreased. 

For comparison, Figure~\ref{fig:corner_100} shows an example triangle plot of the posterior halo parameters from analyzing a single 100-star dataset. Again, there are clear degeneracies between the pairs of parameters. In this case, all of the DM parameters are more poorly constrained (note the wider axis ranges compared to Fig.~\ref{fig:corner_10k}), and in particular the posterior $\gamma$ distribution is almost entirely flat down to the lower edge of our prior range. We emphasize that our choice of parameterization for the DM distribution is simpler than the Hernquist/Zhao parameterization widely employed in the literature \citep[e.g.,][]{Bonnivard:2015xpq,2015MNRAS.446.3002B,Ando:2020yyk}, which has five parameters. Given the extra degrees of freedom in that model, the role of degeneracies would present an even bigger challenge for constraining the inner slope of the DM distribution, especially in the case of statistics-limited datasets.

We can explicitly break the degeneracies in our halo model by holding $\rho_0$ or $r_s$ (or both) fixed to their true values and examining the resulting posterior distributions for $\gamma$. We show the results for parameter set I(III), for a sample size of 100 stars and $\Delta v=2$ km/s, in the top(bottom) panel of Figure~\ref{fig:fix_params_rs_1}. Fixing $\rho_0$ (second column) or fixing $r_s$ (third column) result in a posterior $\gamma$ distribution which is peaked near the true value of $\gamma$, with slightly more constraining power in the case of fixing $\rho_0$. This makes sense intuitively because the inner region of the DM distribution is directly sensitive to $\rho_0$ and $\gamma$, whereas $r_s$ more directly influences the distribution at intermediate radii, and therefore breaking the degeneracy between the former two parameters is more effective at improving the constraint on $\gamma$. If we fix both $\rho_0$ and $r_s$ to their respective true values (fourth column), we recover the true inner slope with high accuracy. 

We have thus demonstrated that, even for our simplified mock datasets and three-parameter DM halo model, the dimensionality of the problem is large enough that constraining the inner slope of the DM density profile for moderately sized stellar samples proves to be difficult. These challenges would be further exacerbated when one takes into account velocity anisotropy, which is difficult to accurately model and is also degenerate with the mass profile~\citep[e.g.,][]{1990AJ.....99.1548M,Wilkinson:2001ut,Lokas:2003ks,DeLorenzi:2008zq,2017MNRAS.471.4541R,Genina:2019job}. 

While it may not be well-motivated to hold DM halo parameters fixed in an analysis on real data, one could ameliorate the effect of parameter degeneracies by setting model-informed priors on the halo parameters~\citep{Strigari:2006rd,Ando:2020yyk}---for example, if one were to assume a specific mass-concentration relation, there would consequently be a specific relation between the normalization $\rho_0$ and scale radius $r_s$, and the priors for those parameters would no longer be independent of each other. Additionally, \cite{Hayashi:2020jze} recently demonstrated that non-spherical mass models can alleviate the effect of parameter degeneracies. 

A separate approach to mitigating the effect of parameter degeneracies is to jointly analyze multiple dwarf galaxies at once, under the assumption that the systems share certain properties---in the simplest case, one could assume that the systems all share the same value of $\gamma$. We discuss the joint analysis approach in more detail in Appendix~\ref{sec:stacking}. We note that a thorough study of the joint analysis method is computationally infeasible within our current analysis framework, because the dimensionality of the model quickly grows with the number of jointly analyzed systems, to the degree that it is highly inefficient to use traditional MCMC or nested sampling methods to sample the posterior probability distributions.

\begin{figure*}[t]
    \centering
    \includegraphics[width=0.47\textwidth]{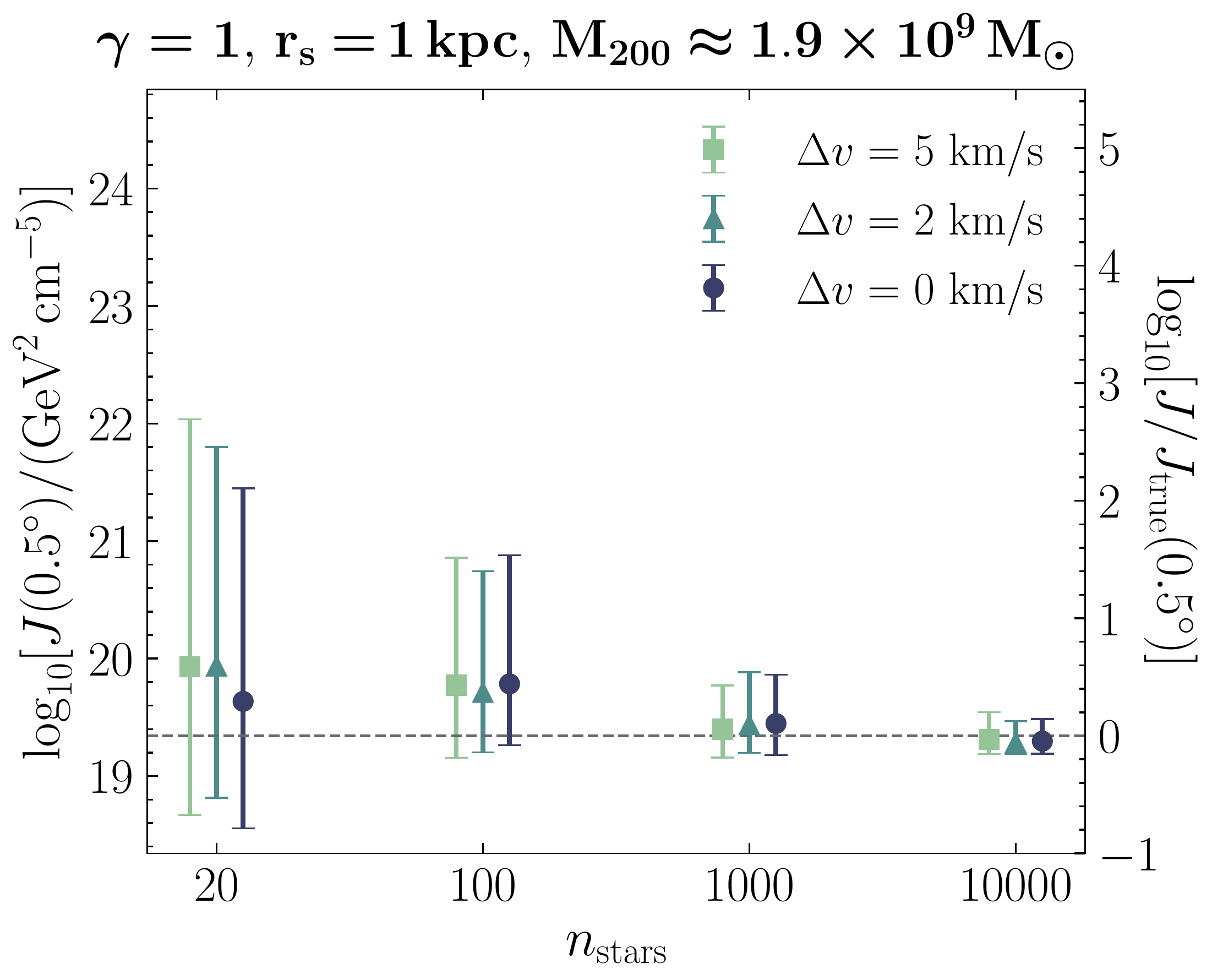}
    \hspace{1.3em}
    \includegraphics[width=0.47\textwidth]{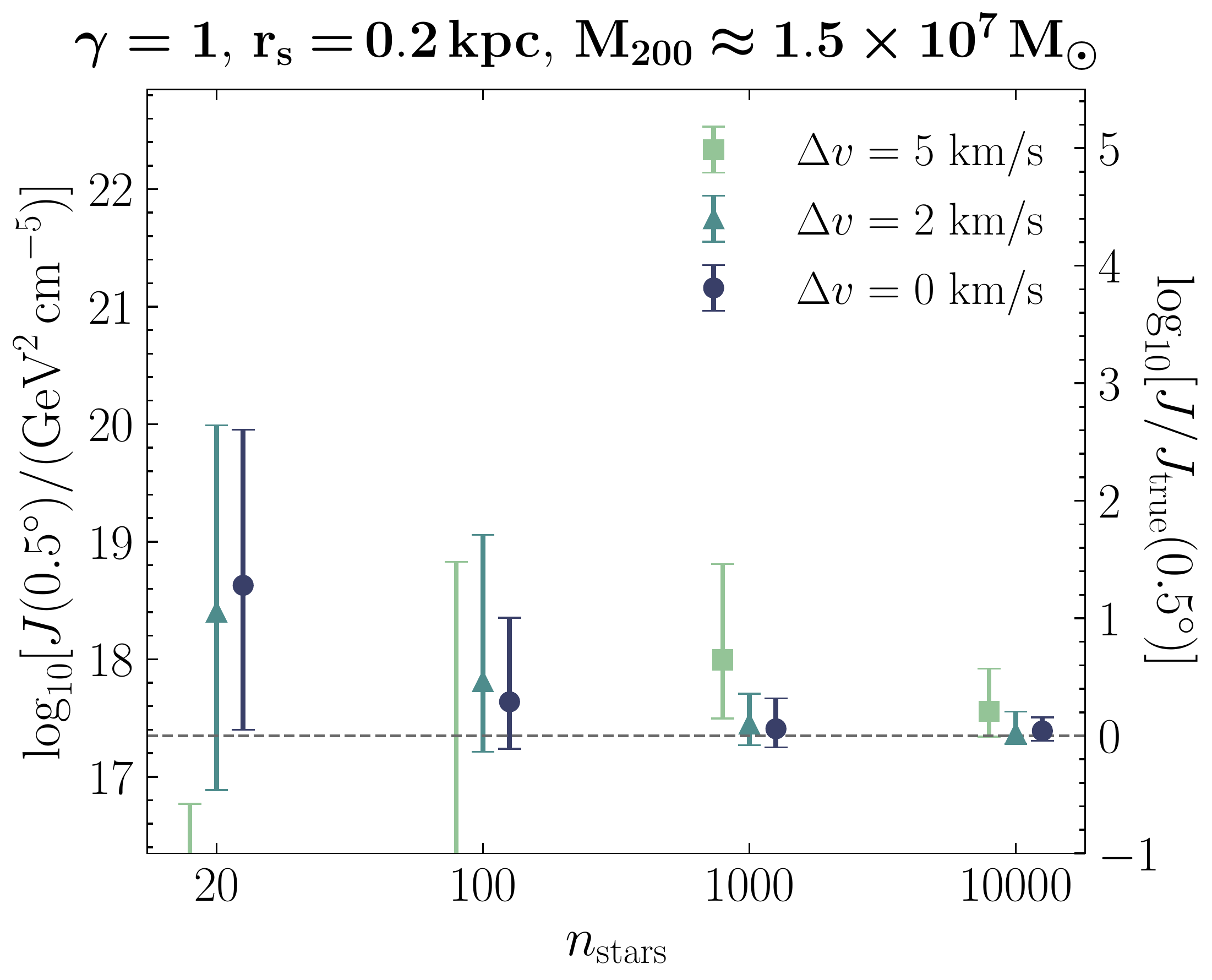}

     \includegraphics[width=0.47\textwidth]{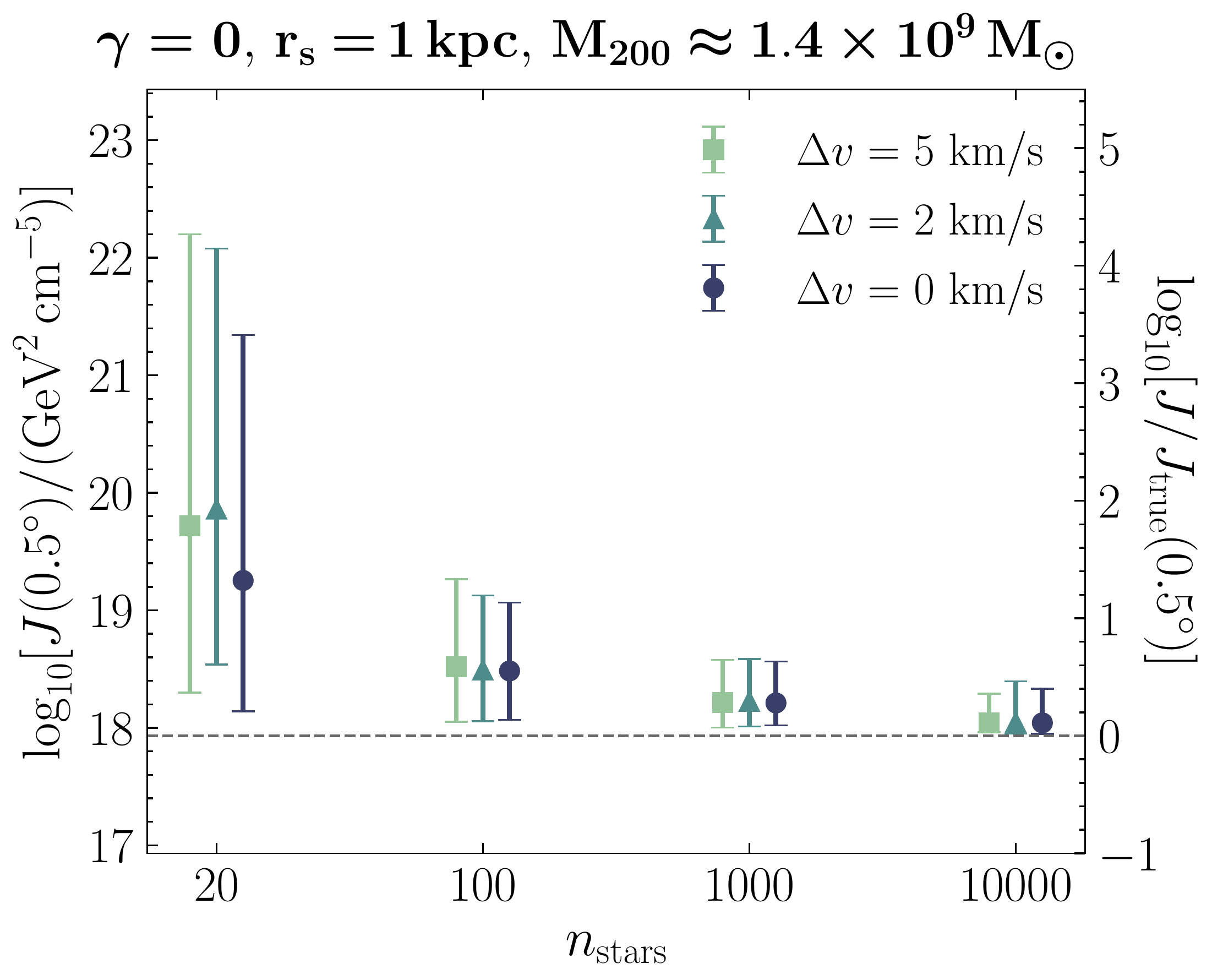}
     \hspace{1.3em}
     \includegraphics[width=0.47\textwidth]{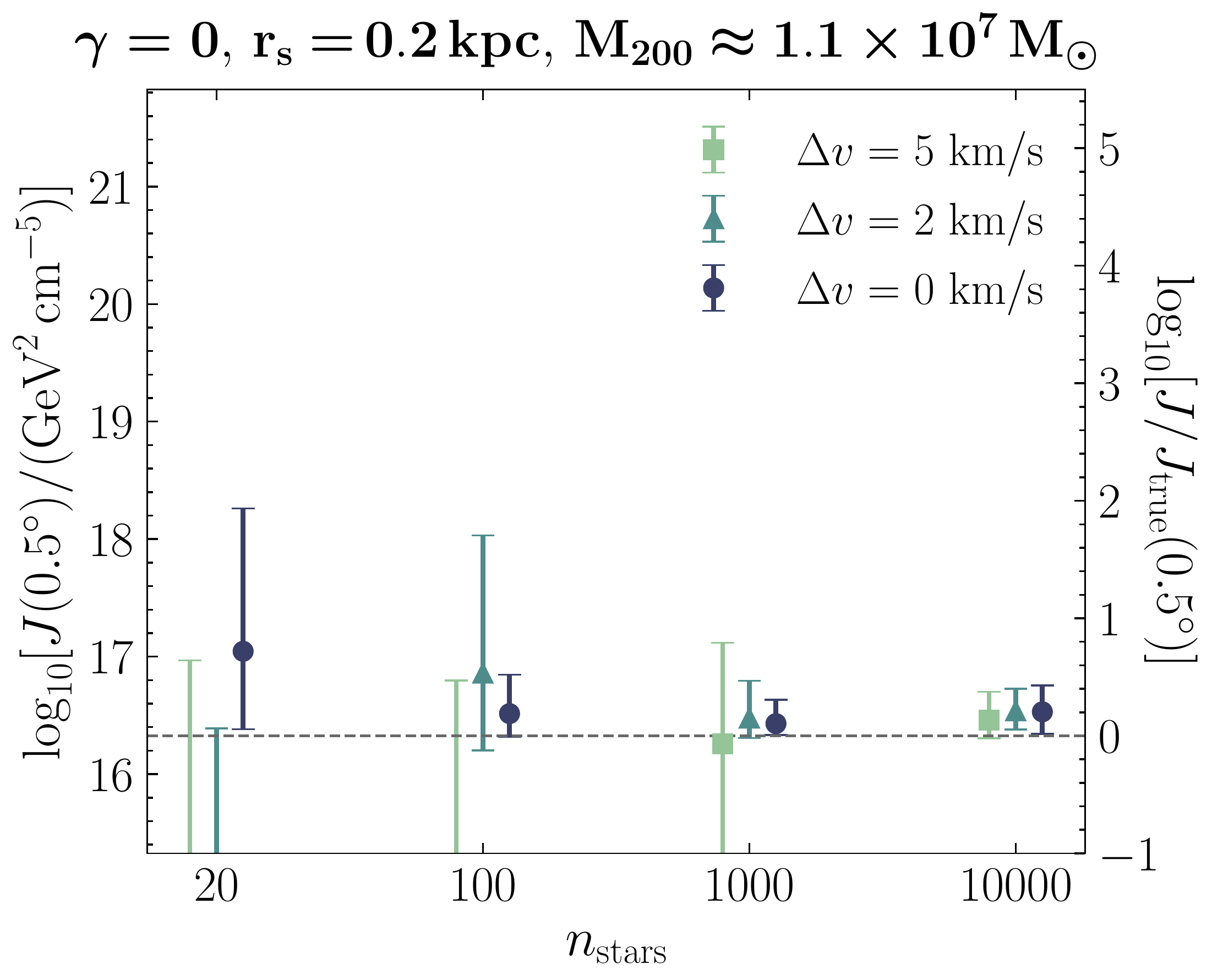}
    \caption{{\jfac}s as a function of the sample size and the measurement error on the line-of-sight velocities, $\Delta v$, for parameter sets I (top left), II (top right), III (bottom left), and IV (bottom right). We take the distance to the dwarf in each case to be 50 kpc and the angle of integration to be $0.5^\circ$. The results for $\Delta v=2\,\mathrm{km/s}$ in the top(bottom) left panel correspond to the recovered DM profiles shown in \Fig{fig:rho_Menc_gamma_1}(\ref{fig:rho_Menc_gamma_0}). For each realization of a given sample size and value of $\Delta v$, we build up a posterior {\jfac} distribution by calculating the {\jfac} for every set of posterior parameters, from which we can calculate the median and middle 68\% containment range of $\log_{10}[J(0.5^\circ)/\mathrm{(GeV^2\,cm^{-5})}]$ for that realization. Each data point shown here denotes the median across the 10 realizations of the median $\log_{10}[J(0.5^\circ)/\mathrm{(GeV^2\,cm^{-5})}]$, and each set of error bars brackets the median across realizations of the 68\% containment. The data points that extend below the range of the right panels correspond to the cases discussed in Section~\ref{sec:errors}, where the recovered DM abundance is significantly underestimated. The values of all plotted {\jfac}s are provided in Table~\ref{tab:mvir_deltav}.}
    \label{fig:Jfacs_all}
\end{figure*}

\section{Implications for Indirect Detection}
\label{sec:indirect}

In this section, we cast the results of our study into the context of indirect detection by calculating the inferred {\jfac}s for the tests discussed in Section~\ref{sec:results}, using the public code \textsc{CLUMPY}~\citep{2012CoPhC.183..656C,Bonnivard:2015pia,2019CoPhC.235..336H} to the perform the {\jfac} computations (as defined in \Eq{eq:jfac}). We examine the effects of sample size and line-of-sight velocity measurement error (Sec.~\ref{sec:jfac_mocks}), choices of priors in the Jeans analysis (Sec.~\ref{sec:GS15}), and spatial selection functions (Sec.~\ref{sec:jfac_selection}) on the inferred {\jfac}s. In Sec.~\ref{sec:observations}, we discuss the implications of our findings on indirect detection results and make recommendations for future observations. 

\subsection{Sample Size and Measurement Error}
\label{sec:jfac_mocks}

\begin{figure*}[t]
    \centering
     \includegraphics[width=0.7\textwidth]{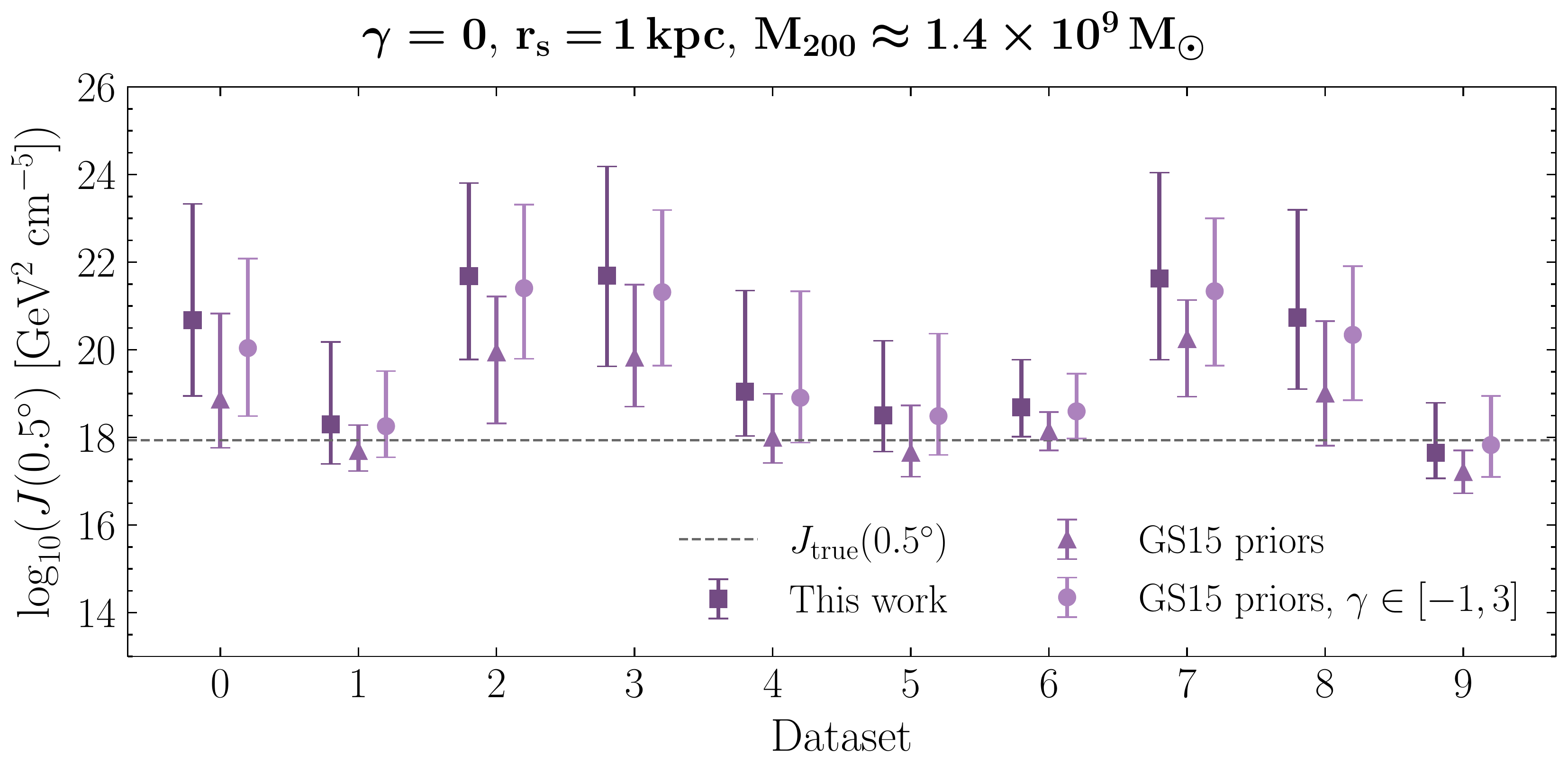}
    \caption{Comparing the {\jfac} results for each of the ten 20-star datasets in parameter set III: using our fiducial analysis setup (dark purple squares), using the DM priors from \cite{Geringer-Sameth:2014yza} (medium purple triangles), and using the priors on the normalization $\rho_0$ and scale radius $r_s$ from \cite{Geringer-Sameth:2014yza} while setting the prior on the inner slope to be $\gamma\in[-1,3]$ (light purple circles). This demonstrates that for these small sample sizes, the results are highly prior-dependent, which is consistent with our findings that the DM profile and inner slope are poorly constrained for the datasets with limited statistics. This additionally demonstrates that the remarkably small uncertainties on the {\jfac}s from \cite{Geringer-Sameth:2014yza} for the ultrafaint dwarfs may be driven by their narrow choice of prior on $\gamma$.}
    \label{fig:compare_Jfacs_gamma_0_rs_1}
\end{figure*}

First, we examine the effects of sample size and velocity measurement error, $\Delta v$, on the inferred {\jfac}s. In the left column of Figure~\ref{fig:Jfacs_all}, we show the inferred {\jfac}s for parameter sets I (top) and III (bottom), for which $M_{200}\sim10^9\msun$, for the different sample sizes and values of $\Delta v$. For an individual scan, we evaluate the {\jfac} for each set of posterior parameters, assuming a distance of $50$ kpc to the dwarf.\footnote{We have verified that qualitatively, our results on the {\jfac} uncertainty are unchanged if we instead assume a distance of $100$ kpc to the dwarf.} Each data point in Fig.~\ref{fig:Jfacs_all} shows the median across our 10 realizations of the median and middle 68\% containment range for the inferred values of $\log_{10}[J(0.5^\circ)/\mathrm{(GeV^2\,cm^{-5})}]$. Within each cluster of three data points corresponding to a particular sample size, the blue circle, teal triangle, and green square show the results for $\Delta v=0,\,2,\,5\,\mathrm{km/s}$, respectively. 

As expected, the uncertainties on the {\jfac} decrease as a function of increasing sample size. Additionally, the {\jfac}s are nearly independent of $\Delta v$, which is expected for parameter sets I and III (see Sec.~\ref{sec:errors} for a discussion on the effects of $\Delta v$). For parameter set I, our estimates of the {\jfac} are on average consistent with the true value for all sample sizes and values of $\Delta v$. For parameter set III, our estimates of the {\jfac} are systematically biased high, although the median values are within a factor of 2 of the true values for the 1000- and 10,000-star samples---this is consistent with the inner density profiles being biased high for the smaller samples, as shown in Fig.~\ref{fig:rho_Menc_gamma_0}. The typical values of the {\jfac} we recover for the different combinations of parameter set, sample size, and $\Delta v$ are tabulated in Table~\ref{tab:mvir_deltav}. 

For parameter sets II and IV (shown in the top right and bottom right panels of Fig.~\ref{fig:Jfacs_all}, respectively), the {\jfac} estimates are highly sensitive to $\Delta v$, in a manner that is consistent with the results discussed in Sec.~\ref{sec:errors} (the corresponding fractional recovered density and enclosed mass profiles are shown in Fig.~\ref{fig:fractional_rho_Menc_gamma_1_rs_0p2} for parameter set II and Fig.~\ref{fig:fractional_rho_Menc_gamma_0_rs_0p2} for parameter set IV). In particular, the data points that extend below the range of the right panels correspond to the cases of larger $\Delta v$ where the recovered DM abundance is significantly underestimated.

\begin{figure}[t]
    \centering
    \includegraphics[width=0.48\textwidth]{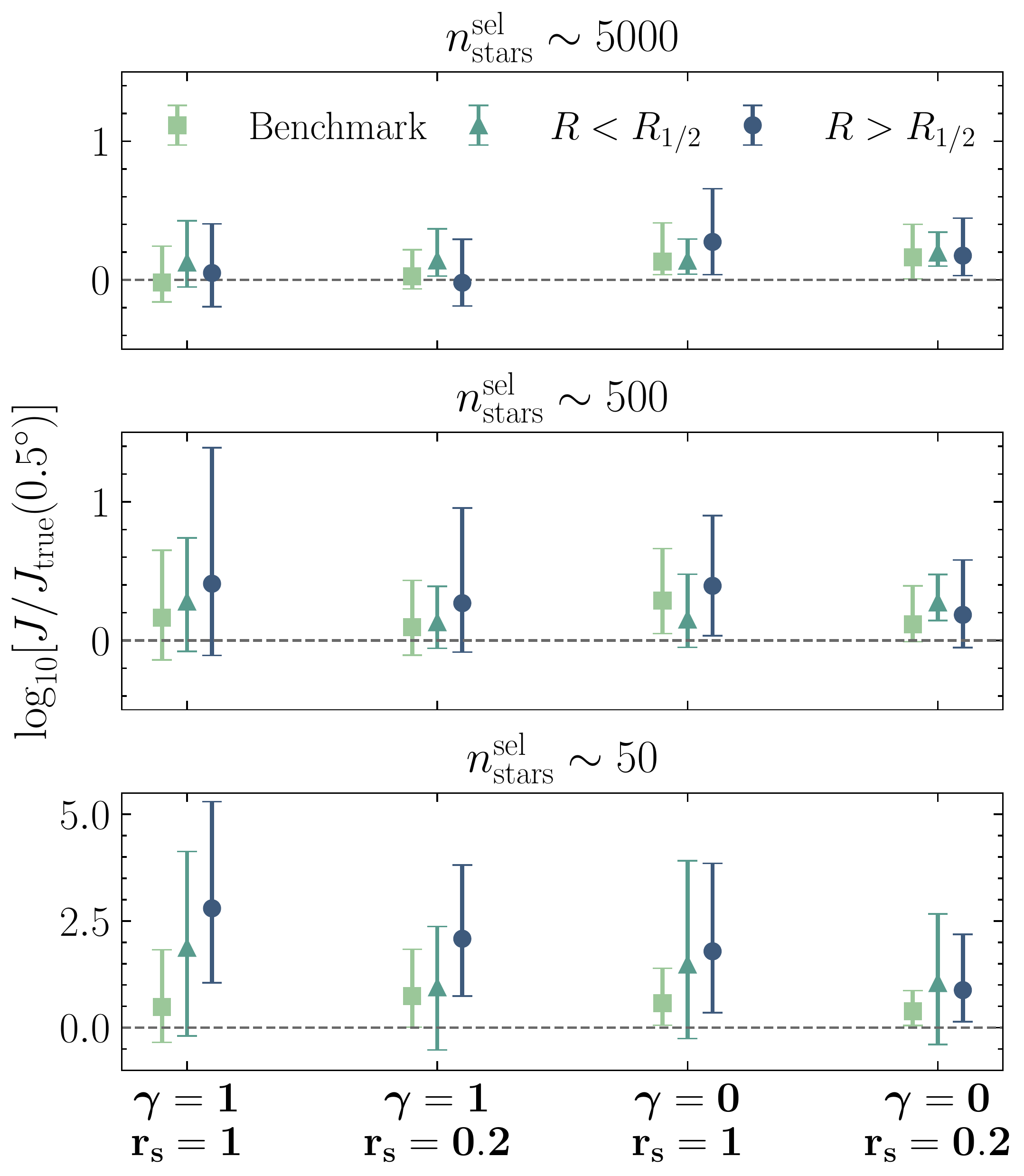}
    \caption{{\jfac}s for a given selected sample size, for all parameter sets and spatial selection functions. From top to bottom, the panels show the results for $\nstarssel\sim5000,\,500,\,50$. From left to right, each cluster of three data points shows the results for parameter set I, II, III, and IV. Each data point denotes the median across the 10 realizations of the median $\log_{10}[J(0.5^\circ)/\mathrm{(GeV^2\,cm^{-5})}]$, and each set of error bars brackets the median of the 68\% containment across the realizations. For the smallest samples, the benchmark case consistently has smaller uncertainties than either the $R>R_{1/2}$ or $R<R_{1/2}$ cases. For some of the larger samples, the uncertainties are slightly reduced ($\mathcal{O}$(0.1 dex) smaller) for $R<R_{1/2}$ relative to the benchmark case. Note the different $y$-axis scale for the bottom panel.}
    \label{fig:Jfacs_drop_stars}
\end{figure}

\subsection{Dependence on Priors}
\label{sec:GS15}

It is crucial to emphasize the dependence of the {\jfac} inference on the priors assumed for the DM halo parameters. The uncertainties on our inferred {\jfac}s are notably larger than values commonly quoted in the literature for the data, such as the ones found in~\cite{Geringer-Sameth:2014yza}, hereafter GS15, which were used to derive the constraints on DM annihilation by the \emph{Fermi}-LAT collaboration in~\cite{Fermi-LAT:2016uux}, hereafter A17. The {\jfac}s from GS15 are listed in Table~\ref{tab:J_n_stars} for reference. For example, Ursa Major II, which has a sample size of 20 stars, is quoted to have a $\sim\pm0.5$ uncertainty on $\log_{10}[J(0.5^\circ)/\mathrm{(GeV^2\,cm^{-5})}]$, whereas on average, the uncertainties on $\log_{10}[J(0.5^\circ)/\mathrm{(GeV^2\,cm^{-5})}]$ for our 20-star samples span $\sim\pm1$--2 (when the values of $\Delta v$ are sufficiently small for the DM to be recovered). This discrepancy is especially surprising because GS15 models the DM density distribution with the Hernquist/Zhao profile, which has two additional slope parameters compared to the gNFW model we use, and additionally models the velocity anisotropy---a model with more free parameters, combined with the added degeneracy between the anisotropy and mass profiles, should give rise to larger uncertainties on the inferred {\jfac}s. 

The primary source of this apparent discrepancy is that in this work, we have assumed wider prior ranges on the halo parameters than what was assumed in GS15---in particular, the analysis in GS15 assumed a prior of $0\leq\gamma\leq1.2$ on the inner DM slope. When we repeated our analysis assuming the same priors on $\rho_0$, $r_s$, and $\gamma$ as the ones used in GS15, the uncertainties on our {\jfac}s decreased significantly. In Figure~\ref{fig:compare_Jfacs_gamma_0_rs_1}, we show the median and middle 68\% range on $\log_{10}[J(0.5^\circ)/\mathrm{(GeV^2\,cm^{-5})}]$ for each of our 10 different 20-star datasets from parameter set III (which is representative of the results for all parameter sets), assuming either our fiducial setup (squares) or the priors from GS15 (triangles). Implementing the GS15 priors reduced the {\jfac} uncertainty in all 10 datasets, by as much as a factor of $\sim3$ in certain cases.

GS15 additionally takes the best-fit Plummer radius from the literature and fixes it in their fit. Analogously, we have also repeated our analysis fixing our light-profile parameters to their best-fit values while assuming the GS15 priors and found the results to be essentially unchanged from the case of GS15 priors without fixing light-profile parameters. Furthermore, GS15 truncates the {\jfac} integration at $r_\mathrm{max}$, the galactocentric distance of the outermost star. We have tested this prescription as well, and found that it makes negligible difference to our values of the {\jfac}. This is expected, because the {\jfac} within the inner $0.5^\circ$ is dominated by the most central regions of the DM halo, and is therefore insensitive to the outer truncation radius of the integration.

As an additional test, we set the priors on the normalization $\rho_0$ and scale radius $r_s$ for the DM profile to the GS15 priors, but rather than using the GS15 prior of $\gamma\in[0,1.2]$ on the inner slope, we assume a wider prior range of $\gamma\in[-1,3]$, which is equivalent to our fiducial prior range with the exclusion of the unphysical values of $\gamma>3$. This directly tests how a wider prior range on $\gamma$ affects the inferred {\jfac}. The results of this test are shown by the circles in Fig.~\ref{fig:compare_Jfacs_gamma_0_rs_1}, and are similar to our fiducial results (squares), indicating that the narrow prior range on $\gamma$ is indeed what primarily drives our fit to reproduce the small {\jfac} uncertainties found in GS15. We have also verified that implementing the GS15 priors (with and without fixing the light profile parameters) on our 1000-star samples decreases the uncertainty on our estimated values of $\log_{10}[J(0.5^\circ)/\mathrm{(GeV^2\,cm^{-5})}]$ by a factor of $\sim2$, making them broadly consistent with the uncertainties quoted in GS15 for the classical dwarfs. 

An important takeaway from this exercise is that the {\jfac}s inferred through the Jeans modeling procedure, for the currently accessible stellar sample sizes, depend sensitively on prior assumptions on $\gamma$, and therefore should be treated with caution. Motivated by the prior-dependence of {\jfac} estimates from Jeans analyses, a complementary method that has been proposed in the literature is a frequentist approach to deriving {\jfac}s~\citep{Chiappo:2016xfs,Chiappo:2018mlt}, which removes the prior-dependence but also loses the ability to construct full posterior probability distributions of the DM inner slope.

\subsection{Spatial Selection}
\label{sec:jfac_selection}
We can revisit the discussion of spatial selection functions detailed in Section~\ref{sec:location}, in the context of {\jfac}s. In \Sec{sec:location}, we found that if we implemented a selection function of $R>R_{1/2}$, i.e., only included stars in the outer regions of the system, the resulting inferred DM density profile was more uncertain in the inner regions of the dwarf than in the benchmark scenario. We also found that if we implemented a selection function of $R<R_{1/2}$, i.e., only included stars in the inner regions of the system, the inferred DM density profile was more uncertain in the outer regions of the dwarf than in the benchmark scenario. Furthermore, for the $R<R_{1/2}$ datasets, we found that the inner profile could be biased high, especially when the sample size was small. The degree of such biases and increased uncertainties on the DM density profile can be quantitatively captured by evaluating the {\jfac}. These results are shown in Figure~\ref{fig:Jfacs_drop_stars} and detailed in Table~\ref{tab:mvir_Jfacs_drop_stars}. 

Overall, we find that for the datasets with $\lesssim50$ observed stars (comparable to the current sample sizes of ultrafaint dwarfs), observing more stars which are distributed across the full range of the radial distribution would have the most potential to decrease the uncertainty on estimates of the {\jfac}s. This is demonstrated in the bottom panel of Fig.~\ref{fig:Jfacs_drop_stars}, in which the uncertainties on the {\jfac} are always smaller in the benchmark case (green squares) than for either of the other two cases (teal triangles and blue circles). For the systems with hundreds or thousands of observed stars, there is room for slight improvement on the accuracy of inferred {\jfac}s by measuring more stars in the inner regions of the systems. This is demonstrated in the top and middle panels of Fig.~\ref{fig:Jfacs_drop_stars}, in which the uncertainties can be somewhat smaller ($\mathcal{O}$(0.1 dex)) for the $R<R_{1/2}$ datasets (teal triangles) than for the benchmark datasets (green squares) or $R>R_{1/2}$ datasets (blue circles). As expected, the behavior of the recovered {\jfac}s is consistent with the ability of the Jeans modeling to accurately recover the inner density profile of the DM, as was discussed in \Sec{sec:results}.

\subsection{Dwarfs in Need of More Measurements}
\label{sec:observations}
Within the literature, there are two approaches to dwarf galaxy indirect detection analyses---individual dwarfs may be analyzed on their own \citep[e.g.,][]{Essig:2010em,2012PhRvD..86f3521B,Hooper:2015ula,Achterbeg:2015dca,Geringer-Sameth:2015lua,Zhao:2017pcz}, or many systems may be stacked to obtain a more competitive limit on DM annihilation~\citep[e.g.,][]{Ackermann:2011wa,2011PhRvL.107x1303G,Ackermann:2015zua,Fermi-LAT:2016uux,Calore:2018sdx,Hoof:2018hyn}. In both cases, achieving robust indirect detection results is dependent upon accurately estimating the {\jfac}s for the dwarfs that dominate the limits. \Tab{tab:J_n_stars} lists the confirmed dwarf galaxies used in the analysis from A17, in order of decreasing {\jfac}. We emphasize that while the dwarfs that give rise to the strongest constraints on DM annihilation are among those with the largest {\jfac}s, having a larger {\jfac} does not necessarily imply that the resulting limit from a given dwarf will be stronger, due to effects such as different levels of background contamination in different regions of the sky. In the following discussion, we will emphasize future observations which are important for obtaining more accurate estimates of the {\jfac}s for the systems that dominate the A17 results.

\begin{table*}[t]
    \begin{center}
    \begin{tabular}{|l|c|c|c|l|}
    \hline
    Dwarf & $n_\mathrm{stars}$ & $\log_{10} J (0.5^\circ)$  & Dispersion  & References \\
     &  &  [GeV$^2$ cm$^{-5}$] &  [km/s] &  \\
    \hline
    \hline
    Ursa Major II &20  & $19.42^{+0.44}_{-0.42}$ & $5.6^{+1.4}_{-1.4}$ & \cite{2019arXiv190105465S} \\
    Segue 1& 70& $19.36^{+0.32}_{-0.35}$ & $3.7^{+1.4}_{-1.1}$ &  \cite{2007ApJ...670..313S} \\
    Coma Berenices & 59 & $19.02^{+0.37}_{-0.41}$ & $4.6^{+0.8}_{-0.8}$ &  \cite{2007ApJ...670..313S} \\
    Ursa Minor & 313 & $18.93^{+0.27}_{-0.19}$ & $9.5^{+1.2}_{-1.2}$ & \cite{2009ApJ...704.1274W} \\
    Draco & 292 & $18.84^{+0.12}_{-0.13}$ & $9.1^{+1.2}_{-1.2}$ &  \cite{2009ApJ...704.1274W} \\
    Sculptor & 1365 & $18.54^{+0.06}_{-0.05}$ & $9.2^{+1.1}_{-1.1}$ &  \cite{2009AJ....137.3100W} \\
    Bootes I & 37 & $18.24^{+0.40}_{-0.37}$ & $4.6^{+0.8}_{-0.6}$ & \cite{2011ApJ...736..146K} \\
    Leo II & 126 & $17.97^{+0.20}_{-0.18}$ & $7.4^{+0.4}_{-0.4}$ &  \cite{2017AJ....153..254S} \\
    Carina & 774 & $17.87^{+0.10}_{-0.09}$ & $6.6^{+1.2}_{-1.2}$ &  \cite{2009AJ....137.3100W} \\
    Ursa Major I & 39& $17.87^{+0.56}_{-0.33}$ & $7.0^{+1.0}_{-1.0}$  & \cite{2019arXiv190105465S} \\
    Leo I & 267 & $17.84^{+0.20}_{-0.16}$ & $9.2^{+0.4}_{-0.4}$ & \cite{2008ApJ...675..201M} \\
    Fornax & 2483 & $17.83^{+0.12}_{-0.06}$ & $11.7^{+0.9}_{-0.9}$ &  \cite{2009AJ....137.3100W}\\
    Canes Venatici II & 25 & $17.65^{+0.45}_{-0.43}$ & $4.6^{+1.0}_{-1.0}$  &  \cite{2007ApJ...670..313S} \\
    Sextans & 441 & $17.52^{+0.28}_{-0.18}$ & $7.9^{+1.3}_{-1.3}$ &  \cite{2009AJ....137.3100W}\\
    Canes Venatici I & 214& $17.43^{+0.37}_{-0.28}$ & $7.6^{+0.4}_{-0.4}$ &  \cite{2007ApJ...670..313S}\\
    Leo T & 19 & $17.11^{+0.44}_{-0.39}$ & $7.5^{+1.6}_{-1.6}$ &  \cite{2007ApJ...670..313S}\\
    Hercules & 30 & $16.86^{+0.74}_{-0.68}$ & $5.1^{+0.2}_{-0.2}$ &  \cite{2007ApJ...670..313S}\\
    Leo V & 5& $16.37^{+0.94}_{-0.87}$ & $2.3^{+3.2}_{-1.6}$ &  \cite{2017MNRAS.467..573C} \\
    Leo IV & 18& $16.32^{+1.06}_{-1.69}$ & $3.3^{+1.7}_{-1.7}$ &  \cite{2007ApJ...670..313S} \\
    Segue 2 & 25 & $16.21^{+1.06}_{-0.98}$ & $<2.2$ &  \cite{2013ApJ...770...16K} \\
    \hline
    \end{tabular}
    \end{center}
    \caption{List of dwarf galaxies used in \cite{Fermi-LAT:2016uux} ordered by decreasing {\jfac}. The observed numbers of stars and {\jfac}s are compiled from \cite{Geringer-Sameth:2014yza}. The dispersions are compiled from \cite{2019arXiv190105465S}.} 
    \label{tab:J_n_stars}
    \end{table*}  

As shown in Fig.~\ref{fig:Jfacs_all} and detailed in Tab.~\ref{tab:mvir_deltav}, the typical uncertainty on $\log_{10}[J(0.5^\circ)/\mathrm{(GeV^2\,cm^{-5})}]$ from our analysis is $\sim\pm1$--$2$ for 20-star systems and $\sim\pm0.5$--$1$ for 100-star systems (excluding the cases of small intrinsic dispersion and large $\Delta v$ where the fit drastically underestimates the abundance of DM), as opposed to $\sim\pm0.5$ and $\sim\pm0.2$, respectively, from GS15 (listed in Tab.~\ref{tab:J_n_stars}). We determined in Sec.~\ref{sec:GS15} that this discrepancy may be due to different prior choices on $\gamma$. To test the effect of larger {\jfac} uncertainties on the resulting indirect detection constraints on DM annihilation, we can use the likelihood functions provided in A17\footnote{\url{http://www-glast.stanford.edu/pub_data/1203/}} to derive limits assuming different values of the {\jfac} uncertainty. Similarly to A17, we use Eq. 3 of \cite{Ackermann:2015zua} to profile over the {\jfac} uncertainty.

The three dwarfs from A17 that provide the strongest limits in the mass range relevant for the DM interpretation of the Galactic Center Excess (GCE) are Ursa Major II, Ursa Minor, and Draco. We first focus on Ursa Major II, which has a sample size of 20 stars. We find that increasing the uncertainty on $\log_{10}[J(0.5^\circ)/\mathrm{(GeV^2\,cm^{-5})}]$ from 0.4 (which was assumed in the A17 analysis) to 1 weakens the limit by a factor of $\sim5$--8 in the 10--100 GeV mass range for the $b\overline{b}$ annihilation channel, resulting in a limit that no longer excludes or is in tension with the regions of parameter space consistent with the GCE from \cite{Gordon:2013vta,Daylan:2014rsa,Calore:2014nla,Abazajian:2014fta}. Similarly, for Draco and Ursa Minor ($\sim$ 300 observed stars each), we find that increasing the uncertainty on $\log_{10}[J(0.5^\circ)/\mathrm{(GeV^2\,cm^{-5})}]$ from their assumed values in A17 of 0.1 and 0.2, respectively, to 0.5(1) weakens the limit by a factor of $\sim2(10)$. We note that a factor of $\sim2$ weakening of the strongest dwarf limits is sufficient to significantly reduce the tension with the DM interpretation of the GCE.

This demonstrates that for the current observed sample sizes, the dwarf galaxy indirect detection limits can be highly sensitive to the assumed priors for the inner DM slope $\gamma$. In order to derive robust indirect detection constraints from the dwarf galaxies, it is crucial to increase the number of observed stars in order to obtain more robust estimates of their {\jfac}s. 
In particular, we emphasize the importance of increasing the sample sizes for Ursa Major II, Ursa Minor, and Draco, which dominate the indirect detection limits. Our results in \Secs{sec:location}{sec:jfac_selection} suggest that measuring more stars spanning the entire spatial extent of the galaxies would be most effective at achieving more accurate estimates of their {\jfac}s (see bottom panel of \Fig{fig:Jfacs_drop_stars}). If sample sizes are increased beyond $\gtrsim500$ stars, our findings suggest that focusing on measuring more stars in the inner regions of the dwarfs may provide additional constraining power on their {\jfac}s (see top two panels of \Fig{fig:Jfacs_drop_stars}).

\section{Conclusions}
\label{sec:conclusions}
In this paper, we performed a systematic study of the spherical Jeans analysis method in the context of inferring the DM content in dwarf galaxies. We focused on simulated data for spherical, isotropic systems, and assessed the performance of the method at accurately recovering the overall dark matter density profile, the virial mass, and the inner slope of the dark matter density profile. Our primary conclusions are the following:

\begin{itemize}
    \item For parameter sets I and III, which describe $M_{200}\sim10^9\,\msun$ halos (intrinsic velocity dispersion $\sim 10$--$15$ km/s) with inner density slopes of $\gamma=1$ and $\gamma=0$, respectively, we find that the virial mass we recover is always consistent with the true value, and is increasingly accurate as the sample size is increased. However, the inner profile of the DM density distribution is less well-constrained---for samples with $\lesssim1000$ stars, the posterior distributions on the inner slope $\gamma$ are typically unconstrained, and there is no statistical preference for a cuspy or cored profile. We recover statistical evidence for the true (cuspy/cored) profile only for samples with $10,000$ stars. For these parameter sets, the results are generally insensitive to varying the measurement error of the line-of-sight velocity, $\Delta v$, over the range $\Delta v=0$--$5$ km/s.
    
    \item For parameter sets II and IV, which describe $M_{200}\sim10^7\,\msun$ halos with inner density slopes of $\gamma=1$ and $\gamma=0$, respectively, we find that the virial mass estimates depend sensitively on $\Delta v$, particularly for the samples with fewer stars. For parameter set II (intrinsic velocity dispersion $\sim3$ km/s), the inferred virial mass for the 20- and 100-star samples is consistent with zero when $\Delta v=5$ km/s. Similarly, for parameter set IV (intrinsic velocity dispersion $\sim2$ km/s), the recovered virial mass for the 20- and 100-star samples is consistent with zero for the cases of $\Delta v=2,\,5$ km/s. This is rectified when the sample size is increased to 1000 or more stars.
    
    \item From our study of spatial selection functions, we conclude based on the overall performance at inferring the DM density profile, the inner slope $\gamma$, and the virial mass, that it is crucial to have measurements of stars across the full radial distribution of the dwarf galaxy, especially for the smallest samples. Doing so allows the fit to anchor the DM distribution across the full radial range, and consistently results in comparable or better performance at accurately reconstructing both the inner and outer profile of the DM distribution, relative to the cases where the data consists purely of stars in either the inner or outer region of the system. For systems with $\lesssim50$ observed stars, measuring more stars across the full radial extent can reduce the uncertainties on $\log_{10}[J(0.5^\circ)/\mathrm{(GeV^2\,cm^{-5})}]$ by a factor of $\sim3$ compared to measuring the same number of stars only within the half-light radius.
    
    \item Degeneracy between the DM halo parameters in our model makes it difficult to constrain the inner slope, $\gamma$, especially when sample sizes are small. We emphasize that this is separate from the issue of the velocity anisotropy profile being degenerate with the enclosed mass profile. While datasets with larger sample size can help resolve these parameter degeneracies, it is unfeasible to measure upwards of 10,000 stars---the sample size required for constraining $\gamma$---in the dwarf galaxies in the near future. Instead, a potential method for increasing the constraining power of Jeans analyses on the core-cusp problem is to jointly fit to many dwarf galaxies simultaneously. This is computationally challenging to implement using standard MCMC or nested sampling techniques, so a thorough study of joint fits requires the use of other methods for approximating posterior distributions.
    
    \item Cast in the context of indirect detection, we find that for the 20-star samples across all parameter sets (in the cases of sufficiently small $\Delta v$ for the DM to be recovered), the median $1\sigma$ uncertainty on $\log_{10}[J(0.5^\circ)/\mathrm{(GeV^2\,cm^{-5})}]$ across our 10 realizations is $\sim\pm1$--2, in contrast with the uncertainties of $\sim\pm0.5$ quoted for some of the current ultrafaint dwarf measurements (with $\sim20$ stars) in GS15 (see Table~\ref{tab:J_n_stars}), which were used to derive the dwarf galaxy constraints on DM annihilation in A17. We find that this discrepancy may be driven by the more restrictive prior ranges for the DM profile parameters in GS15---in particular the prior range on the inner slope $\gamma$---and note that the resulting indirect detection results should be interpreted with this prior-dependence in mind.
\end{itemize}

In our study, we have focused on the case of spherical, isotropic systems with the goal of understanding the limitations of Jean analyses even in the absence of challenges that are known to complicate studies that use this method, such as background contamination~\citep[e.g.,][]{2016MNRAS.462..223B,Ichikawa:2016nbi,Ichikawa:2017rph,Horigome:2020kyj}, the effect of assuming equilibrium for systems which are not in equilibrium~\citep{2017ApJ...835..193E}, the assumption of sphericity for systems which are non-spherical~\citep{2015MNRAS.446.3002B,Klop:2016lug}, the degeneracy between the enclosed mass and velocity anisotropy~\citep[e.g.,][]{1990AJ.....99.1548M,Wilkinson:2001ut,Lokas:2003ks,DeLorenzi:2008zq,2017MNRAS.471.4541R,Genina:2019job}, and the presence of potentially large fractions of binary stars in the dwarf galaxies~\citep[e.g.,][]{2010ApJ...722L.209M,Minor:2010vp,2013ApJ...771...29G,2017AJ....153..254S,2018AJ....156..257S,2019MNRAS.487.2961M}.

With regard to the core-cusp problem, we have found that even for the idealized systems we consider, and a relatively simple three-parameter halo model, the Jeans modeling method is severely limited in its ability to constrain the inner slope $\gamma$ of the dark matter density profile. A crucial reason behind this is that there are degeneracies between the three parameters that describe our DM profiles. The fact that $\gamma$ is difficult to pinpoint is consistent with previous Jeans modeling results in the literature \cite[e.g.,][]{2009ApJ...704.1274W,Read:2018pft,Genina:2019job}; we have additionally determined that, in order to constrain $\gamma$ within this framework, it is necessary to measure $\sim10,000$ stars within a single dwarf galaxy, which is not practical within the near future. We therefore need to search for alternative methods for addressing the core-cusp problem using Jeans analysis methods. 

While complementary mass modeling methods have claimed preference for cores or cusps in the dwarf galaxies, important caveats when interpreting such results have been identified in the literature. For example, while many rotation curve analyses have shown preference for cored DM distributions, studies have shown that systematic effects in rotation curve analyses can erroneously bias the inferred DM distribution towards a centrally cored profile \citep[see, e.g.,][and references within]{2004ApJ...617.1059R,2007ApJ...657..773V,2017MNRAS.466...63P}. It has also been demonstrated in \cite{2013MNRAS.431.2796K,2018MNRAS.474.1398G} that results using mass estimator methods such as the ones proposed in \cite{2010MNRAS.406.1220W,2011ApJ...742...20W} can depend sensitively on the specific line of sight that is chosen, and can result, for example, in predicting a cored profile when the true halo is cuspy.

The parameter degeneracy that limits our ability to reconstruct $\gamma$ is a distinct from the well-known mass-anisotropy degeneracy which plagues Jeans analyses, for which a number of proposed solutions exist in the literature: using higher order moments of the velocity distribution~\citep{1990AJ.....99.1548M,2013MNRAS.432.3361R,2014MNRAS.441.1584R,2017MNRAS.471.4541R,Genina:2019job} and incorporating proper motion measurements of stars~\citep{Strigari:2007vn,Lazar_2020} are among the methods that have been demonstrated to ameliorate the mass-anisotropy degeneracy. It is worth exploring whether or not these methods would also lead to improved constraints on the inner slope of the DM density profile, the answer to which is not intuitively obvious. \cite{Alvarez:2020cmw} recently used the framework described in \cite{2017MNRAS.471.4541R}, which parameterizes the DM density profile as a multiply-broken power law and employs higher order moments, to derive {\jfac}s for the classical dwarfs. They obtained {\jfac} estimates which are consistent with the ones from GS15, but with reduced uncertainties. Additionally, jointly fitting to multiple dwarf galaxies at once is a potential method for leveraging moderately-sized datasets to achieve better constraints on $\gamma$. While we have not yet explored this avenue systematically, due to computational challenges, it is a promising direction for future work. 

Finally, we have used our results to make recommendations for future observations. For the purpose of achieving more accurate, less prior-dependent {\jfac} estimates for the systems that dominate the indirect detection results presented in A17, we identify Ursa Major II, Ursa Minor, and Draco as the dwarf galaxies that would most benefit from more stars being measured. Our preliminary analyses show that if we assume the typical {\jfac} uncertainties that we find in our work, the DM annihilation limits for these systems may be weakened to the degree of significantly affecting their implications on the DM interpretation of the GCE. 

\section*{Acknowledgements}
We are particularly grateful to M. Geha and M. Lisanti for their helpful insight on the topic. We also thank P. Hopkins, A. Ji, E. Kirby, J. Read, J. Simon, and M. Walker for helpful discussions. LJC thanks S. Mishra-Sharma for fruitful discussions and moral support. LJC is supported by a Paul \& Daisy Soros Fellowship and an NSF Graduate Research Fellowship under Grant Number DGE-1656466. LN is supported by the DOE under Award Number
DESC0011632, the Sherman Fairchild fellowship, and the University of California Presidential Fellowship. The work presented in this paper was performed on computational resources managed and supported by Princeton Research Computing, a consortium of groups including the Princeton Institute for Computational Science and Engineering (PICSciE) and the Office of Information Technology's High Performance Computing Center and Visualization Laboratory at Princeton University.

\def\bibsection{} 
\bibliographystyle{aasjournal}
\bibliography{dwarfs}

\appendix 

\setcounter{figure}{0} \renewcommand{\thefigure}{A\arabic{figure}} \renewcommand{\theHfigure}{A\arabic{figure}}
\setcounter{table}{0} \renewcommand{\thetable}{A\arabic{table}} \renewcommand{\theHtable}{A\arabic{table}}
\setcounter{equation}{0} \renewcommand{\theequation}{A\arabic{equation}} \renewcommand{\theHequation}{A\arabic{equation}}

\section{Enclosed Mass Functions}
\label{sec:enclosed_mass}
We list here for reference the closed-form expressions for the enclosed mass functions corresponding to the density profiles given by \Eq{eq:gnfw}--\Eq{eq:nfw_cored}. 

\begin{align}
M_{\rm{DM}}^{\rm{gNFW}} (r)& = \frac{4 \pi}{3 - \gamma} \rho_0~r^3  \left( \frac{r}{r_s}\right)^{-\gamma} \,_2F_1 \left( 3 -\gamma, 3 - \g; 4 - \g; - \frac{r}{r_s}  \right) \label{eq:M_gNFW}\\
M_{\rm{DM}}^{\rm{NFW}} (r)& = 4 \pi \rho_0~ r_s^3 \left( \frac{-r}{r + r_s} + \log \left( 1 + \frac{r}{r_s} \right) \right) \\
M_{\rm{DM}}^{\rm{NFWc}} (r)& =  4\pi \rho_0~ r_s^3 \left[ - \frac{r(3r + r_s)}{2(r+ r_s)^2} + \log \left( 1 + \frac{r}{r_s}\right) \right] 
\end{align}
\vspace{0.5cm}


\setcounter{figure}{0} \renewcommand{\thefigure}{B\arabic{figure}} \renewcommand{\theHfigure}{B\arabic{figure}}
\setcounter{table}{0} \renewcommand{\thetable}{B\arabic{table}} \renewcommand{\theHtable}{B\arabic{table}}
\setcounter{equation}{0} \renewcommand{\theequation}{B\arabic{equation}} \renewcommand{\theHequation}{B\arabic{equation}}

\section{Additional Figures}

\begin{figure*}[h]
    \centering
    \includegraphics[width=0.7\textwidth]{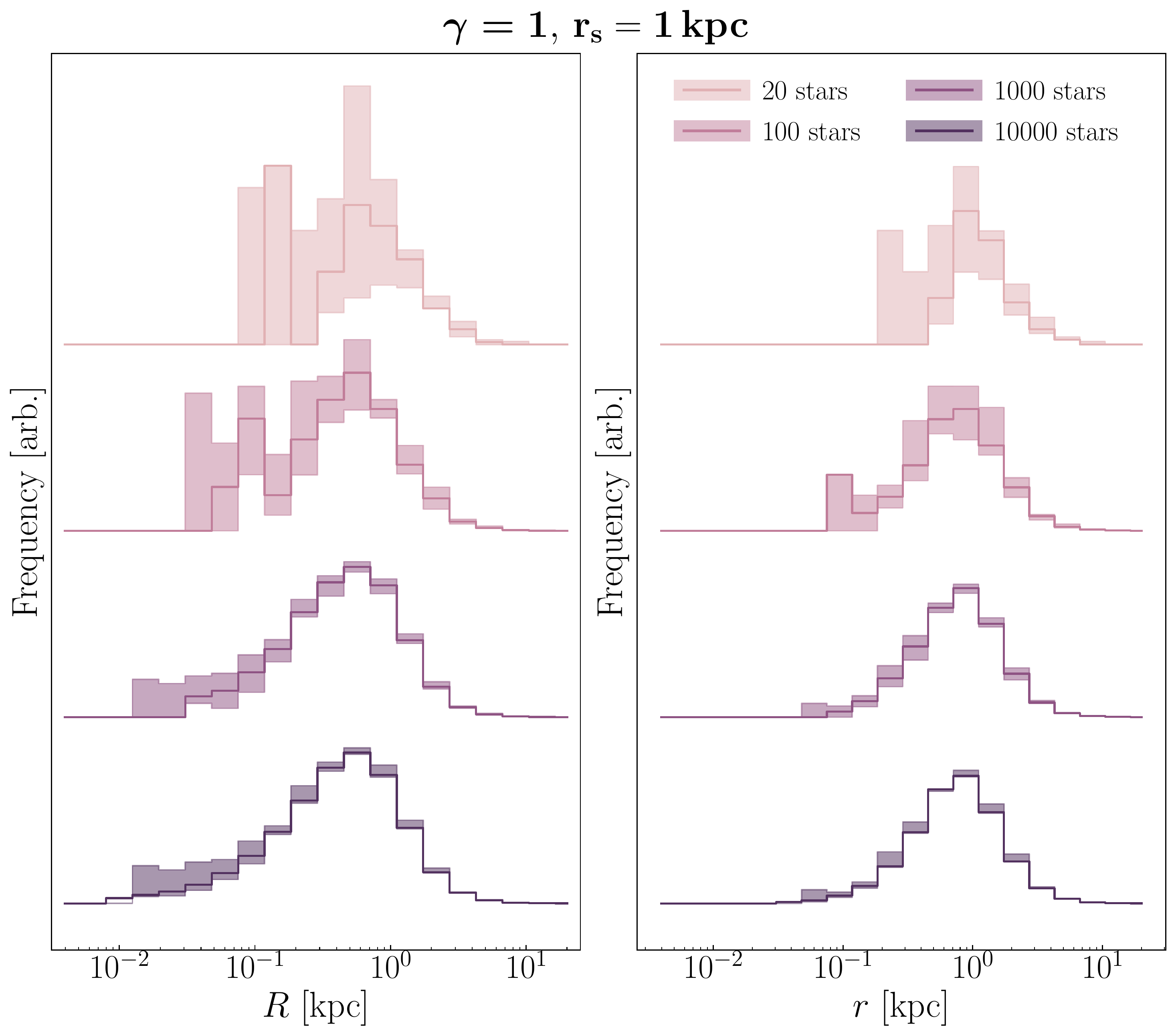}
    \caption{Distributions of the projected radii (left panel) and 3d galactocentric radii (right panel) in our mock datasets for parameter set I. From top to bottom (lightest to darkest color), we show the distributions for sample sizes of 20, 100, 1000, and 10,000 stars. For each histogram, the solid line denotes the median counts per radial bin and the band shows the 68\% containment across 10 realizations.}
    \label{fig:hist_stars}
\end{figure*}

\begin{figure*}[h]
    \centering
    \includegraphics[width=0.95\textwidth]{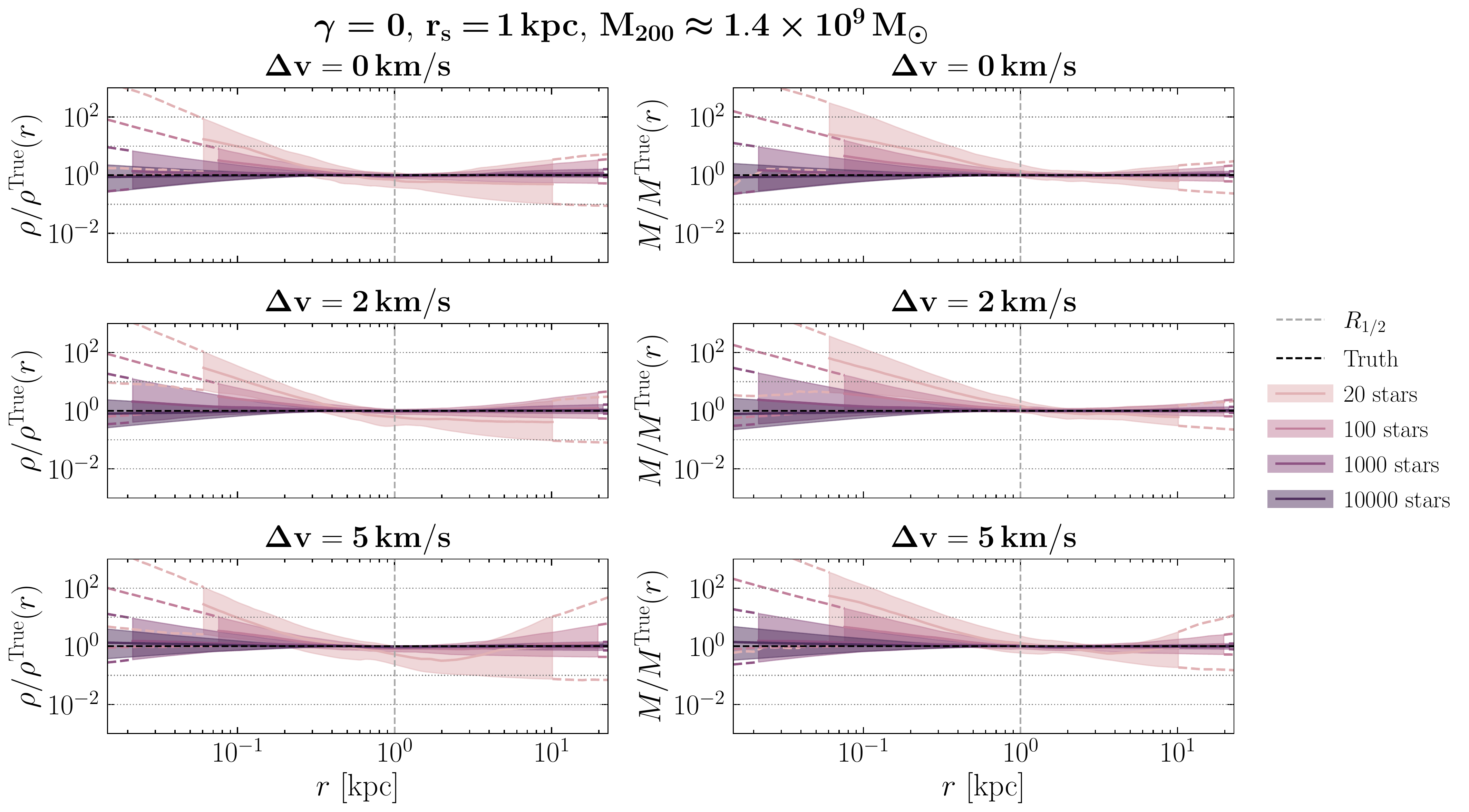}
    \caption{Same as Figure~\ref{fig:fractional_rho_Menc_gamma_0_rs_0p2}, but for parameter set III.}
    \label{fig:fractional_rho_Menc_gamma_0_rs_1}
\end{figure*}

\begin{figure*}[h]
    \centering
    \includegraphics[width=0.95\textwidth]{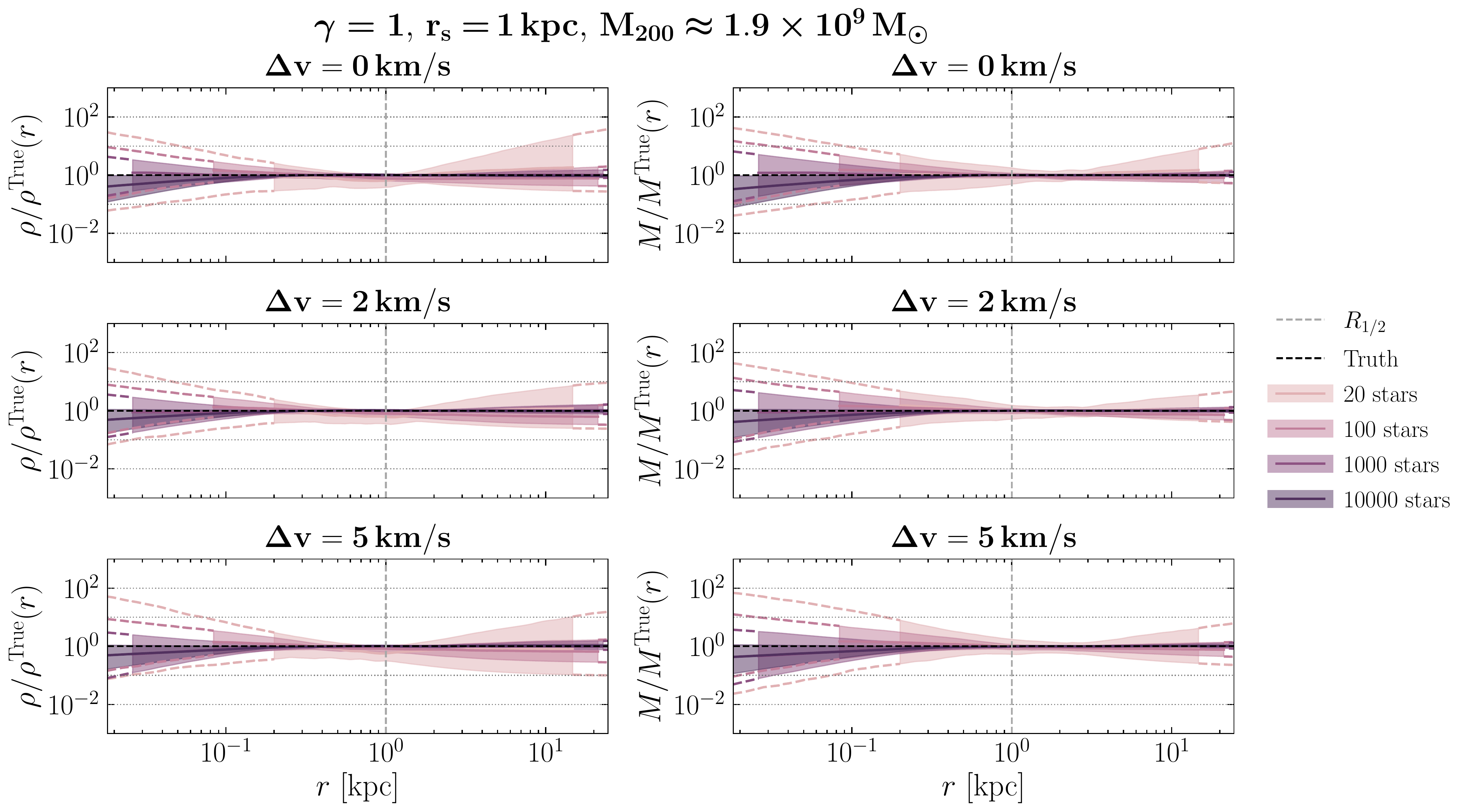}
    \caption{Same as Figure~\ref{fig:fractional_rho_Menc_gamma_0_rs_0p2}, but for parameter set I.}
    \label{fig:fractional_rho_Menc_gamma_1_rs_1}
\end{figure*}

\begin{figure*}[h]
    \centering
    \includegraphics[width=0.95\textwidth]{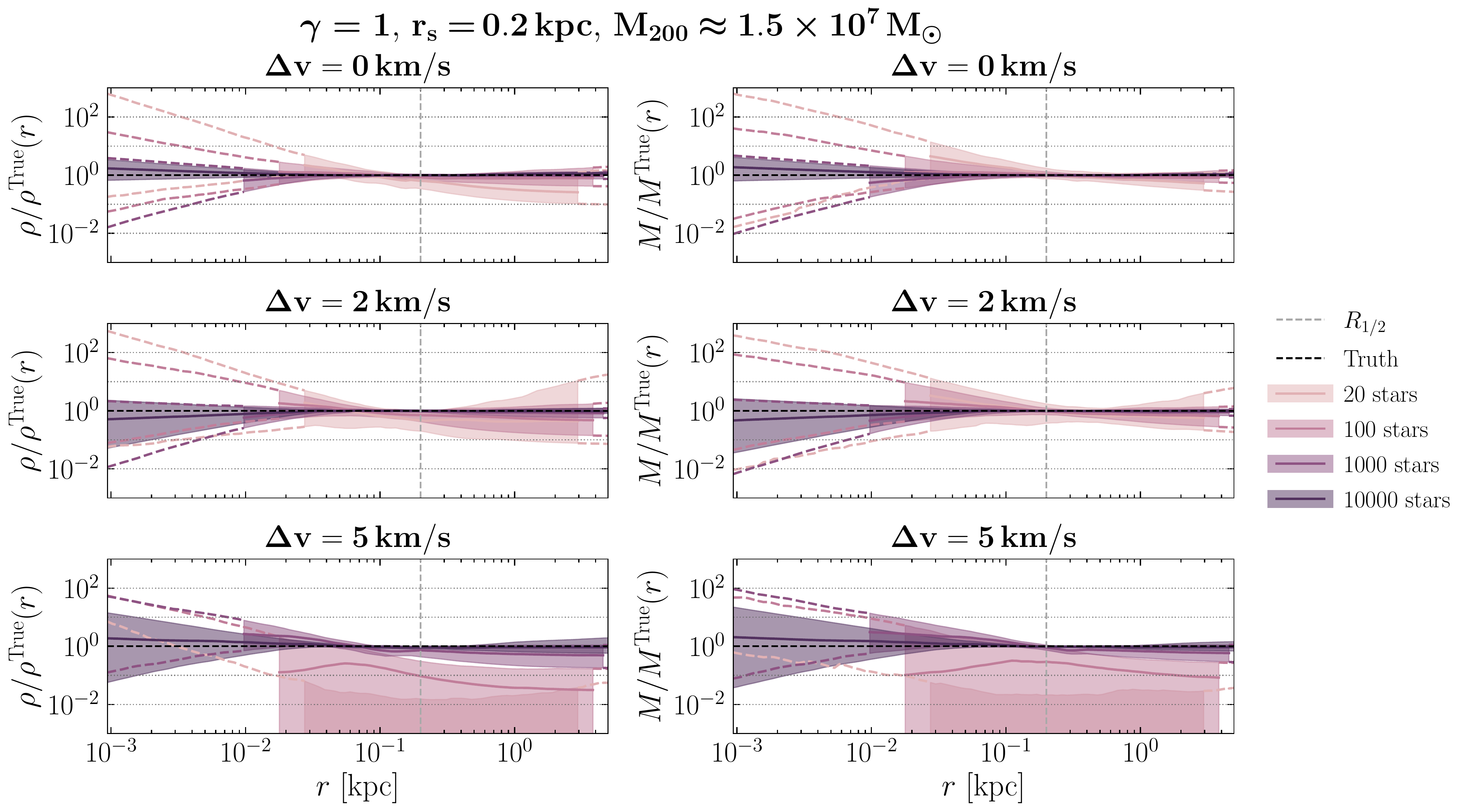}
    \caption{Same as Figure~\ref{fig:fractional_rho_Menc_gamma_0_rs_0p2}, but for parameter set II.}
    \label{fig:fractional_rho_Menc_gamma_1_rs_0p2}
\end{figure*}

\begin{figure*}[h]
    \centering
    \includegraphics[width=0.95\textwidth]{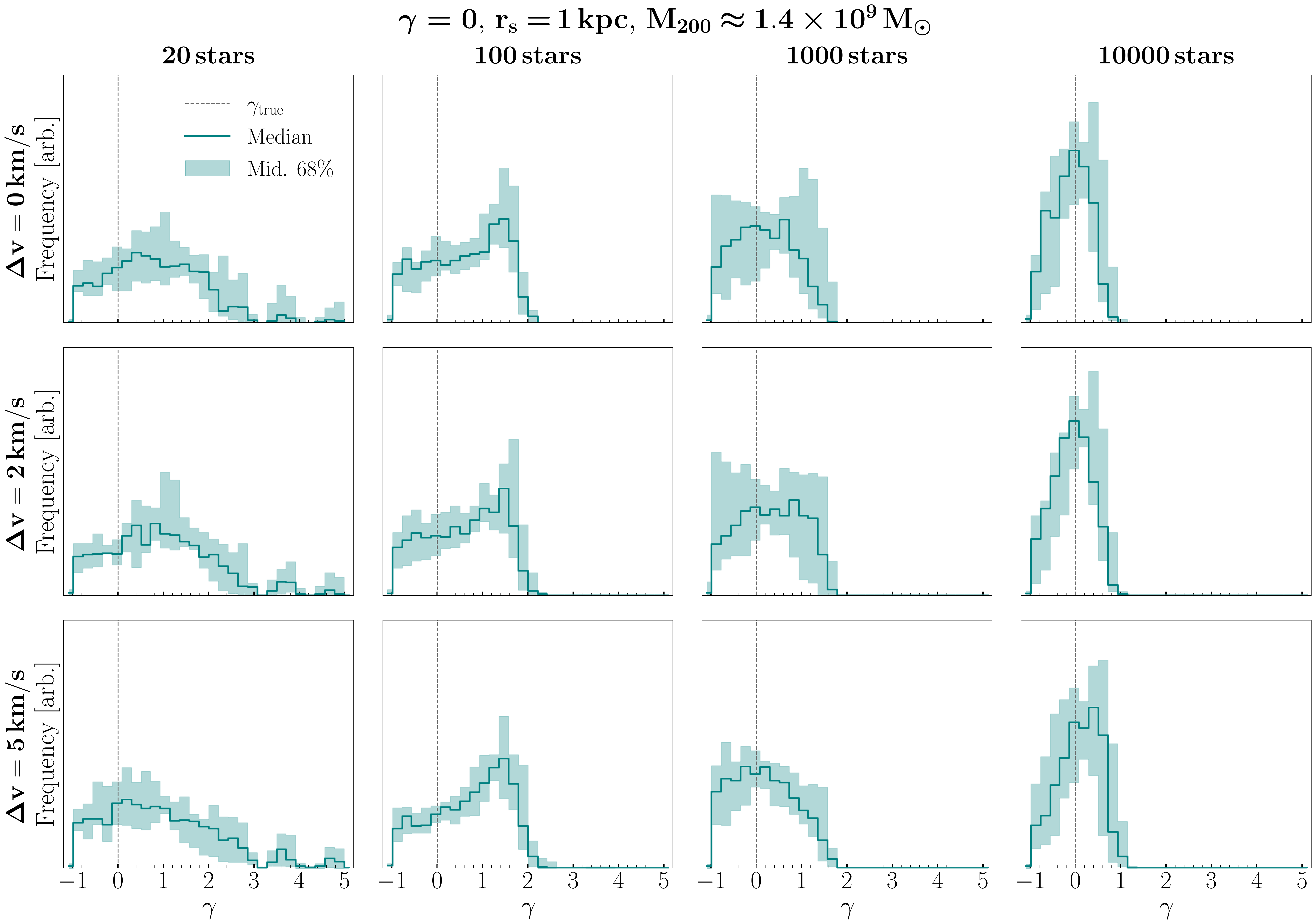}
    \caption{Same as Figure~\ref{fig:gamma_post_all_verr_gamma_0_rs_0p2}, but for parameter set III, with vertical scale adjusted for presentation. All panels in this figure share the same vertical scale.}
    \label{fig:gamma_post_all_verr_gamma_0_rs_1}
\end{figure*}

\begin{figure*}[h]
    \centering
    \includegraphics[width=0.95\textwidth]{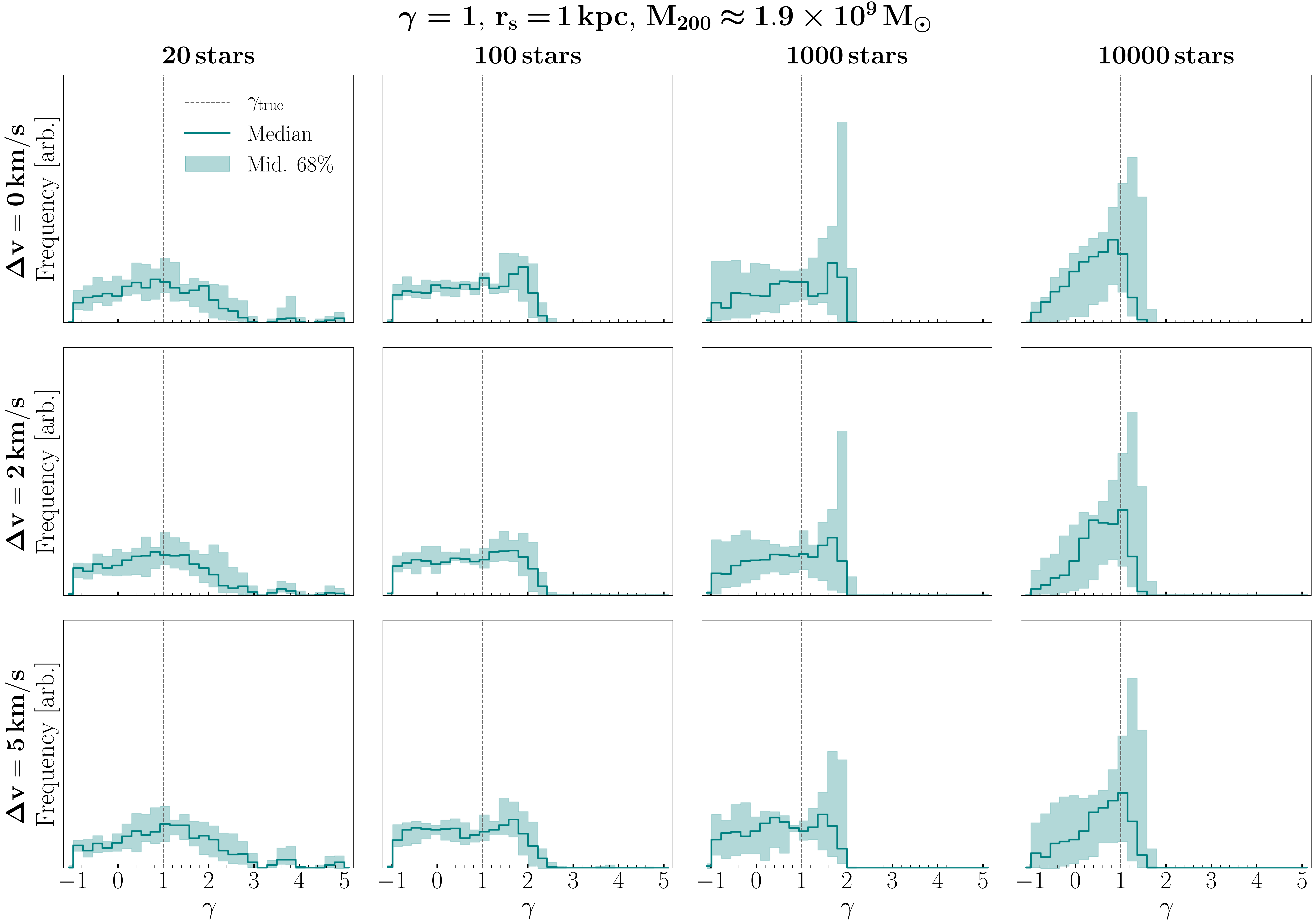}
    \caption{Same as Figure~\ref{fig:gamma_post_all_verr_gamma_0_rs_0p2}, but for parameter set I, with vertical scale adjusted for presentation. All panels in this figure share the same vertical scale.}
    \label{fig:gamma_post_all_verr_gamma_1_rs_1}
\end{figure*}

\begin{figure*}[h]
    \centering
    \includegraphics[width=0.95\textwidth]{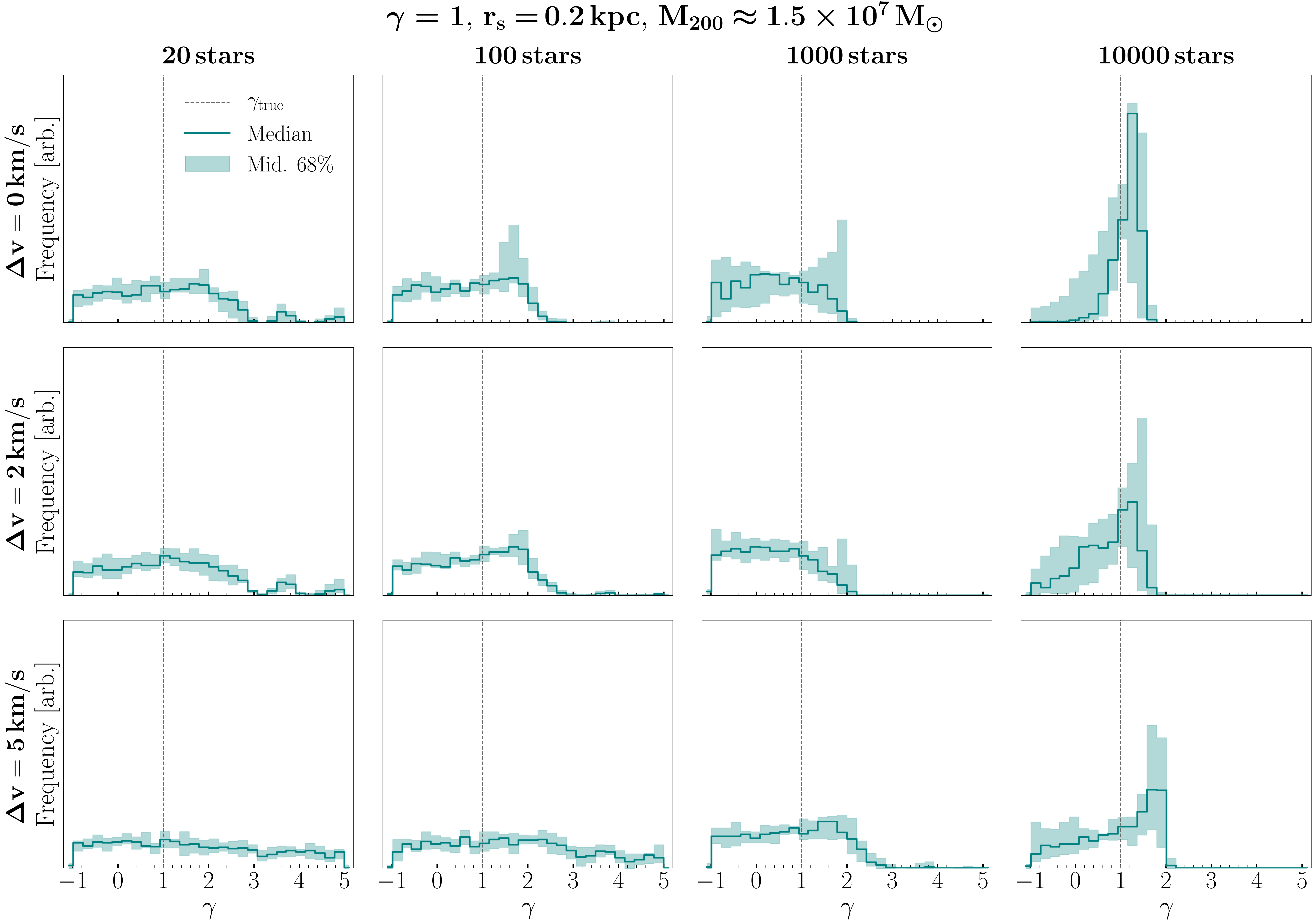}
    \caption{Same as Figure~\ref{fig:gamma_post_all_verr_gamma_0_rs_0p2}, but for parameter set II, with vertical scale adjusted for presentation. All panels in this figure share the same vertical scale.}
    \label{fig:gamma_post_all_verr_gamma_1_rs_0p2}
\end{figure*}
\clearpage
\begin{figure*}[h]
    \centering
    \includegraphics[width=0.8\textwidth]{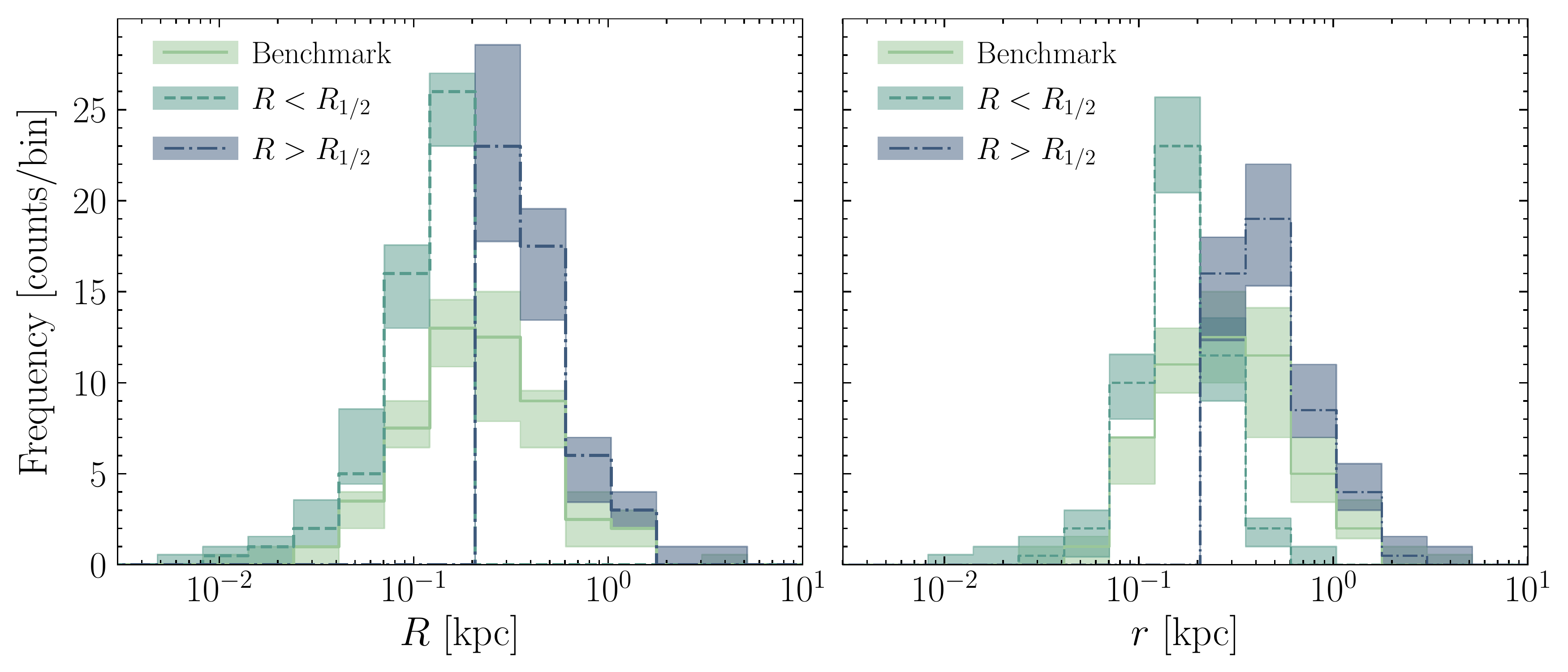}
    \caption{Histograms of the projected radius $R$ (left panel) and the 3d radius $r$ (right panel) for parameter set III starting with a sample size of $\nstars=100$ stars, resulting in selected samples of $\nstarssel\sim50$ stars, for the three different spatial selection functions. For each selection function, the line denotes the median counts per bin and the band shows the 68\% containment across 10 realizations. The recovered DM density and enclosed mass profiles corresponding to these datasets are shown in Figure~\ref{fig:rho_Menc_gamma_0_rs_0p2_drop_stars}.}
    \label{fig:hist_gamma_0_drop_stars}
\end{figure*}

\begin{figure*}[h]
    \centering
    \includegraphics[width=0.8\textwidth]{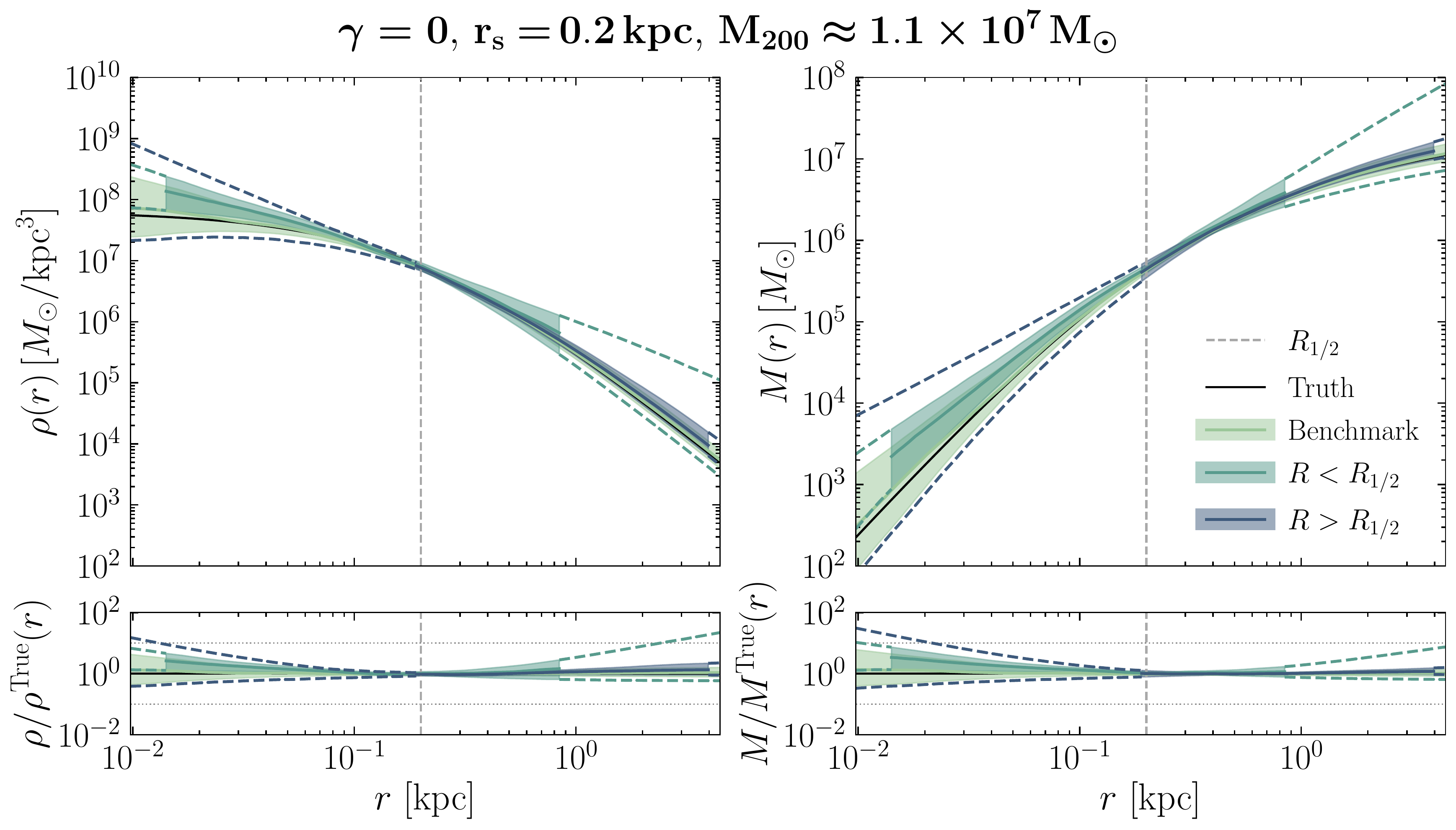}
    \caption{Same as Figure~\ref{fig:rho_Menc_gamma_0_rs_0p2_drop_stars}, but starting with a sample size of $\nstars=1000$ stars, resulting in selected samples of $\nstarssel\sim500$ stars. In this case, the bias on the density profile for the $R<R_{1/2}$ case is less severe than in Fig.~\ref{fig:rho_Menc_gamma_0_rs_0p2_drop_stars}, and the uncertainty on the inner density profile is smaller for the $R<R_{1/2}$ case than for the benchmark case.}
    \label{fig:rho_Menc_gamma_0_rs_0p2_drop_stars_100}
\end{figure*}

\clearpage

\setcounter{figure}{0} \renewcommand{\thefigure}{C\arabic{figure}} \renewcommand{\theHfigure}{B\arabic{figure}}
\setcounter{table}{0} \renewcommand{\thetable}{C\arabic{table}} \renewcommand{\theHtable}{B\arabic{table}}
\setcounter{equation}{0} \renewcommand{\theequation}{C\arabic{equation}} \renewcommand{\theHequation}{B\arabic{equation}}

\section{Light Profile Fitting Procedure}
\label{sec:light_profile_fit}

We take a binned likelihood approach to fit the stellar light profile in the initial step of our analysis, modeling the light profile as a projected Plummer profile (\Eq{eq:surface_density}). For a sample size of $n_\mathrm{stars}$, we bin the data in $\sim\sqrt{n_\mathrm{stars}}$ logarithmically-spaced bins in the projected radius $R$. Because the measurement errors on the stellar positions are small---largely driven by the uncertainties on the distance to the galaxy, given the accurate measurements on the angular positions of stars on the sky---we take the uncertainty on the number of stars in each bin to be the Poisson uncertainty corresponding to the mean number of stars in that bin. For a $100(1-\alpha)\%$ confidence level, the lower and upper bound of the Poisson uncertainty are given by~\citep{Tanabashi:2018oca}
\begin{align}
&\mu_\mathrm{lo}=\frac{1}{2}F_{\chi^2}^{-1}\left(\frac{\alpha}{2};2\hat{n}\right)\\
&\mu_\mathrm{up}=\frac{1}{2}F_{\chi^2}^{-1}\left(1-\frac{\alpha}{2};2(\hat{n}+1)\right)\,, 
\end{align}
where $F_{\chi^2}^{-1}$ is the inverse of the $\chi^2$ cumulative distribution function and $\hat{n}$ is the mean number of counts. We then have $\sigma_\mathrm{lo}=\hat{n}-\mu_\mathrm{lo}$ and $\sigma_\mathrm{up}=\mu_\mathrm{up}-\hat{n}$, which need to be modeled in our likelihood. In order to account for the asymmetric uncertainties that arise from this prescription, we use the following approximation to a Gaussian log-likelihood for $\hat{n}_i$ observed counts and $n_i(\theta)$ predicted counts in the $i^\mathrm{th}$ bin, where $\theta$ are the model parameters~\citep{Barlow:2004wg}:

\es{eq:asymll}{\ln\mathcal{L}(\hat{n}|\theta)=-\frac{1}{2}\sum_{i}\frac{(\hat{n}_i-n_i(\theta))^2}{V_i-V_i'(\hat{n}_i-n_i(\theta))}\,,}
where $V=\sigma_\mathrm{lo}\sigma_\mathrm{up}$ and $V'=\sigma_\mathrm{up}-\sigma_\mathrm{lo}$. 

\begin{figure}[b]
    \centering
    \includegraphics[width=0.4\textwidth]{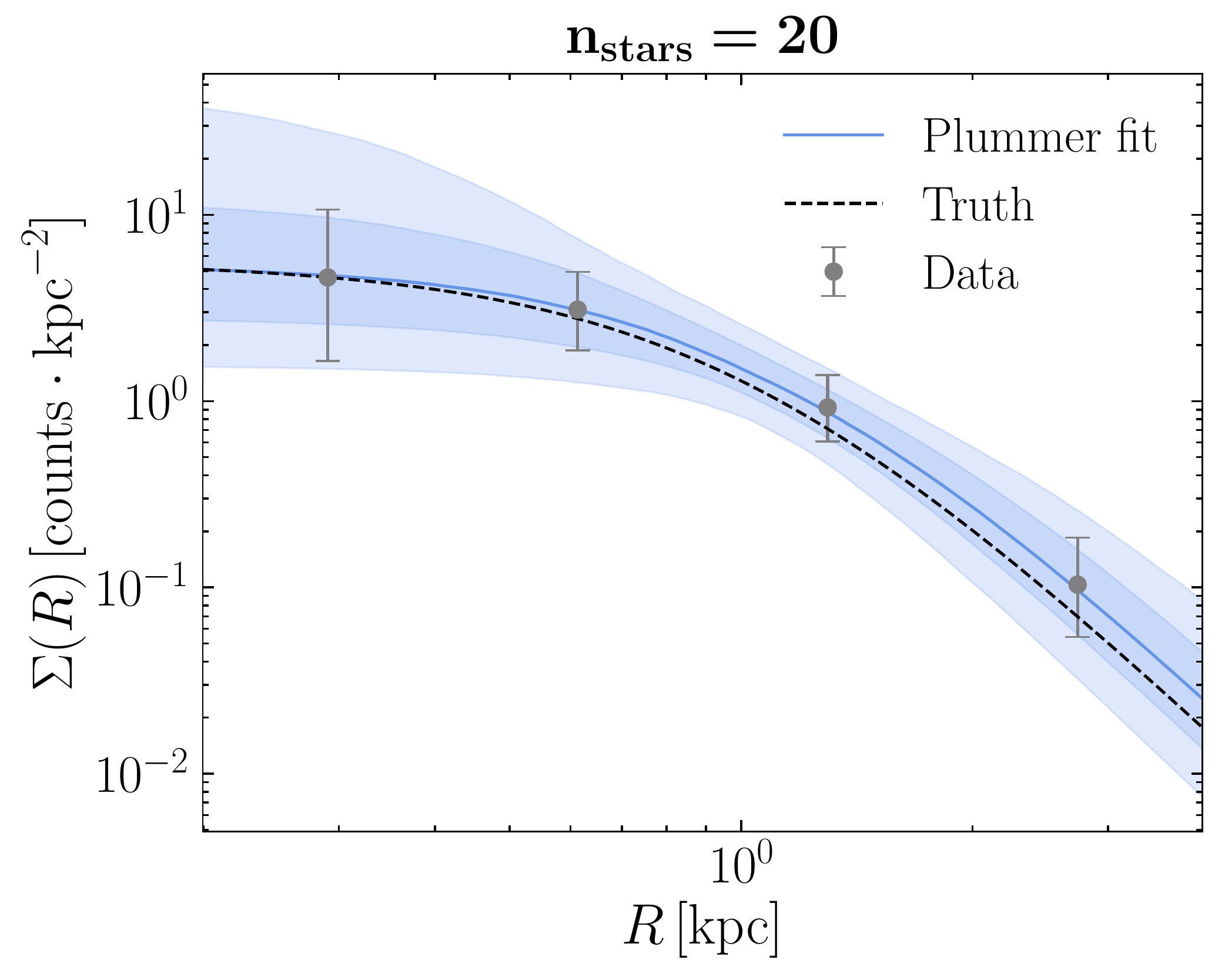}
    \includegraphics[width=0.3\textwidth]{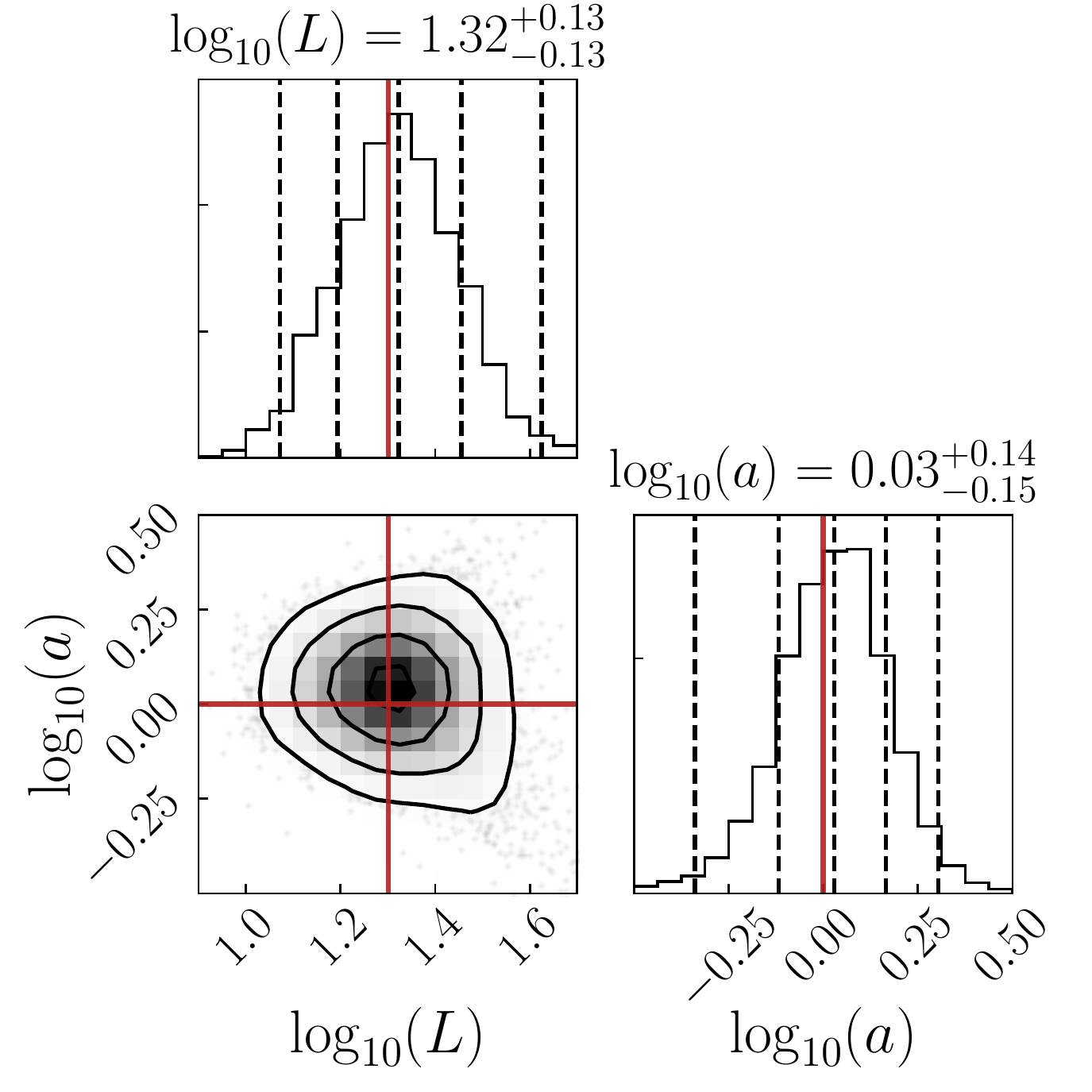}
    \caption{An example light profile fit for a single 20-star sample. In the left panel, the data points in the show the binned data, with error bars corresponding to the Poisson error for each bin; the blue line shows the median recovered profile, and the blue bands show the middle 68 and 95\% containment. The right panel shows the corresponding triangle plot on the light profile parameters, with the true value of $a$ indicated by the red lines. To convert the units of $L$ from star counts to luminosity, we have assumed that each star has luminosity $L_\odot$.}
    \label{fig:lp_20stars}
\end{figure}

\begin{figure}[h]
    \centering
    \includegraphics[width=0.4\textwidth]{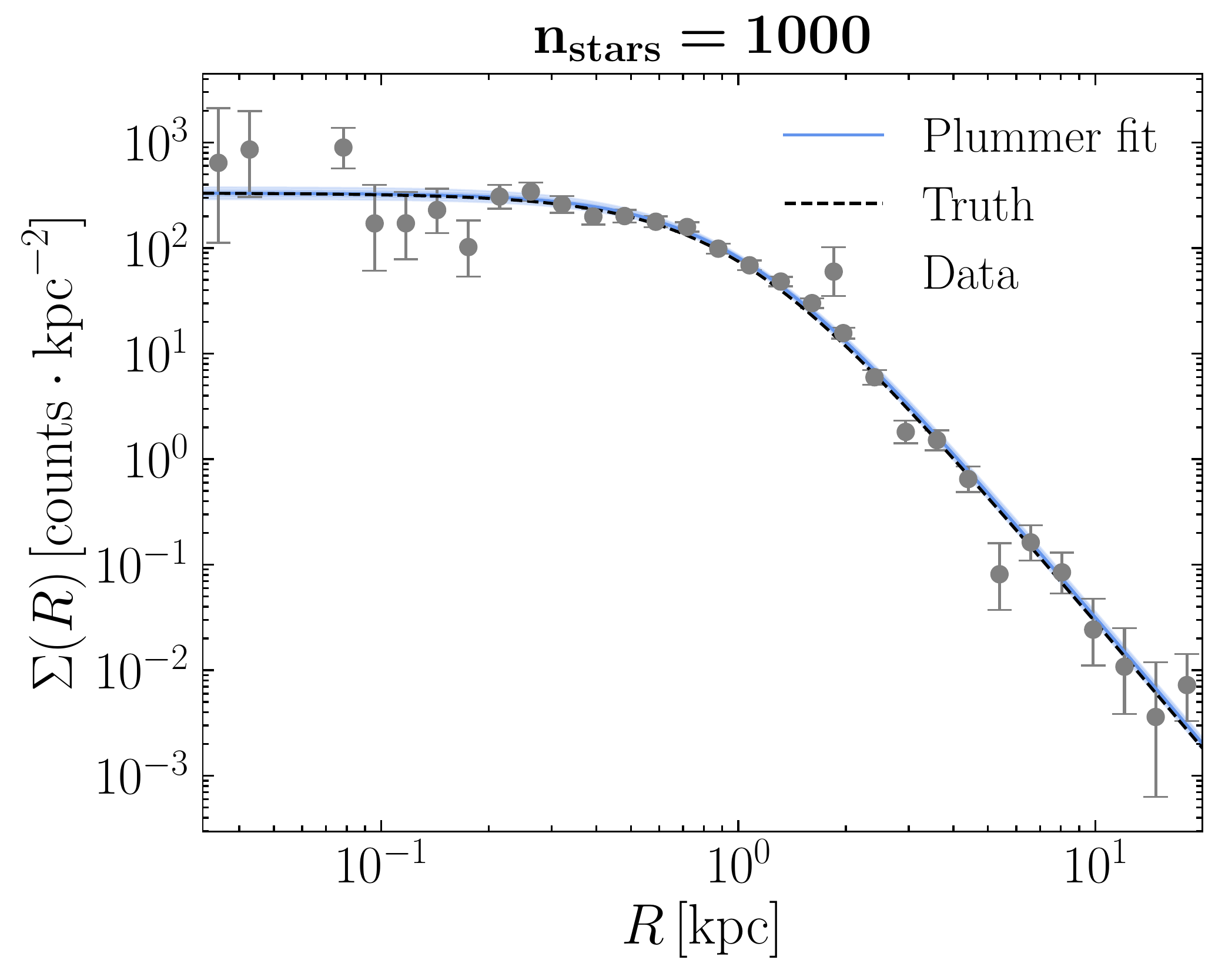}
    \includegraphics[width=0.3\textwidth]{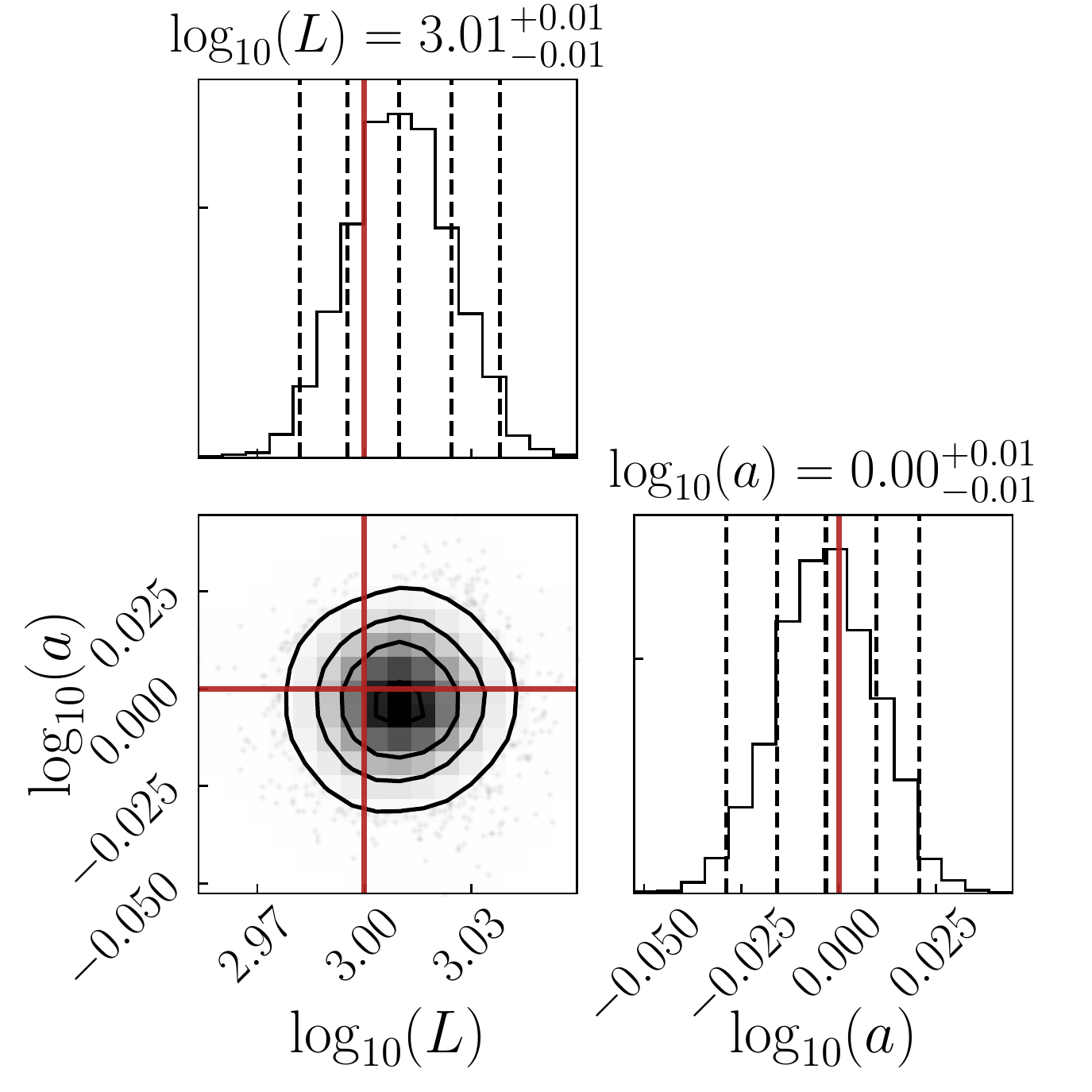}
    \caption{Same as Figure~\ref{fig:lp_20stars}, but for a 1000-star sample. Compared to the case of the 20-star sample, the light profile is significantly better constrained (note the different axes scales on the triangle plot compared to Fig.~\ref{fig:lp_20stars}).}
    \label{fig:lp_1000stars}
\end{figure}

We find that this approximation works well for our purposes, and we can generally fit the light profile extremely well. We note that in practice, it can be numerically easier to fit for the stellar surface density in each bin rather than the star counts themselves, but the principles remain unchanged. We use the results of the light profile fit to set the priors on the surface brightness parameters in our Jeans analysis---conservatively, we set the prior ranges of the surface brightness parameters to be the middle 95\% containment range on their posteriors from the light profile fit, similar to the procedure in \cite{2017MNRAS.471.4541R}. We show an example light profile fit for a 20-star sample in Figure~\ref{fig:lp_20stars} and a 1000-star sample in Figure~\ref{fig:lp_1000stars}. The fit results are generally in excellent agreement with the data, and become increasingly well-constrained as the sample size is increased.

\setcounter{figure}{0} \renewcommand{\thefigure}{D\arabic{figure}} \renewcommand{\theHfigure}{C\arabic{figure}}
\setcounter{table}{0} \renewcommand{\thetable}{D\arabic{table}} \renewcommand{\theHtable}{C\arabic{table}}
\setcounter{equation}{0} \renewcommand{\theequation}{D\arabic{equation}} \renewcommand{\theHequation}{C\arabic{equation}}

\section{Prior Selection and Joint Analysis}
\subsection{Implementing Narrow Priors}
\label{sec:prior_tests}

In our fiducial analysis, we choose conservative priors on the DM halo parameters. Specifically, we impose a wide prior of $\gamma\in[-1,5]$ due to the large theoretical uncertainty on the inner slopes of DM halos. While values of $\gamma<0$ are not physically-motivated, because they predict a density profile that dips down in the central region of the halo, we choose the lower bound of $-1$ to allow $\gamma$ the freedom to converge at $0$---this would not be possible if the lower bound on $\gamma$ were set exactly at $0$. The values of $\gamma$ on the highest end are also unphysical, because for $\gamma\geq3$, the enclosed mass (\Eq{eq:M_gNFW}) diverges at finite radius. Based on the posterior $\gamma$ distributions from our fiducial scans (\Fig{fig:gamma_post_1_0}), we do not expect that assuming a prior range of $\gamma\in[-1,3]$ instead would qualitatively change our results, because the posterior probability for values of $\gamma\geq3$ tend to be negligible.

Additionally, we have assumed a wide prior on the DM scale radius $r_s$ of $\ln(r_s/\mathrm{kpc})\in[-10,10]$ for the purpose of being fully agnostic. However, we can follow the example of GS15 and set the more physically-motivated prior range of on $r_s$ of $1\,\mathrm{pc}$ to $100\,\mathrm{kpc}$, i.e., $\ln(r_s/\mathrm{kpc})\in[\ln(10^{-3}),\ln(10^2)]\sim[-6.9,4.6]$. For reference, a commonly used value for the NFW scale radius of the Milky Way DM halo is $\sim20\,\mathrm{kpc}$~\citep[e.g.][]{Goodenough:2009gk, Daylan:2014rsa, Calore:2014xka, TheFermi-LAT:2015kwa, Chang:2018bpt, Ackermann:2012rg}---because we expect the dwarf galaxy DM halos to have smaller scale radii than the Milky Way halo, the GS15 priors are still fairly conservative.

\begin{figure}[t]
    \centering
    \includegraphics[width=0.8\textwidth]{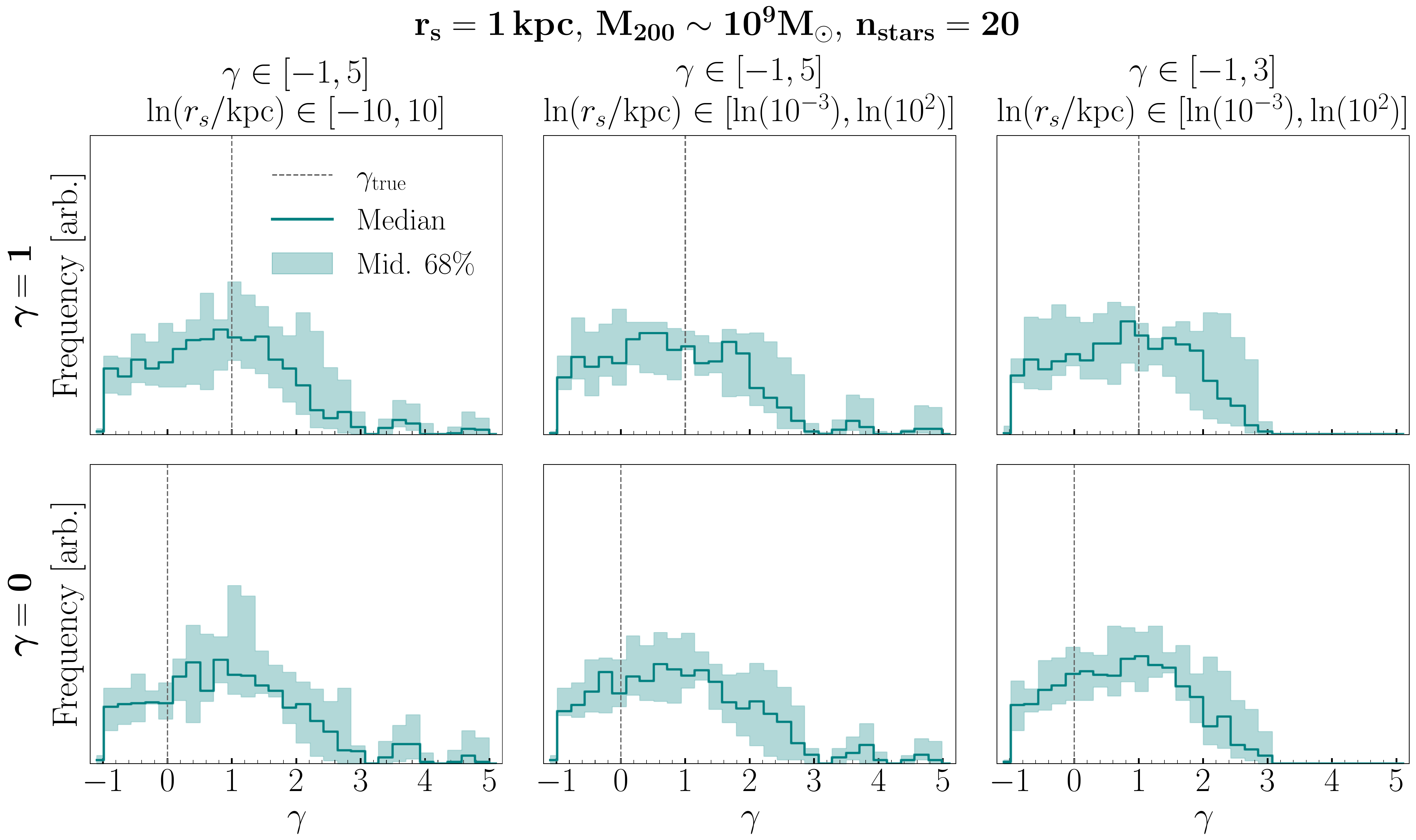}
    \caption{Posterior $\gamma$ distributions for the 20-star samples for parameter sets I (top row) and III (bottom row). From left to right, the columns show the results for the fiducial priors, the narrow prior on $r_s$ and fiducial prior on $\gamma$, and the narrow priors on both $r_s$ and $\gamma$.}
    \label{fig:gamma_post_1_0_rs_1_priors}
\end{figure}

\begin{figure}[t]
    \centering
    \includegraphics[width=0.85\textwidth]{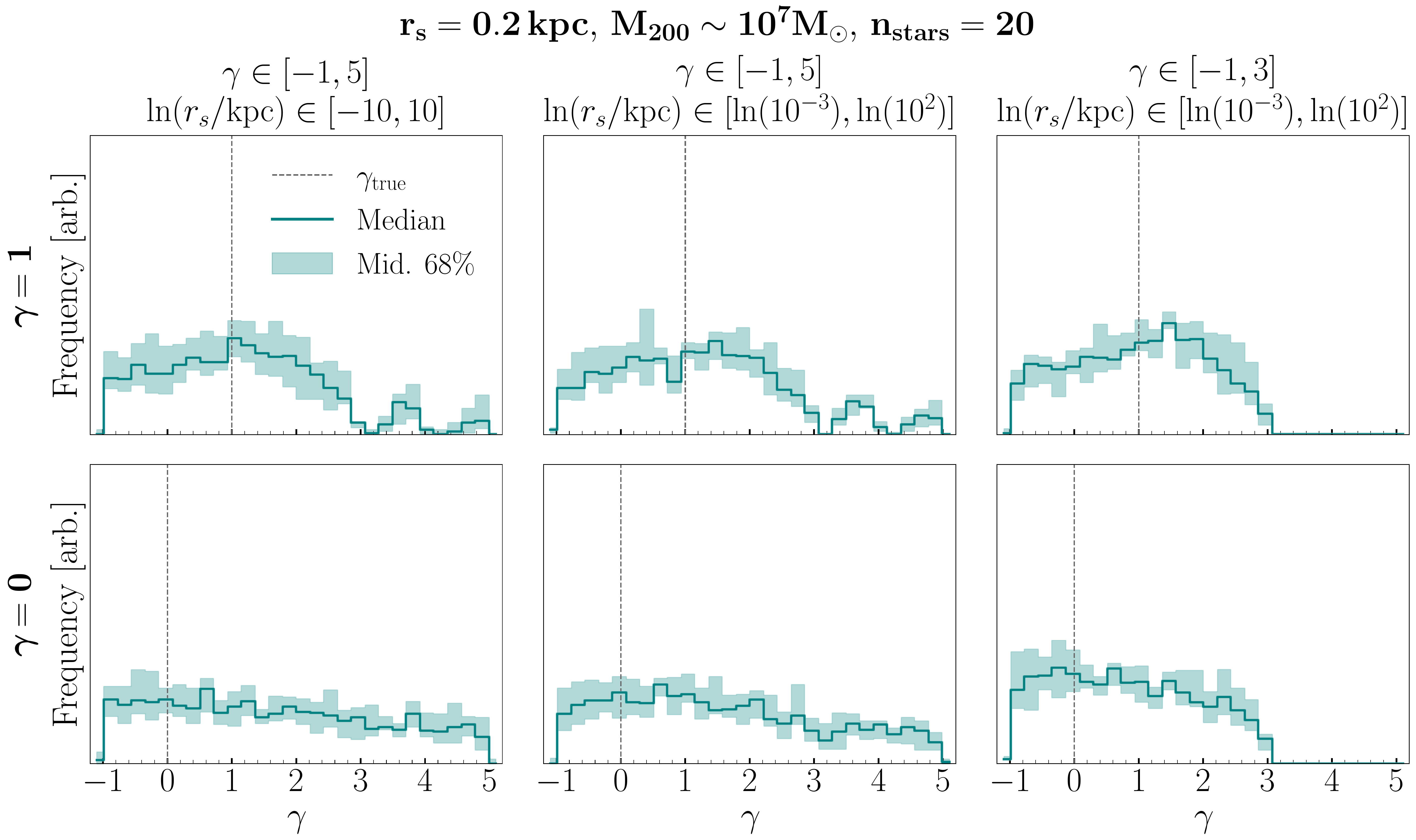}
    \caption{Same as Figure~\ref{fig:gamma_post_1_0_rs_1_priors}, but for parameter sets II (top row) and IV (bottom row).}
    \label{fig:gamma_post_1_0_rs_0p2_priors}
\end{figure}

We explicitly verify that implementing a narrower choice of priors on $\gamma$ and on $r_s$ does not qualitatively affect the results of our study, focusing on the 20-star samples because the smallest samples are most sensitive to prior choices. In Figure~\ref{fig:gamma_post_1_0_rs_1_priors}, we show the posterior $\gamma$ distributions for the 20-star samples for parameter sets I (top row) and III (bottom row). From left to right, the columns show the results for the fiducial priors, the narrow prior on $r_s$ and fiducial prior on $\gamma$, and the narrow priors on both $r_s$ and $\gamma$. We assume fiducial priors on all other parameters and the fiducial velocity error, $\Delta v=2\,\mathrm{km/s}$. The left column of Fig.~\ref{fig:gamma_post_1_0_rs_1_priors} corresponds to the left column of Fig.~\ref{fig:gamma_post_1_0} (for ease of comparison between the different sets of priors, the vertical scale here is zoomed in compared to Fig.~\ref{fig:gamma_post_1_0}). While there are slight quantitative changes, the key result---that the posterior $\gamma$ distributions are unconstrained, and therefore do not give rise to statistical evidence for a cusp or a core---remains unchanged. We show the analogous results for parameter sets II and IV in Figure~\ref{fig:gamma_post_1_0_rs_0p2_priors}.

\begin{figure}[t]
    \centering
    \includegraphics[width=0.8\textwidth]{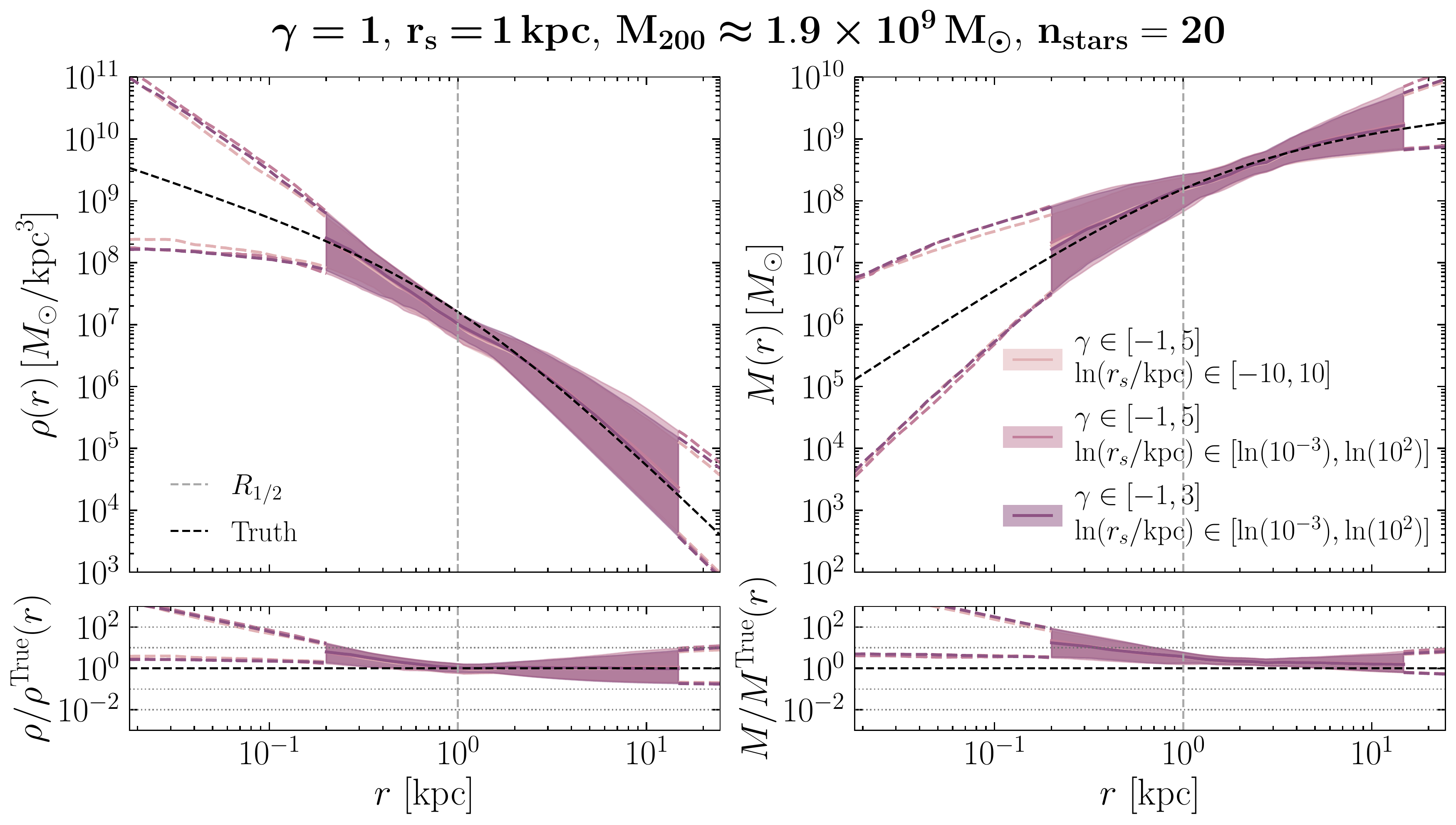}
    \caption{Inferred DM density profiles $\rho(r)$ (left panels) and corresponding enclosed mass profiles $M(r)$ (right panels) for 20-star samples parameter set I. From lightest to darkest color, we show the results for the fiducial priors, the narrow prior on $r_s$ and fiducial prior on $\gamma$, and the narrow priors on both $r_s$ and $\gamma$. The recovered distributions are overall insensitive to the prior choices on $r_s$ and $\gamma$. The recovered virial mass is $M_{200}\sim\errorbars{2.3}{10.2}{1.5}\times10^9M_\odot$ for the case of narrow prior on $r_s$ and fiducial prior on $\gamma$ and $M_{200}\sim\errorbars{2.2}{7.5}{1.4}\times10^9M_\odot$ for the case of narrow priors on both $r_s$ and $\gamma$. For the fiducial analysis, this value is $M_{200}\sim\errorbars{2.1}{6.7}{1.3}\times10^9M_\odot$. In each case, the recovered virial mass is consistent within uncertainty with the true value of $M_{200}\sim1.9\times10^9M_\odot$.}
    \label{fig:rho_Menc_gamma_1_priors}
\end{figure}

\begin{figure}[t]
    \centering
    \includegraphics[width=0.8\textwidth]{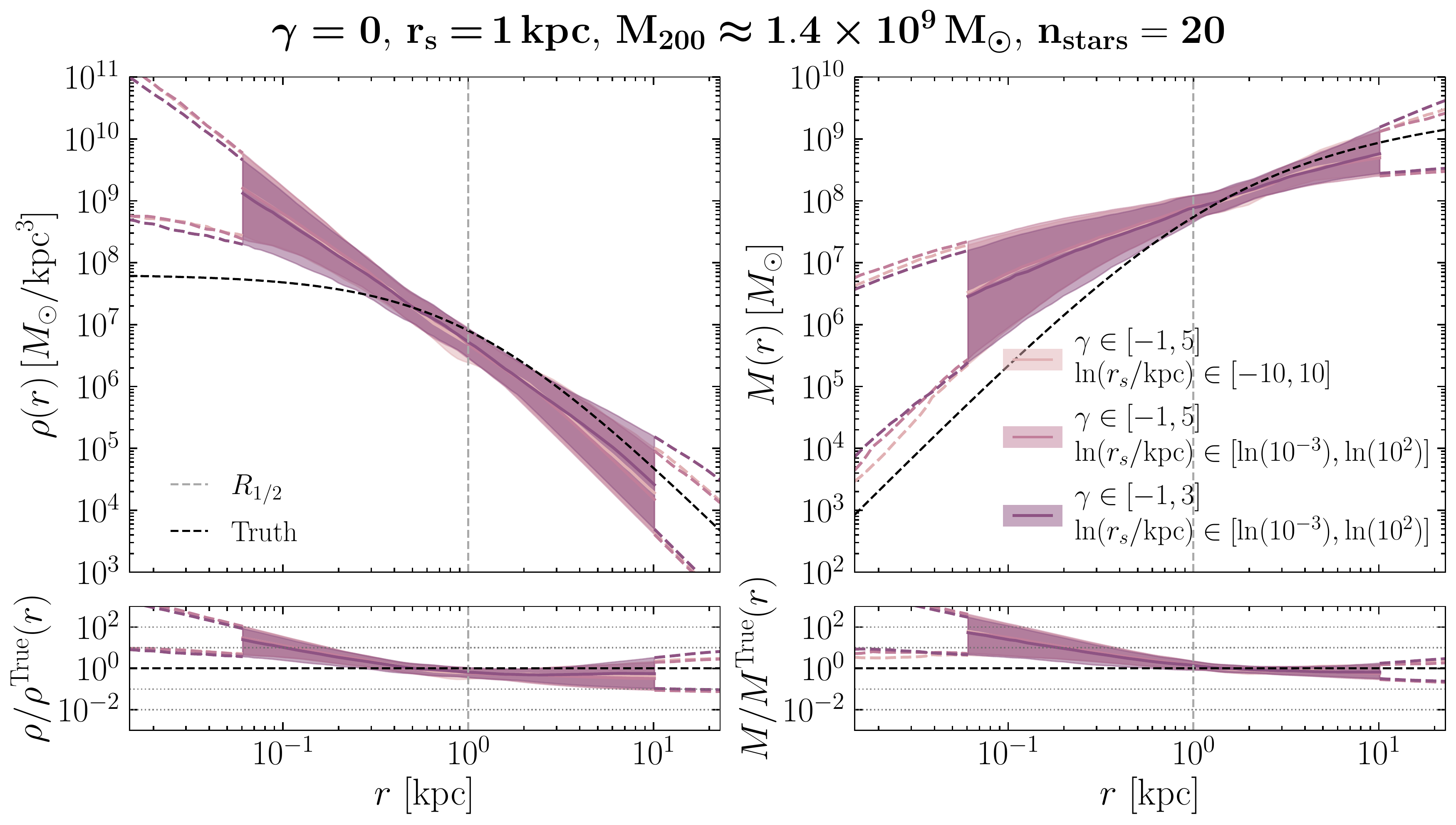}
    \caption{Same as Figure~\ref{fig:rho_Menc_gamma_1_priors}, but for parameter set III. The recovered virial mass is $M_{200}\sim\errorbars{0.7}{2.0}{0.4}\times10^9M_\odot$ for the case of narrow prior on $r_s$ and fiducial prior on $\gamma$ and $M_{200}\sim\errorbars{1.0}{3.3}{0.6}\times10^9M_\odot$ for the case of narrow priors on both $r_s$ and $\gamma$. For the fiducial analysis, this value is $M_{200}\sim\errorbars{0.7}{2.3}{0.4}\times10^9M_\odot$. In each case, the recovered virial mass is consistent within uncertainty with the true value of $M_{200}\sim1.4\times10^9M_\odot$.}
    \label{fig:rho_Menc_gamma_0_priors}
\end{figure}

Similarly, we can examine the recovered density and enclosed mass profiles that result from the narrow prior choices and compare them to our fiducial results. We show this comparison for the 20-star samples from parameter set I in Figure~\ref{fig:rho_Menc_gamma_1_priors}. Qualitatively, we find that the recovered distributions are insensitive to the prior choices on $r_s$ and $\gamma$. Quantitatively, the recovered virial mass is $M_{200}\sim\errorbars{2.3}{10.2}{1.5}\times10^9M_\odot$ for the case of narrow prior on $r_s$ and fiducial prior on $\gamma$ and $M_{200}\sim\errorbars{2.2}{7.5}{1.4}\times10^9M_\odot$ for the case of narrow priors on both $r_s$ and $\gamma$. For the fiducial analysis, this value is $M_{200}\sim\errorbars{2.1}{6.7}{1.3}\times10^9M_\odot$. In each case, the recovered virial mass is consistent within uncertainty with the true value of $M_{200}\sim1.9\times10^9M_\odot$. Additionally, imposing narrow priors on $r_s$ and $\gamma$ does not result in smaller uncertainties on the inferred virial mass. We show the analogous results for parameter set III in Figure~\ref{fig:rho_Menc_gamma_0_priors}. The results for parameter sets II and IV are qualitatively similar.

For brevity, we only present selected representative results here. We have verified that, for our spatially selected samples (see Section~\ref{sec:location} for detailed discussion), the choice of narrow priors on $r_s$ and $\gamma$ also results in qualitatively unchanged results from the fiducial ones presented in the paper. We have found, however, that the narrow priors have a regulating effect in our preliminary study of jointly analyzing multiple dwarfs simultaneously, relative to our fiducial priors---we therefore employ the narrow priors in our discussion of the joint analysis in App.~\ref{sec:stacking}. 

\vspace{1cm}
\subsection{Joint Analysis}
\label{sec:stacking}

In lieu of obtaining much larger datasets (up to $\sim\mathcal{O}(10,000)$ stars) for the measured dwarf galaxies, one potential way to gain more constraining power on the DM halo parameters using moderately-sized datasets is to jointly analyze multiple dwarf galaxies at once. While it may not be feasible in the near future to increase the stellar sample sizes within measured dwarf galaxies by orders of magnitude, with the advent of digital surveys, the number of discovered dwarf galaxies has exploded over the past five years~\citep[see, e.g., Figure 1 of][]{2019arXiv190105465S}, and is expected to continue to grow drastically in the era of future surveys such as the Vera C. Rubin Observatory~\citep[formerly known as LSST, see, e.g., Table 1 of][]{Hargis:2014kaa}. We could therefore try to leverage a large number of measured dwarf galaxies, even if within the individual systems the number of observed stars is small.

Within our analysis framework, we can in principle perform a joint analysis on $N_\mathrm{dwarfs}$ of our simulated dwarfs. For simplicity, we assume all $N_\mathrm{dwarfs}$ systems are from the same parameter set and have the same number of stars, and we analyze them simultaneously, under the prior assumption that they all share the same value of $\gamma$ but are otherwise independent. This simulates the scenario of having a group of similarly-sized dwarf galaxies for which one might expect \emph{a priori}, based on the specifics of the DM and baryonic feedback models, to have the same inner DM profile shape. 

The joint likelihood is the product of \Eq{eq:likelihood} over each dwarf, 
\begin{equation}
\mathcal{L}_\mathrm{joint} = \prod_{j=1}^{N_\mathrm{dwarfs}}\prod_{i=1}^{N_{\rm{stars}}} \frac{(2\pi)^{-1/2}}{ \sqrt{ \sigma_{p,j}^2 (R_{ij}) + \Delta_{v_{ij}}^2}}\times\exp \left[ - \frac{1}{2} \left( \frac{(v_{ij} - \ov{v}_j)^2}{\sigma_{p,j}^2(R_{ij}) + \Delta_{v_{ij}}^2} \right) \right]\,.
\label{eq:likelihood_stacked}
\end{equation}
As in the case of the individual analyses, we model each dwarf with a Plummer light profile and gNFW DM distribution, but fit for only one value of $\gamma$ for all the dwarfs, i.e., $\gamma_j=\gamma$. The joint analysis model therefore has $(N_\mathrm{dwarfs}\times5+1)$ free parameters. 

We have tested this method by taking five 20-star samples from the same parameter set and maximizing their joint likelihood. We note that for the results shown in this section, we have used the narrow priors on $r_s$ and $\gamma$ described in App.~\ref{sec:prior_tests} and assumed a velocity error of $\Delta v=0\,\mathrm{km/s}$ for cleanliness. All other priors are the same as in our fiducial setup. We choose to focus on the narrow priors because we have found that, for the cases we have tested, the joint analysis results can be biased more often towards incorrect values of $\gamma$ when using our fiducial priors.

\begin{figure}[t]
    \centering
    \includegraphics[width=0.95\textwidth]{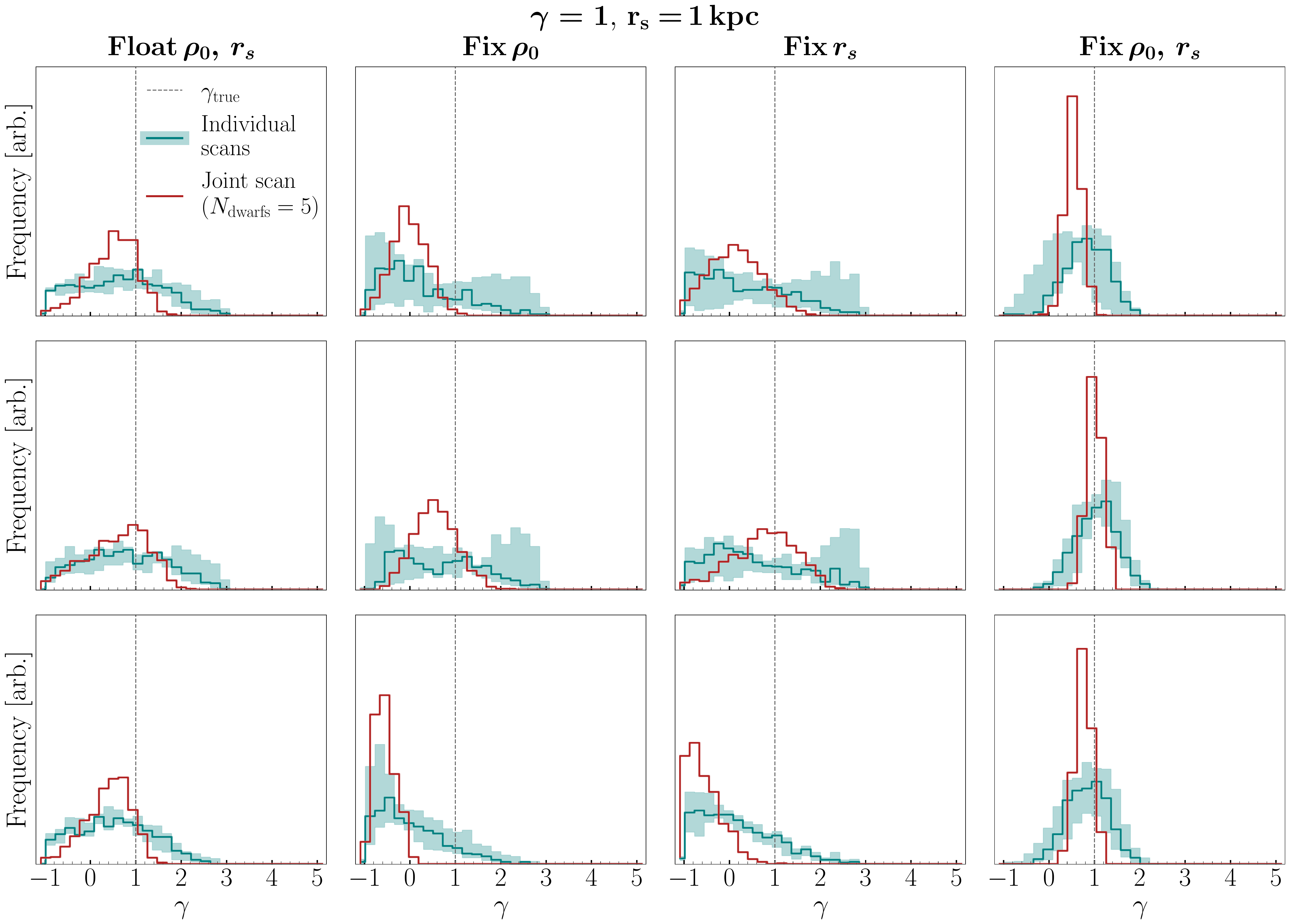}
    \caption{Example results from jointly analyzing five 20-star samples from parameter set I, for which $\gamma=1$. Each row corresponds to a different set of five jointly analyzed samples. From left to right, we show the results for floating all free parameters (26 free parameters), fixing $\rho_0$ for each sample to the true value (21 free parameters), fixing $r_s$ for each sample to the true value (21 free parameters), and fixing both $\rho_0$ and $r_s$ for each sample to their respective true values (16 free parameters). Within each panel, we show the posterior $\gamma$ distributions resulting from the individual as well as the joint analyses---the teal line(band) shows the median of the median(middle 68\%) in each $\gamma$ bin across the five individual scans, while the red line shows the posterior $\gamma$ distribution from the joint scan.}
    \label{fig:stacking_g_1}
\end{figure}

In Figure~\ref{fig:stacking_g_1}, we show example results for parameter set I (for which $\gamma=1$), with each row corresponding to a different set of five jointly analyzed samples. In the first column, we show the results floating all 26 free parameters. Within each panel, we show the posterior $\gamma$ distributions resulting from the individual as well as the joint analyses---the teal line(band) shows the median of the median(middle 68\%) in each $\gamma$ bin across the five individual scans, while the red line shows the posterior $\gamma$ distribution from the joint scan. While the posterior $\gamma$ distribution from the joint scan is more constrained and peaked near $\gamma=1$, the posterior probability at $\gamma=0$ tends to be non-negligible. The degeneracy between the DM halo parameters still has a strong effect on these particular results, as demonstrated by the fourth column, in which we fix $\rho_0$ and $r_s$ to their respective true values for each of the five samples (i.e., we now float a total of 16 parameters). In this case, the posterior distribution is narrowly peaked and the posterior probability at $\gamma=0$ is negligible in all cases (although depending on the specific set of samples, the location of the peak may be shifted away from the true value of $\gamma=1$). If we fix either $\rho_0$ or $r_s$ individually, we find that the joint analysis can accentuate biases that are present in the underlying samples (most clearly demonstrated by the middle two panels of the bottom row).

\begin{figure}[t]
    \centering
    \includegraphics[width=0.95\textwidth]{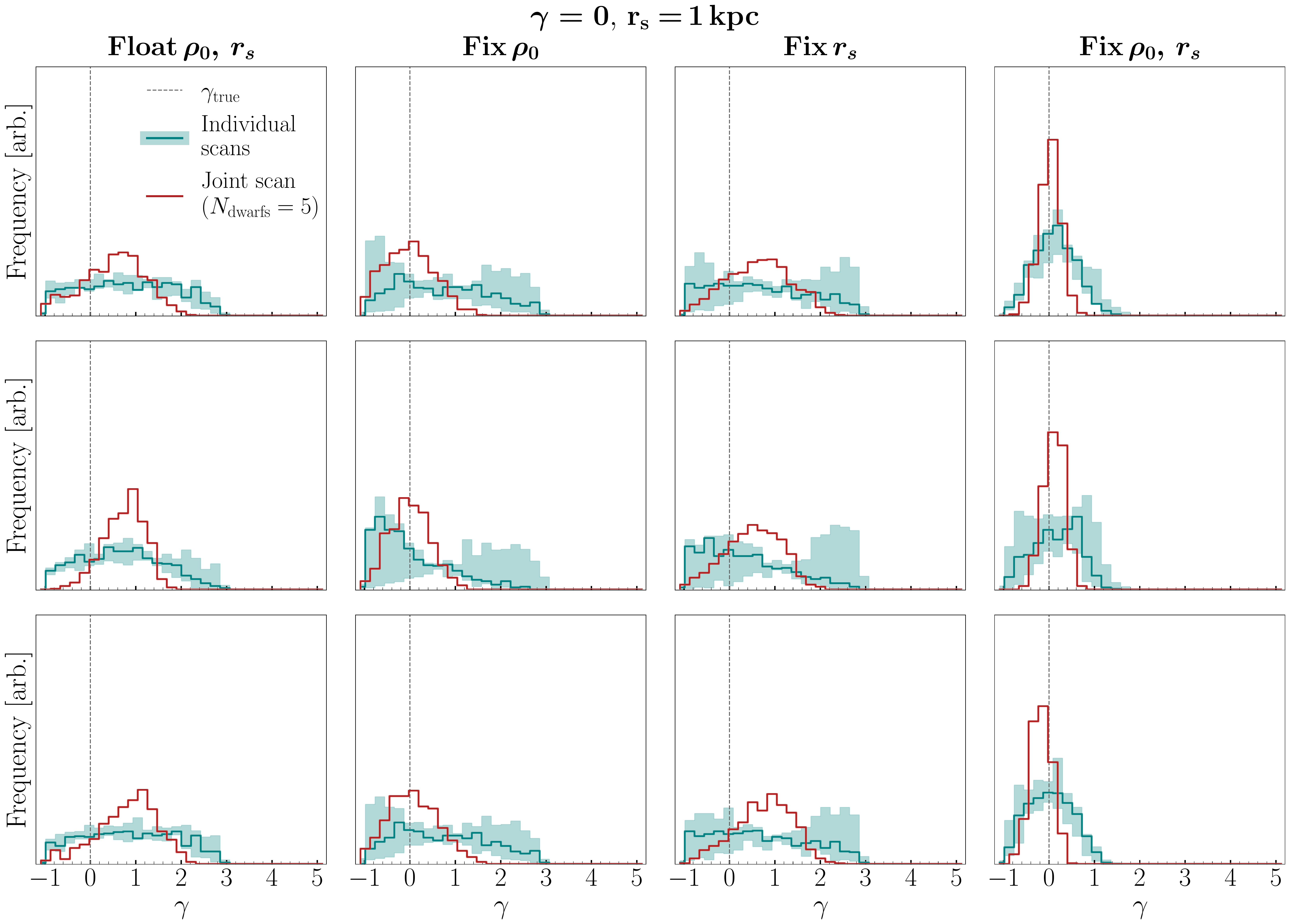}
    \caption{Same as Figure~\ref{fig:stacking_g_1} but for parameter set III, for which $\gamma=0$.}
    \label{fig:stacking_g_0}
\end{figure}

In Figure~\ref{fig:stacking_g_0}, we show analogous example results for parameter set III (for which $\gamma=0$). In this case, when all 26 free parameters are floated, the posterior $\gamma$ distributions from the joint analysis tend to also be peaked near $\gamma=1$. In the examples shown here, fixing $\rho_0$ for all the samples in the joint analysis resolves this bias, resulting in posterior $\gamma$ distributions which are peaked near $\gamma=0$ and better-constrained than the corresponding posteriors from the individual scans. When both $\rho_0$ and $r_s$ are fixed to their respective true values for each of the five samples, the posterior distributions from the joint scans are peaked cleanly near $\gamma=0$ in all three cases; however, the bias towards $\gamma=1$ is again present if we only fix $r_s$.

Further detailed study is required in order to understand the source of the biases we see, and also to characterize how the output of a joint analysis depends on factors such as the value of $N_\mathrm{dwarfs}$, the sample size and measurement precision in each dwarf, and the relaxation of the assumption that the dwarfs all share the same value of $\gamma$ (for example, by assuming a central value of $\gamma$ and some scatter about it for the population of dwarfs being analyzed). However, the dimensionality of the model quickly grows as $N_\mathrm{dwarfs}$ is increased, making a joint analysis difficult to efficiently implement using standard MCMC or nested sampling techniques. In particular, the number of \textsc{Multinest} evaluations required for convergence scales exponentially above $\sim30$ dimensions~\citep{2015MNRAS.453.4384H}, making it computationally infeasible to perform a detailed study using the analysis framework presented here. Nevertheless, our preliminary results suggest that a joint analysis approach is a promising method for making the most of the data moving forward, and deserves its own dedicated study. This would require the use of newer inference techniques which are designed to approximate posterior distributions for high-dimensional likelihoods, such as stochastic variational inference~\citep{2012arXiv1206.7051H}. 

\clearpage

\setcounter{figure}{0} \renewcommand{\thefigure}{E\arabic{figure}} \renewcommand{\theHfigure}{E\arabic{figure}}
\setcounter{table}{0} \renewcommand{\thetable}{E\arabic{table}} \renewcommand{\theHtable}{E\arabic{table}}
\setcounter{equation}{0} \renewcommand{\theequation}{E\arabic{equation}} \renewcommand{\theHequation}{E\arabic{equation}}

\section{Virial Mass and $J$-Factor Estimates}
\begin{table}[h]
    \begin{center}
    \renewcommand{\arraystretch}{1.2}
    \begin{tabular}{|c||c|c||c|c||c|c|}
    \multicolumn{7}{c}{$\boldsymbol{\mathbf{I.\,\,\gamma=1,\,r_s=1\,kpc,\,M_{200}\approx1.9\times10^9\,M_\odot,\,\log_{10}\left[J(0.5^\circ)/(GeV^2cm^{-5})\right]\approx 19.3}}$}
    \vspace{1.5mm} \\
    \hline
    & \multicolumn{2}{c||}{$\Delta v=0$ km/s} & \multicolumn{2}{c||}{$\Delta v=2$ km/s} & \multicolumn{2}{c|}{$\Delta v=5$ km/s}\\
    \hline
    $\nstars$ & $M_{200}\,[10^9\,M_\odot]$ & $\log_{10}[J(0.5^\circ)]$ & $M_{200}\,[10^9\,M_\odot]$ & $\log_{10}[J(0.5^\circ)]$ & $M_{200}\,[10^9\,M_\odot]$ & $\log_{10}[J(0.5^\circ)]$ \\    
    \hline
    20 & $3.0^{+22.0}_{-2.1}$ & $\errorbars{19.6}{1.8}{1.1}$ & $2.1^{+6.7}_{-1.3}$ & $\errorbars{19.9}{1.9}{1.1}$ & $1.6^{+9.9}_{-1.1}$ & $\errorbars{19.9}{2.1}{1.3}$ \\
    100 & $1.5^{+1.2}_{-0.5}$ & $\errorbars{19.8}{1.1}{0.5}$ & $1.3^{+1.0}_{-0.5}$ & $\errorbars{19.7}{1.0}{0.5}$ & $1.4^{+1.2}_{-0.6}$ & $\errorbars{19.8}{1.1}{0.6}$ \\
    1000 & $1.9^{+0.5}_{-0.3}$ & $\errorbars{19.4}{0.4}{0.3}$ & $1.9^{+0.6}_{-0.3}$ & $\errorbars{19.4}{0.4}{0.2}$ & $1.9^{+0.6}_{-0.3}$ & $\errorbars{19.4}{0.4}{0.2}$ \\
    10,000 & $1.8^{+0.1}_{-0.1}$ & $\errorbars{19.3}{0.2}{0.1}$ & $1.9^{+0.1}_{-0.1}$ & $\errorbars{19.3}{0.2}{0.1}$ & $1.9^{+0.2}_{-0.1}$ & $\errorbars{19.3}{0.2}{0.1}$ \\
    \hline
    \end{tabular}
    \end{center}
    
    
    \begin{center}
    \renewcommand{\arraystretch}{1.2}
    \begin{tabular}{|c||c|c||c|c||c|c|}
    \multicolumn{7}{c}{$\boldsymbol{\mathbf{II.\,\,\gamma=1,\,r_s=0.2\,kpc,\,M_{200}\approx1.5\times10^7\,M_\odot,\,\log_{10}\left[J(0.5^\circ)/(GeV^2cm^{-5})\right]\approx 17.3}}$}
    \vspace{1.5mm} \\
    \hline
    & \multicolumn{2}{c||}{$\Delta v=0$ km/s} & \multicolumn{2}{c||}{$\Delta v=2$ km/s} & \multicolumn{2}{c|}{$\Delta v=5$ km/s}\\
    \hline
    $\nstars$ & $M_{200}\,[10^7\,M_\odot]$ & $\log_{10}[J(0.5^\circ)]$ & $M_{200}\,[10^7\,M_\odot]$ & $\log_{10}[J(0.5^\circ)]$ & $M_{200}\,[10^7\,M_\odot]$ & $\log_{10}[J(0.5^\circ)]$ \\
    \hline
    20 & $0.7^{+1.0}_{-0.3}$ & $\errorbars{18.6}{1.3}{1.2}$ & $0.8^{+8.5}_{-0.6}$ & $\errorbars{18.4}{1.6}{1.5}$ & $0.00001^{+0.06}_{-0.00001}$ & $\errorbars{11.2}{5.6}{7.4}$ \\
    100 & $1.2^{+0.9}_{-0.4}$ & $\errorbars{17.6}{0.7}{0.4}$ & $1.0^{+1.4}_{-0.5}$ & $\errorbars{17.8}{1.2}{0.6}$ & $0.1^{+0.3}_{-0.1}$ & $\errorbars{15.3}{3.5}{7.6}$ \\
    1000 & $1.5^{+0.3}_{-0.2}$ & $\errorbars{17.4}{0.3}{0.2}$ & $1.3^{+0.4}_{-0.3}$ & $\errorbars{17.4}{0.3}{0.2}$ & $0.9^{+0.7}_{-0.4}$ & $\errorbars{18.0}{0.8}{0.5}$ \\
    10,000 & $1.6^{+0.1}_{-0.1}$ & $\errorbars{17.4}{0.2}{0.1}$ & $1.4^{+0.2}_{-0.1}$ & $\errorbars{17.4}{0.2}{0.1}$ & $1.4^{+0.8}_{-0.3}$ & $\errorbars{17.6}{0.4}{0.2}$ \\
    \hline
    \end{tabular}
    \end{center}
    
    
    \begin{center}
    \renewcommand{\arraystretch}{1.2}
    \begin{tabular}{|c||c|c||c|c||c|c|}
    \multicolumn{7}{c}{$\boldsymbol{\mathbf{III.\,\,\gamma=0,\,r_s=1\,kpc,\,M_{200}\approx1.4\times10^9\,M_\odot,\,\log_{10}\left[J(0.5^\circ)/(GeV^2cm^{-5})\right]\approx 17.9}}$}
    \vspace{1.5mm} \\
    \hline
    & \multicolumn{2}{c||}{$\Delta v=0$ km/s} & \multicolumn{2}{c||}{$\Delta v=2$ km/s} & \multicolumn{2}{c|}{$\Delta v=5$ km/s}\\
    \hline
    $\nstars$ & $M_{200}\,[10^9\,M_\odot]$ & $\log_{10}[J(0.5^\circ)]$ & $M_{200}\,[10^9\,M_\odot]$ & $\log_{10}[J(0.5^\circ)]$ & $M_{200}\,[10^9\,M_\odot]$ & $\log_{10}[J(0.5^\circ)]$ \\  
    \hline
    20 & $0.8^{+3.6}_{-0.5}$ & $\errorbars{19.3}{2.1}{1.1}$ & $0.7^{+2.3}_{-0.4}$ & $\errorbars{19.9}{2.2}{1.3}$ & $1.4^{+16.5}_{-1.0}$ & $\errorbars{19.7}{2.5}{1.4}$ \\
    100 & $1.5^{+1.6}_{-0.6}$ & $\errorbars{18.5}{0.6}{0.4}$ & $1.6^{+1.8}_{-0.7}$ & $\errorbars{18.5}{0.6}{0.4}$ & $1.6^{+2.3}_{-0.9}$ & $\errorbars{18.5}{0.7}{0.5}$ \\
    1000 & $1.6^{+0.4}_{-0.3}$ & $\errorbars{18.2}{0.4}{0.2}$ & $1.5^{+0.5}_{-0.2}$ & $\errorbars{18.2}{0.4}{0.2}$ & $1.4^{+0.4}_{-0.2}$ & $\errorbars{18.2}{0.4}{0.2}$ \\
    10,000 & $1.4^{+0.1}_{-0.1}$ & $\errorbars{18.0}{0.3}{0.1}$ & $1.5^{+0.1}_{-0.1}$ & $\errorbars{18.0}{0.3}{0.1}$ & $1.5^{+0.2}_{-0.1}$ & $\errorbars{18.0}{0.3}{0.1}$ \\
    \hline
    \end{tabular}
    \end{center}
    
    \begin{center}
    \renewcommand{\arraystretch}{1.2}
    \begin{tabular}{|c||c|c||c|c||c|c|}
    \multicolumn{7}{c}{$\boldsymbol{\mathbf{IV.\,\,\gamma=0,\,r_s=0.2\,kpc,\,M_{200}\approx1.1\times10^7\,M_\odot,\,\log_{10}\left[J(0.5^\circ)/(GeV^2cm^{-5})\right]\approx 16.3}}$}
    \vspace{1.5mm} \\
    \hline
    & \multicolumn{2}{c||}{$\Delta v=0$ km/s} & \multicolumn{2}{c||}{$\Delta v=2$ km/s} & \multicolumn{2}{c|}{$\Delta v=5$ km/s}\\
    \hline
    $\nstars$ & $M_{200}\,[10^7\,M_\odot]$ & $\log_{10}[J(0.5^\circ)]$ & $M_{200}\,[10^7\,M_\odot]$ & $\log_{10}[J(0.5^\circ)]$ & $M_{200}\,[10^7\,M_\odot]$ & $\log_{10}[J(0.5^\circ)]$ \\
    \hline
    20 & $0.4^{+1.7}_{-0.3}$ & $\errorbars{17.0}{1.2}{0.7}$ & $0.00004^{+0.1}_{-0.00004}$ & $\errorbars{11.6}{4.8}{7.8}$ & $0.00004^{+0.3}_{-0.00004}$ & $\errorbars{11.7}{5.3}{8.2}$ \\
    100 &  $1.9^{+2.3}_{-0.8}$ & $\errorbars{16.5}{0.3}{0.2}$ & $0.4^{+1.7}_{-0.3}$ & $\errorbars{16.9}{1.2}{0.7}$ & $0.0001^{+0.6}_{-0.0001}$ & $\errorbars{12.1}{4.7}{7.9}$ \\
    1000 & $1.4^{+0.3}_{-0.2}$ & $\errorbars{16.4}{0.2}{0.1}$ & $1.5^{+0.8}_{-0.4}$ & $\errorbars{16.5}{0.3}{0.2}$ & $0.6^{+3.5}_{-0.6}$ & $\errorbars{16.3}{0.9}{3.4}$ \\
    10,000 & $1.1^{+0.08}_{-0.06}$ & $\errorbars{16.5}{0.2}{0.2}$ & $1.2^{+0.2}_{-0.1}$ & $\errorbars{16.5}{0.2}{0.2}$ & $1.9^{+1.6}_{-0.9}$ & $\errorbars{16.5}{0.2}{0.2}$ \\
    \hline
    \end{tabular}
    \end{center}
    \caption{Inferred values of the virial mass $M_{200}$ and {\jfac} for the different parameter sets, sample sizes, and values of $\Delta v$. The {\jfac}s are in units of $\log_{10}[J(0.5^\circ)/(\mathrm{GeV}^2\mathrm{cm}^{-5})]$. Each entry in this table represents the median across 10 realizations of the median and $\pm1\sigma$ values.}
    \label{tab:mvir_deltav}
\end{table}

\begin{table}[h]
    \begin{center}
    \renewcommand{\arraystretch}{1.2}
    \begin{tabular}{|c||c|c||c|c||c|c|}
    \multicolumn{7}{c}{$\boldsymbol{\mathbf{I.\,\,\gamma=1,\,r_s=1\,kpc,\,M_{200}\approx1.9\times10^9\,M_\odot,\,\log_{10}\left[J(0.5^\circ)/(GeV^2cm^{-5})\right]\approx 19.3}}$}
    \vspace{1.5mm} \\
    \hline
    & \multicolumn{2}{c||}{Benchmark} & \multicolumn{2}{c||}{$R<R_{1/2}$} & \multicolumn{2}{c|}{$R>R_{1/2}$}\\
    \hline
    $\nstarssel$ & $M_{200}\,[10^9\,M_\odot]$ & $\log_{10}[J(0.5^\circ)]$ & $M_{200}\,[10^9\,M_\odot]$ & $\log_{10}[J(0.5^\circ)]$ & $M_{200}\,[10^9\,M_\odot]$ & $\log_{10}[J(0.5^\circ)]$ \\
    \hline
    $\sim50$ & $\errorbars{1.2}{1.2}{0.5}$ & $\errorbars{19.8}{1.3}{0.8}$ & $\errorbars{0.4}{0.7}{0.2}$ & $\errorbars{21.2}{2.3}{2.1}$ & $\errorbars{1.2}{0.7}{0.4}$ & $\errorbars{22.1}{2.5}{1.7}$ \\
    $\sim500$ & $\errorbars{2.0}{0.7}{0.5}$ & $\errorbars{19.5}{0.5}{0.3}$ & $\errorbars{1.6}{3.0}{0.6}$ & $\errorbars{19.6}{0.5}{0.4}$ & $\errorbars{1.7}{0.5}{0.4}$ & $\errorbars{19.8}{1.0}{0.5}$ \\
    $\sim5000$ & $\errorbars{1.8}{0.2}{0.1}$ & $\errorbars{19.3}{0.3}{0.1}$ & $\errorbars{1.8}{1.4}{0.3}$ & $\errorbars{19.5}{0.3}{0.2}$ & $\errorbars{1.8}{0.1}{0.1}$ & $\errorbars{19.4}{0.4}{0.2}$ \\
    \hline
    \end{tabular}
    \end{center}
    
    \begin{center}
    \renewcommand{\arraystretch}{1.2}
    \begin{tabular}{|c||c|c||c|c||c|c|}
    \multicolumn{7}{c}{$\boldsymbol{\mathbf{II.\,\,\gamma=1,\,r_s=0.2\,kpc,\,M_{200}\approx1.5\times10^7\,M_\odot,\,\log_{10}\left[J(0.5^\circ)/(GeV^2cm^{-5})\right]\approx 17.3}}$}
    \vspace{1.5mm} \\
    \hline
    & \multicolumn{2}{c||}{Benchmark} & \multicolumn{2}{c||}{$R<R_{1/2}$} & \multicolumn{2}{c|}{$R>R_{1/2}$}\\
    \hline
    $\nstarssel$ & $M_{200}\,[10^7\,M_\odot]$ & $\log_{10}[J(0.5^\circ)]$ & $M_{200}\,[10^7\,M_\odot]$ & $\log_{10}[J(0.5^\circ)]$ & $M_{200}\,[10^7\,M_\odot]$ & $\log_{10}[J(0.5^\circ)]$ \\
    \hline
    $\sim50$ & $\errorbars{1.0}{1.7}{0.5}$ & $\errorbars{18.1}{1.1}{0.7}$ & $\errorbars{0.5}{1.5}{0.2}$ & $\errorbars{18.3}{1.4}{1.5}$ & $\errorbars{0.8}{0.6}{0.2}$ &  $\errorbars{19.4}{1.7}{1.3}$ \\
    $\sim500$ & $\errorbars{1.4}{0.5}{0.3}$ & $\errorbars{17.4}{0.3}{0.2}$ & $\errorbars{1.9}{6.8}{0.9}$ & $\errorbars{17.5}{0.3}{0.2}$ & $\errorbars{1.4}{0.4}{0.3}$ & $\errorbars{17.6}{0.7}{0.4}$ \\
    $\sim5000$ & $\errorbars{1.5}{0.2}{0.1}$ & $\errorbars{17.4}{0.2}{0.1}$ & $\errorbars{1.2}{1.9}{0.2}$ & $\errorbars{17.5}{0.2}{0.1}$ & $\errorbars{1.5}{0.1}{0.1}$ & $\errorbars{17.3}{0.3}{0.2}$ \\
    \hline
    \end{tabular}
    \end{center}
    
    \begin{center}
    \renewcommand{\arraystretch}{1.2}
    \begin{tabular}{|c||c|c||c|c||c|c|}
    \multicolumn{7}{c}{$\boldsymbol{\mathbf{III.\,\,\gamma=0,\,r_s=1\,kpc,\,M_{200}\approx1.4\times10^9\,M_\odot,\,\log_{10}\left[J(0.5^\circ)/(GeV^2cm^{-5})\right]\approx 17.9}}$}
    \vspace{1.5mm} \\
    \hline
    & \multicolumn{2}{c||}{Benchmark} & \multicolumn{2}{c||}{$R<R_{1/2}$} & \multicolumn{2}{c|}{$R>R_{1/2}$}\\
    \hline
    $\nstarssel$ & $M_{200}\,[10^9\,M_\odot]$ & $\log_{10}[J(0.5^\circ)]$ & $M_{200}\,[10^9\,M_\odot]$ & $\log_{10}[J(0.5^\circ)]$ & $M_{200}\,[10^9\,M_\odot]$ & $\log_{10}[J(0.5^\circ)]$ \\
    \hline
    $\sim50$ & $\errorbars{1.6}{2.5}{0.9}$ & $\errorbars{18.5}{0.8}{0.5}$ & $\errorbars{0.2}{0.8}{0.1}$ & $\errorbars{19.4}{2.4}{1.7}$ & $\errorbars{0.9}{1.7}{0.5}$ & $\errorbars{19.7}{2.1}{1.4}$ \\
    $\sim500$ & $\errorbars{1.6}{0.8}{0.4}$ & $\errorbars{18.2}{0.4}{0.2}$ & $\errorbars{4.0}{13.5}{2.9}$ & $\errorbars{18.1}{0.3}{0.2}$ & $\errorbars{1.6}{0.5}{0.4}$ & $\errorbars{18.3}{0.5}{0.4}$ \\
    $\sim5000$ & $\errorbars{1.5}{0.2}{0.1}$ & $\errorbars{18.1}{0.3}{0.1}$ & $\errorbars{3.6}{8.3}{2.3}$ & $\errorbars{18.1}{0.2}{0.1}$ & $\errorbars{1.5}{0.2}{0.1}$ & $\errorbars{18.2}{0.4}{0.2}$ \\
    \hline
    \end{tabular}
    \end{center}
    
    \begin{center}
    \renewcommand{\arraystretch}{1.2}
    \begin{tabular}{|c||c|c||c|c||c|c|}
    \multicolumn{7}{c}{$\boldsymbol{\mathbf{IV.\,\,\gamma=0,\,r_s=0.2\,kpc,\,M_{200}\approx1.1\times10^7\,M_\odot,\,\log_{10}\left[J(0.5^\circ)/(GeV^2cm^{-5})\right]\approx 16.3}}$}
    \vspace{1.5mm} \\
    \hline
    & \multicolumn{2}{c||}{Benchmark} & \multicolumn{2}{c||}{$R<R_{1/2}$} & \multicolumn{2}{c|}{$R>R_{1/2}$}\\
    \hline
    $\nstarssel$ & $M_{200}\,[10^7\,M_\odot]$ & $\log_{10}[J(0.5^\circ)]$ & $M_{200}\,[10^7\,M_\odot]$ & $\log_{10}[J(0.5^\circ)]$ & $M_{200}\,[10^7\,M_\odot]$ & $\log_{10}[J(0.5^\circ)]$ \\
    \hline
    $\sim50$ & $\errorbars{1.6}{3.2}{0.9}$ & $\errorbars{16.7}{0.5}{0.3}$ & $\errorbars{0.2}{0.7}{0.1}$ & $\errorbars{17.4}{1.6}{1.4}$ & $\errorbars{0.9}{1.0}{0.4}$ & $\errorbars{17.2}{1.3}{0.7}$ \\
    $\sim500$ & $\errorbars{1.2}{0.4}{0.2}$ & $\errorbars{16.4}{0.3}{0.1}$ & $\errorbars{2.1}{6.8}{1.3}$ & $\errorbars{16.6}{0.2}{0.1}$ & $\errorbars{1.4}{0.4}{0.3}$ & $\errorbars{16.5}{0.4}{0.2}$ \\
    $\sim5000$ & $\errorbars{1.1}{0.1}{0.1}$ & $\errorbars{16.5}{0.2}{0.2}$ & $\errorbars{2.3}{5.4}{1.3}$ & $\errorbars{16.5}{0.1}{0.1}$ & $\errorbars{1.2}{0.1}{0.1}$ & $\errorbars{16.5}{0.3}{0.1}$ \\
    \hline
    \end{tabular}
    \end{center}
    \caption{Inferred values of the virial mass $M_{200}$ and {\jfac} for the different parameter sets, selected sample sizes, and spatial selection functions, with $\Delta v=0$ in all cases. The {\jfac}s are in units of $\log_{10}[J(0.5^\circ)/(\mathrm{GeV}^2\mathrm{cm}^{-5})]$. Each entry in this table represents the median across 10 realizations of the median and $\pm1\sigma$ values. }
    \label{tab:mvir_Jfacs_drop_stars}
\end{table}

\end{document}